\documentclass[twocolumn]{aastex62}
\pdfoutput=1 
\usepackage{epstopdf} 
\usepackage{url}

\usepackage{tabularx}
\usepackage{comment}
\usepackage[caption=false]{subfig}  
\usepackage{graphicx}
\usepackage{xcolor}

\usepackage{array,multirow}

\usepackage[T1]{fontenc}

\graphicspath{{./}{paprfigs/}}

\shorttitle{A classification algorithm for time-domain novelties}
\shortauthors{Soraisam et al.}

\begin{document}

\title{A classification algorithm for time-domain novelties in preparation for LSST alerts\\Application to variable stars and transients detected with DECam in the Galactic Bulge}

\correspondingauthor{Monika Soraisam}
\email{soraisam@illinois.edu}

\author{Monika D.~Soraisam}
\affiliation{National Center for Supercomputing Applications, University of Illinois at Urbana-Champaign, Urbana, IL 61801, USA}
\affiliation{Department of Astronomy, University of Illinois at Urbana-Champaign, Urbana, IL 61801, USA}
\author{Abhijit Saha}
\affiliation{NSF's National Optical-Infrared Astronomy Research Laboratory, Tucson, AZ 85719, USA}
\author{Thomas Matheson}
\affiliation{NSF's National Optical-Infrared Astronomy Research Laboratory, Tucson, AZ 85719, USA}
\author{Chien-Hsiu Lee}
\affiliation{NSF's National Optical-Infrared Astronomy Research Laboratory, Tucson, AZ 85719, USA}
\author{Gautham Narayan}
\affiliation{Department of Astronomy, University of Illinois at Urbana-Champaign, Urbana, IL 61801, USA}
\author{A.~Katherina Vivas}
\affiliation{Cerro Tololo Inter-American Observatory, NSF's National Optical-Infrared Astronomy Research Laboratory, Casilla 603, La Serena, Chile}
\author{Carlos Scheidegger}
\affiliation{Department of Computer Science, The University of Arizona, 1040 E. 4th St., Tucson, AZ 85721, USA}
\author{Niels Oppermann}
\affiliation{Tucson, AZ 85719, USA}
\author{Edward W.~Olszewski}
\affiliation{Steward Observatory, The University of Arizona, Tucson, AZ 85721, USA}
\author{Sukriti Sinha}
\affiliation{The University of Arizona, Tucson, USA}
\author{Sarah R.~DeSantis}
\affiliation{The University of Arizona, Tucson, USA}
\collaboration{(ANTARES collaboration)}

\begin{abstract}

With the advent of the Large Synoptic Survey Telescope (LSST), time-domain astronomy will be faced with an unprecedented volume and rate of data. Real-time processing of variables and transients detected by such large-scale surveys is critical to identifying the more unusual events and allocating scarce follow-up resources efficiently. We develop an algorithm to identify these novel events within a given population of variable sources. We determine the distributions of magnitude changes ($dm$) over time intervals ($dt$) for a given passband $f$, $p_f(dm|dt)$, and use these distributions to compute the likelihood of a test source being consistent with the population, or an outlier. We demonstrate our algorithm by applying it to the DECam multi-band time-series data of more than 2000 variable stars identified by Saha et al. (2019) in the Galactic Bulge that are largely dominated by long-period variables and pulsating stars. Our algorithm discovers 18 outlier sources in the sample, including a microlensing event, a dwarf nova, and two chromospherically active RS~CVn stars, as well as sources in the Blue Horizontal Branch region of the color-magnitude diagram without any known counterparts. We compare the performance of our algorithm for novelty detection with multivariate KDE and Isolation Forest on the simulated PLAsTiCC dataset. We find that our algorithm yields comparable results despite its simplicity. Our method provides an efficient way for flagging the most unusual events in a real-time alert-broker system.

\end{abstract}


\section{Introduction} \label{sec:intro}
Fast, wide-field ground-based optical surveys are redefining time-domain astronomy as a field driven by Big Data. The Zwicky Transient Facility (ZTF; \citealt{Bellm-2019, Masci-2019}) and the upcoming Large Synoptic Survey Telescope (LSST; \citealt{Ivezic-2019}) are notable surveys poised to revolutionize time-domain astronomy in this way. The data deluge from these surveys, particularly that expected from LSST, has driven the community to automate all aspects of its processing. This includes difference imaging for detection of varying sources or {\em alerts} \citep[e.g.,][]{Alard-1998,Bramich-2008, Zackay-2016}, different machine-learning algorithms to remove artifacts in the detected sources~\citep{Brink-2013,Goldstein-2015,Masci-2017,Cabrera-2017}, as well as classification of the sources themselves~\citep[e.g.,][]{Richards-2011,Bloom-2012,Mahabal-2017,Narayan-2018}. Even follow-up of interesting sources can now be automated using tools like the Target and Observation Manager \citep{Street-2018}.

The need for real-time processing of the alerts at a rate of around 10,000 {\it per minute} expected for LSST{\footnote{\url{https://dmtn-102.lsst.io/DMTN-102.pdf}}} motivates the design of alert-broker systems. These systems seek to enable time-critical science, including early-time detection of supernovae that would inform different progenitor models, lensed supernovae, short-lived rare events such as kilonovae, and yet-unknown transient and variable phenomena. We have developed the Arizona-NOAO Temporal Analysis and Response to Events System (ANTARES; \citealt{Saha-2016}) as a community alert-broker to accommodate the real-time needs of the various time-domain astrophysical applications. ANTARES\footnote{\url{https://antares.noao.edu/}} is already online, processing the public alert stream of the ZTF survey, annotating the alert data with contextual information via cross-identification with different legacy surveys and characterizing alerts based on basic statistical properties such as amplitude, detection significance, etc.

Different authors have worked on building efficient machine-learning algorithms, particularly aimed toward real-time classification of alerts in the upcoming LSST era. For example, \citet{Muthukrishna-2019} developed the Real-time Automated Photometric IDentification tool based on a deep recurrent neural network for early-time classification of transients. They achieved an accuracy of more than $95\%$ on classifying 12 types of transients using simulated data with ZTF observing characteristics, and also accurately identified three real transients from ZTF.

Most of the machine-learning algorithms in time-domain astronomy developed to-date, however, focus on {\em classification} of the sources, which requires well-labeled training data sets. This requirement makes such algorithms suboptimal for two applications---(1) identification of unusual events, and (2) situations in which no training data are available, for example, because the survey is exploring a new parameter space. The latter is the case for LSST, which will be observing more deeply than many of its preceding surveys. One of the data products that the community most desires from broker systems is an output stream of candidate novel events for immediate follow-up.

In this paper, we present a new algorithm to identify novel events from a population of variable stars. The peculiarity of the source may relate to its intrinsic rarity (in terms of rates and numbers) or to its unusual appearance as an outlier member of a larger population of sources. We demonstrate our algorithm using a real astronomical time-series data set in multiple passbands similar to the LSST passbands and adopting computationally inexpensive features that make it well-suited for incorporation into the broker ecosystem. The paper is organized as follows. In Sect.~\ref{sec:data}, we present the time-series data used in our study. In Sect.~\ref{sec:method}, we describe our algorithm and show its results when applied to astronomical time-series data. We discuss the characteristics of the identified unusual sources in Sect.~\ref{sec:char}. We assess the performance of our algorithm in Sect.~\ref{sec:eff} and end with a summary of the results in Sect.~\ref{sec:summary}.

 \begin{figure*}
\includegraphics[width=88mm]{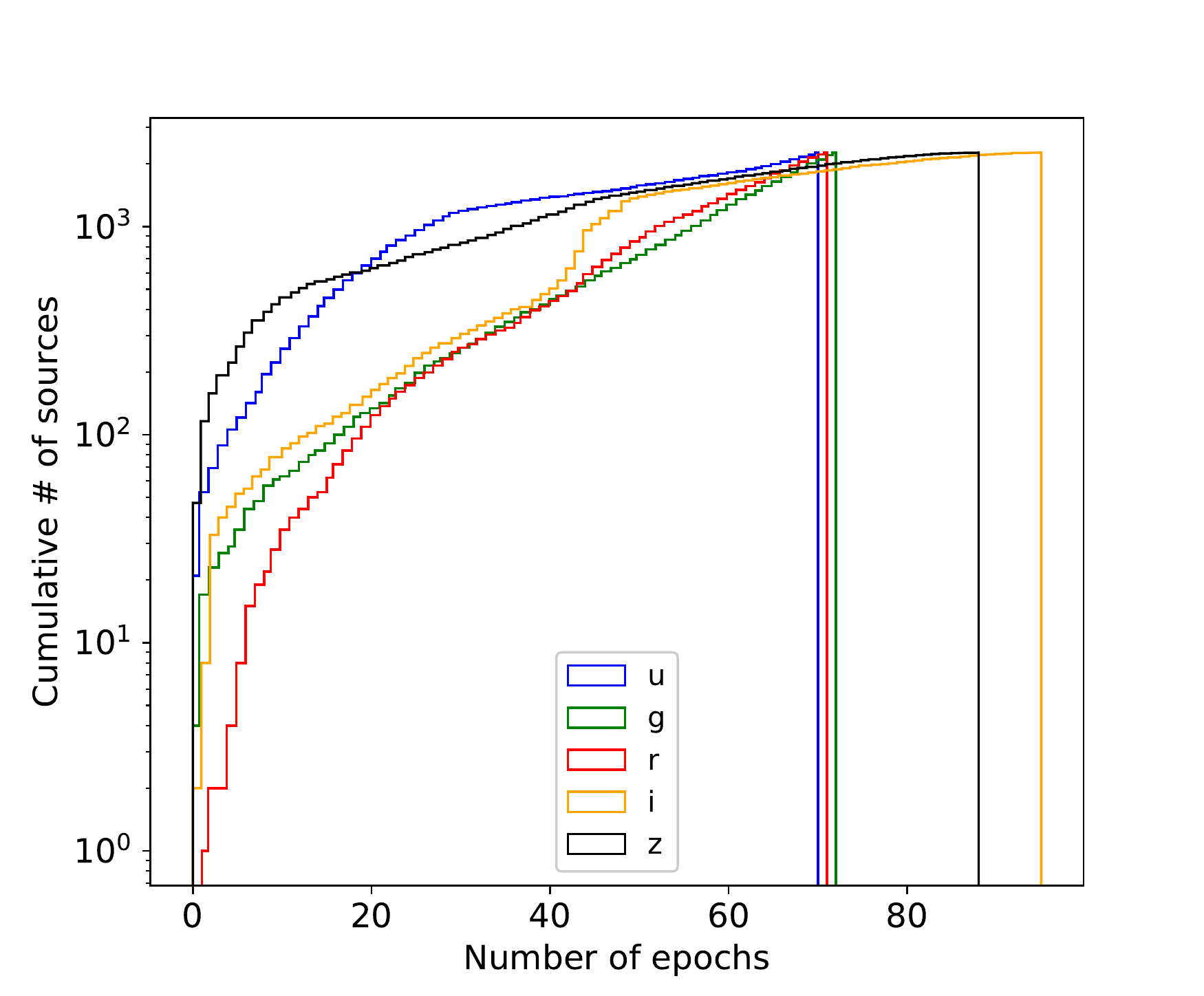}\hfil\includegraphics[width=88mm]{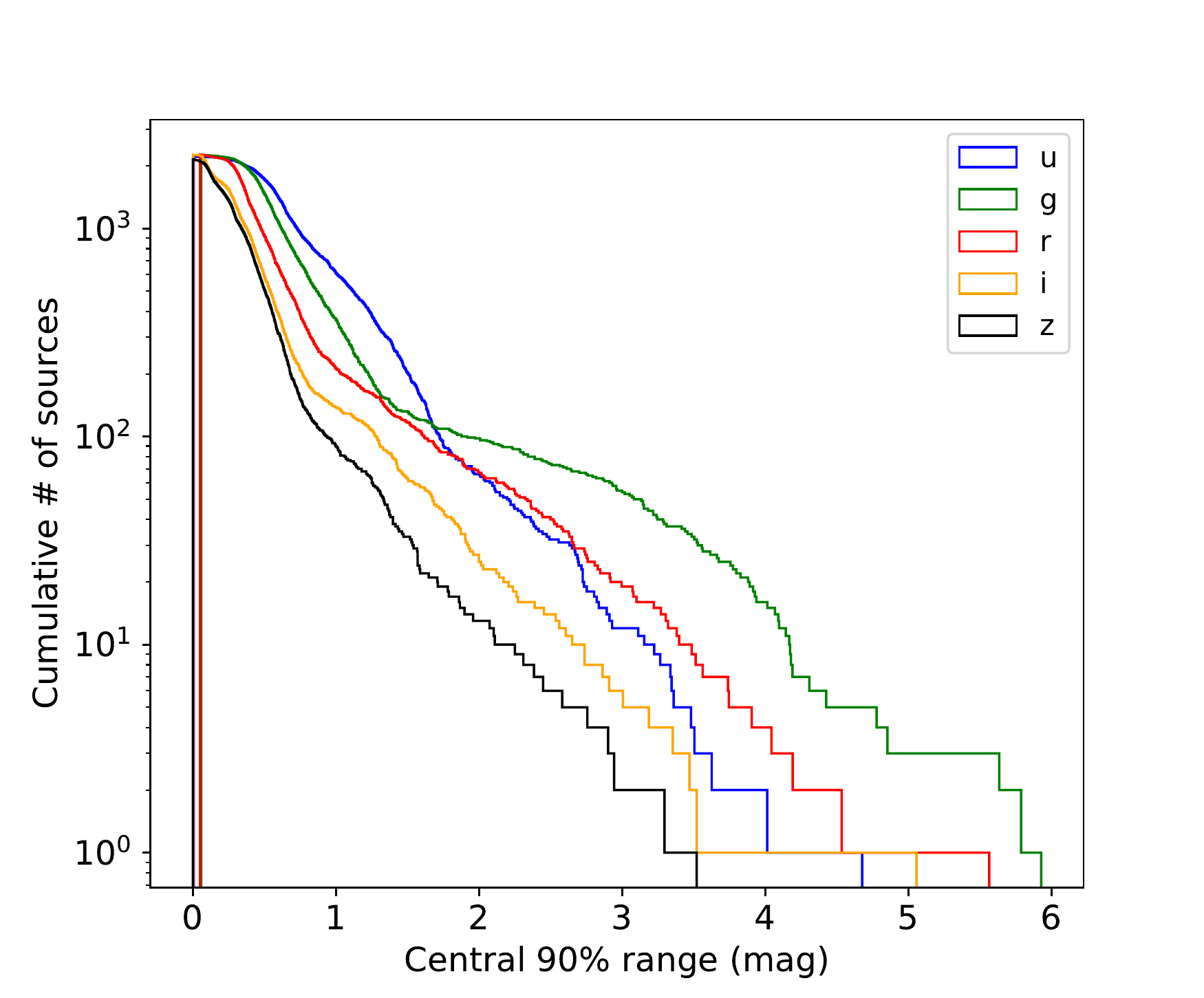}
\caption{{\it Left}: Cumulative distribution of the number of observations obtained for each variable star. 
{\it Right}: Cumulative distribution of the central 90\% ranges of the magnitude distributions for the 2266 sources studied here. The different colors correspond to different passbands as indicated by the legends in the panels.}
\label{fig:epochs}
\end{figure*}

\section{Time-domain data sample} \label{sec:data}
The time-series data we use here are obtained from an NOAO observing program for conducting a deep synoptic survey of the Galactic bulge (2013A-0719; PI: A.~Saha) using the DECam imager \citep{Flaugher-2015} mounted atop the Victor~M.~Blanco 4\,m telescope at the Cerro Tololo Inter-American Observatory. The program imaged six separate fields (each of area $3\,{\rm deg}^{2}$, corresponding to the imager's field of view) toward the Galactic center, named B1--B6, in five passbands {\it u}, {\it g}, {\it r}, {\it i}, {\it z} over multiple epochs extending from 2013 until 2015 (seven nights in 2013 and three nights in 2015). A mix of two different exposure times---short (several seconds) and long (hundreds of seconds)---were used to accrue a large dynamical range of the magnitudes, reaching $i>23$~mag, ultimately limited by confusion noise due to crowding. 
In particular, the field B1 is centered on ``Baade's Window''---RA 18h03m34.0s, DEC -30d02m02.0s (J2000)---which provides a low-extinction window to probe near the center of our Galaxy. \citet{Saha-2019} presented an analysis of the data for the field B1, including source detection, photometry, and variability assessment for the detected sources. Their work has derived a new reddening law different from the standard $R_{V}=3.1$ toward the Galactic Bulge and produced a sub-arcminute-resolution line-of-sight reddening map.

The criteria used by Saha et al.\ to flag a star as variable include cleaning pathological measurements for a given passband and then performing a reduced chi-square ($\chi^{2}_{\nu}$) analysis following \citet{Saha-1990}. Generally, the measurement errors do not strictly follow a Gaussian distribution and are subject to bias such that the expectation value of $\chi^{2}_{\nu}$ becomes a function of the star's brightness. Saha et al. accounted for this by computing the mode of $\log \chi^{2}_{\nu}$ for a given mean magnitude of the star and set 1.3 times this mode value as the threshold for flagging it as variable. In total, they found 4877 stars to be variable in the B1 field at least in one passband; of these, 2266 stars are variable in two or more passbands. They also computed the individual extinction corrections for these stars using the new high-resolution reddening map. We use the cleaner subsample of 2266 variable stars for our study, which guarantees that variability is real, whether or not it is classifiable. The multi-band light curve data for all these variable stars are given in Table~\ref{tab:lc_data} and also available via the Data Lab science platform of the NSF's OIR Lab.{\footnote{\url{https://datalab.noao.edu/}}}. The magnitudes of these stars used throughout this paper are extinction-corrected assuming they are all located in the Galactic bulge. It is to be noted that, except for RR Lyrae stars, only variability has been established for these stars and not the types they belong to or their ``labels'' (see below).

\begin{table*}
\caption{Light curve data}\label{tab:lc_data}
\renewcommand\arraystretch{1.0}
\renewcommand{\tabcolsep}{14pt}
\centering
\begin{tabularx}{\textwidth}{clllcll}
\hline
Name	&RA (deg)		&DEC (deg)	&${\rm HJD}-2,400,000.0$		&Passband 	&Mag 	&Mag\_error\\
\hline
B1-3063005  &270.67905  &-29.90554  &56423.665618  &u  &16.896  &0.018\\
B1-3063005  &270.67905  &-29.90554   &56423.814222  &u  &15.559  &0.012\\
B1-3063005  &270.67905  &-29.90554   &56423.895484  &u  &15.871  &0.012\\
B1-3063005  &270.67905  &-29.90554   &56424.718744  &u  &15.23  &0.013\\
B1-3063005  &270.67905  &-29.90554   &56424.810501  &u  &16.117  &0.011\\
B1-3063005  &270.67905  &-29.90554   &56424.888096  &u  &16.59  &0.014\\
B1-3063005  &270.67905  &-29.90554  &56450.6106  &u  &16.258  &0.01\\
B1-3063005  &270.67905  &-29.90554  &56451.614623  &u  &16.814  &0.014\\
B1-3063005  &270.67905  &-29.90554   &56451.796018  &u  &16.407  &0.012\\
B1-3063005  &270.67905  &-29.90554   &56452.735363  &u  &15.318  &0.01\\
\hline
\end{tabularx}
\tablenotetext{}{(This is only a part of the table to demonstrate its form and content; the full version is available in machine-readable form online.)}
\end{table*}

The typical baseline in each of the passbands is around 1.7--1.8~years, with the interval between any two observations ranging from approximately 1~hour to 1.7~years. The cumulative distribution of the number of epochs logged in for each variable star is shown in Fig.~\ref{fig:epochs}. The plot shows that there are typically tens of measurements in the light curves of the stars. It can also be seen that the distributions of epochs for the different passbands are dissimilar, and in particular, the bluer passbands have fewer epochs on the whole, as may be expected since observations in these passbands were done only during the dark lunar phases of the observing runs. On the other hand, observations in the $z$-band for many of the sources were affected by confusion and/or saturation.  We  also  give an  estimate  of  the  covered  variability  amplitudes.   For  this,  we  assemble  the magnitude measurements for a given star and calculate the size of the central 90-percent range of these values, i.e., the difference between their 95th and 5th percentiles. In Fig.~\ref{fig:epochs} (right panel) we show the cumulative distributions of this quantity across all sources, separately for the different passbands. As can be seen from the plot, a wide range of amplitudes is covered for the different passbands extending beyond 5~mag. The non-uniform time sampling and the rather sparse sampling of the time-series of these variable stars, complemented by the panchromatic information, make them apt for use in designing algorithms for the characterization of time-variable sources aimed toward LSST.

\begin{figure*}
\includegraphics[width=88mm]{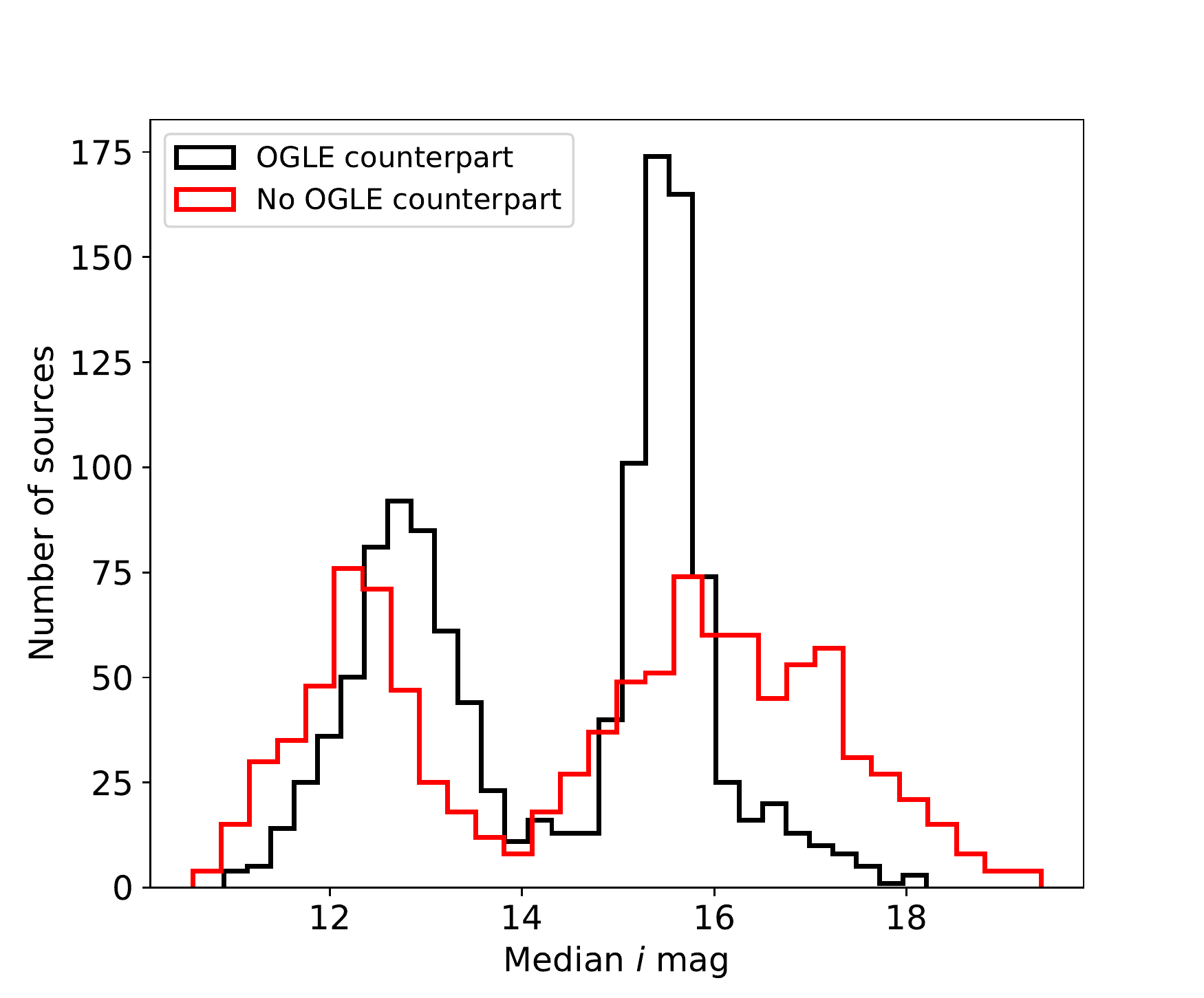}\hfil\includegraphics[width=88mm]{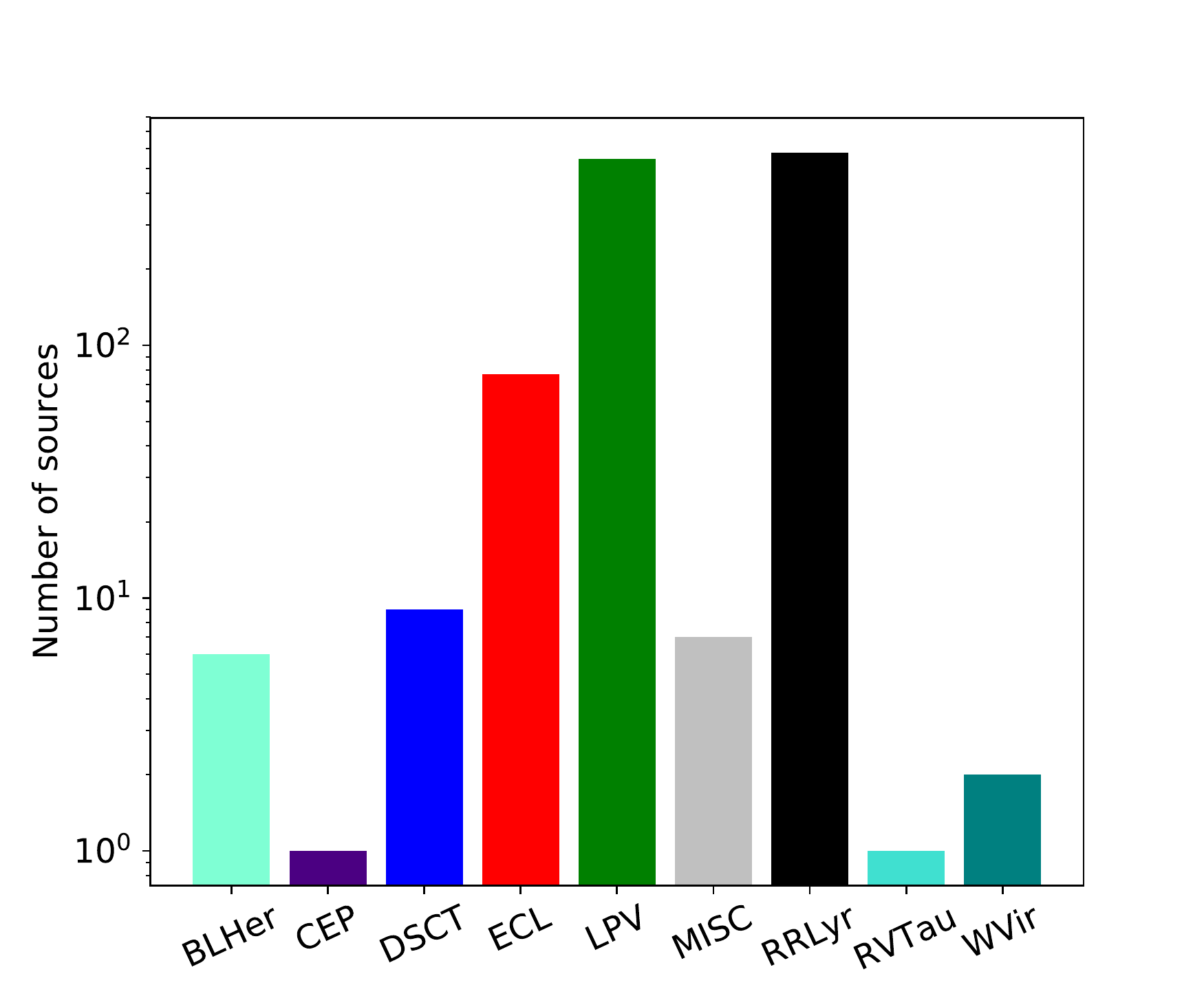}
\caption{{\it Left}: Distribution of the {\it i}-band magnitudes of the variable stars, separated by the availability of labels from OGLE. {\it Right}: Distribution of available labels from OGLE for the variable stars. Note the logarithmic scale on the vertical axis.}
\label{fig:dif_ogle_abi}
\end{figure*}

\begin{figure*}[t]
\includegraphics[width=60mm]{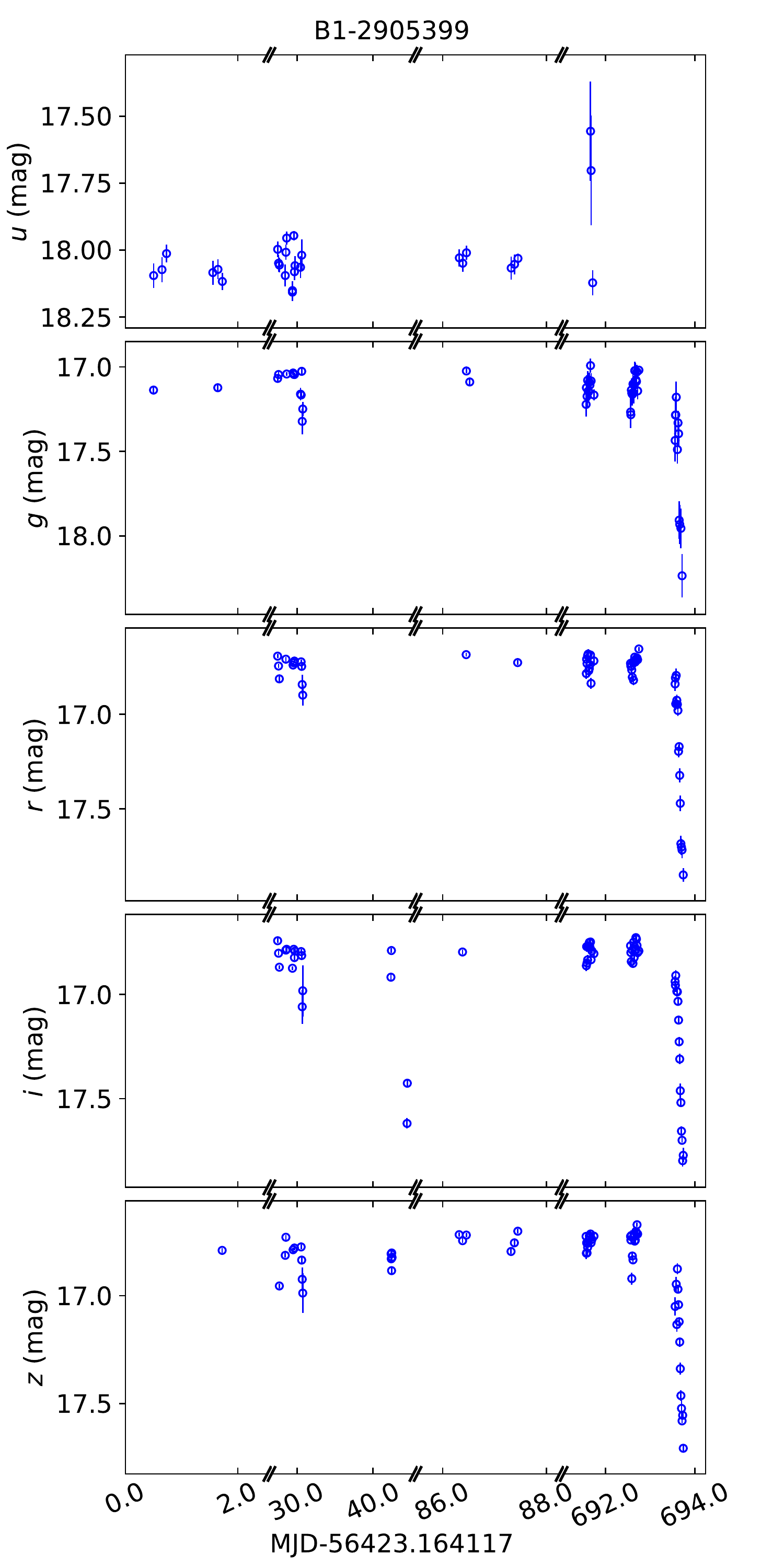}\hfill\includegraphics[width=60mm]{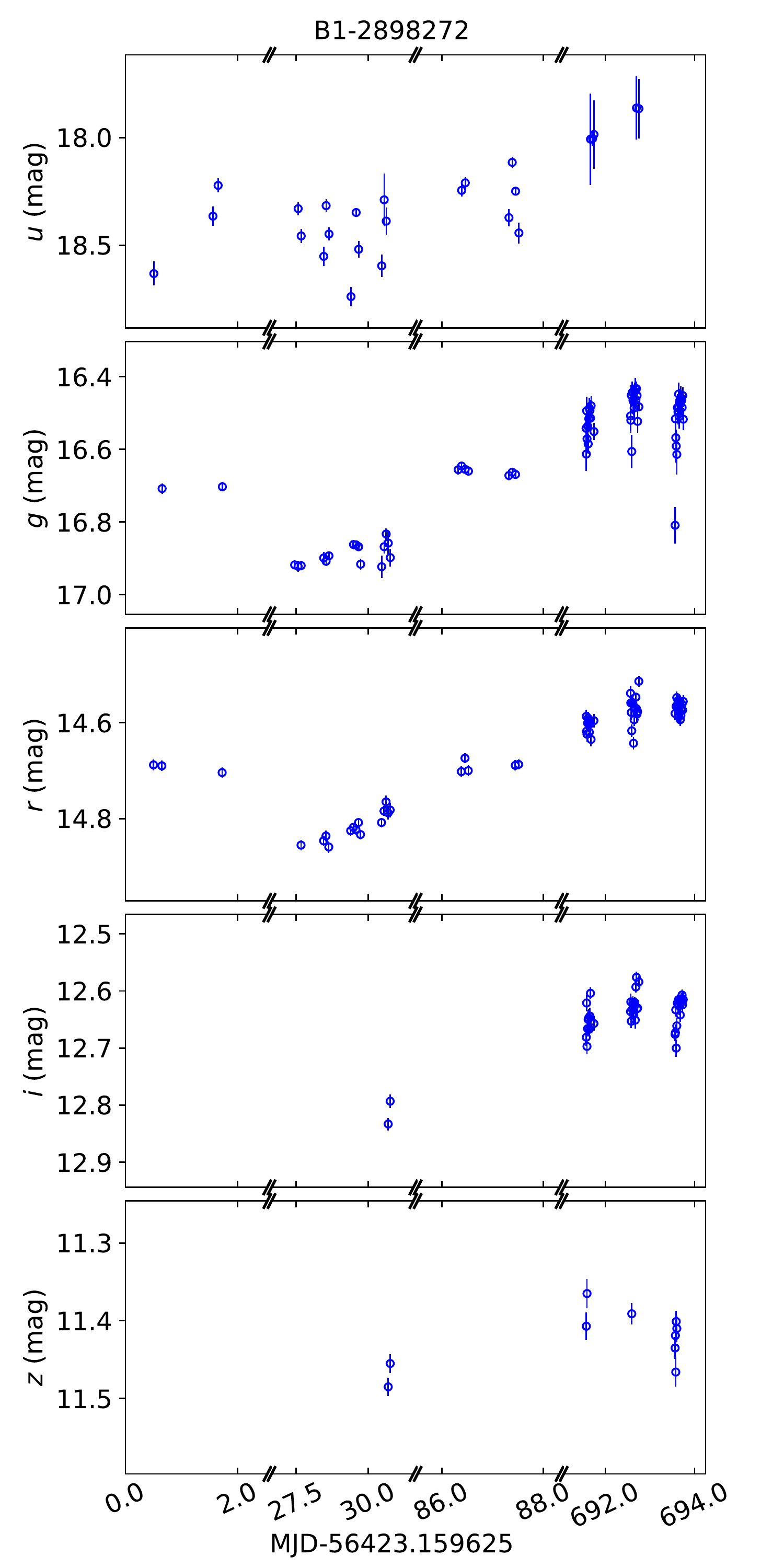}\hfill\includegraphics[width=60mm]{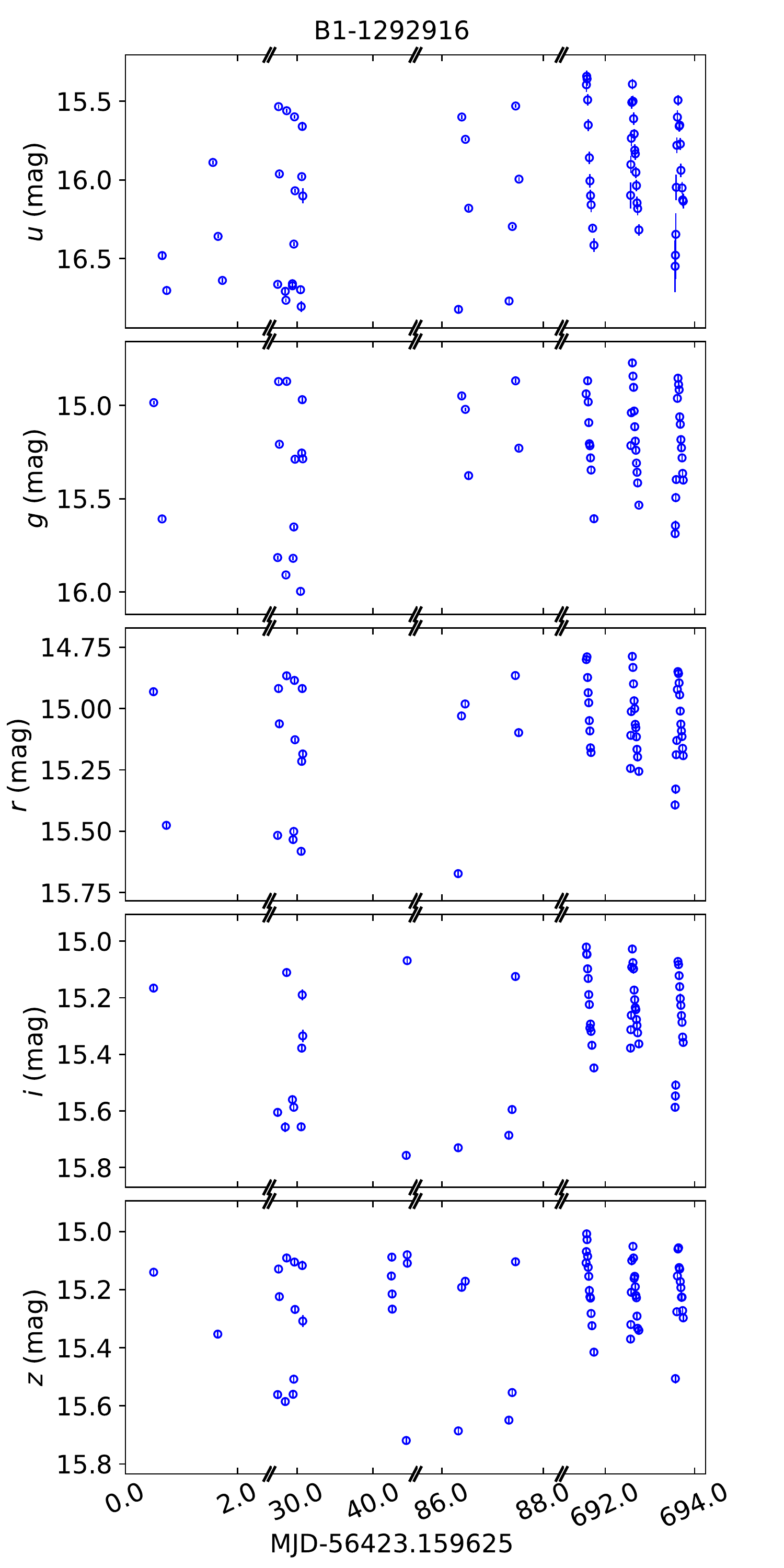}
\caption{Example multi-wavelength light curves of B1 variable stars belonging to the three most populated labels in Fig.~\ref{fig:dif_ogle_abi} (right panel), namely, ECL (left), LPV (middle) and RRLyr (right). The IDs of the stars are shown on top of each plot, and the reported magnitudes are in the AB system. Note that the horizontal axes showing the MJD of observations are broken due to  large temporal gaps. The cadence increased in the last few nights in 2015, as a result of which structures in the light curves are smoother for later epochs.}\label{fig:egs}
\end{figure*}

A good fraction of these stars have been monitored and classified by the Optical Gravitational Lensing Experiment (OGLE; \citealt{Udalski-1992}) survey, which observed mainly in a single passband ({\it I}-band) but with a more dense temporal sampling and a longer baseline. 
Cross-matching with the labeled variable stars in the Galactic Bulge from the OGLE collaboration,\footnote{\url{http://ogle.astrouw.edu.pl/}} specifically the variable-star catalogs from \citet{Udalski-1994, Udalski-1995a, Udalski-1995b, Pietrukowicz-2015, Soszynski-2011, Soszynski-2011b, Soszynski-2013, Soszynski-2014, Soszynski-2015} covering OGLE I, III and IV, we found matches for 1228 (approximately half the sample) of the B1 variables. To avoid misidentification in very crowded field like the Galactic bulge, we use a conservative search radius of $1''$ for this cross-matching. The magnitude distribution in {\it i}-band (most similar to the passband used in most OGLE observations) for these variables are shown along with that for variables without matches in Fig.~\ref{fig:dif_ogle_abi}.

The DECam monitoring program probes deeper than OGLE by around 1--2~mags (see the color-magnitude diagrams in \citealt{Saha-2019}), and is also more sensitive to variability at the bright end through short exposures avoiding saturation. Both of these facts are evident in Fig.~\ref{fig:dif_ogle_abi}. The distribution of the OGLE labels for the cross-matched sources is shown in the right panel of Fig.~\ref{fig:dif_ogle_abi}. We do not distinguish between sub-labels from OGLE (if any); for example, RR Lyrae of types ab, c and d are all simply labeled as RR Lyrae. The OGLE labels represented in the plot include pulsating star types---BL Herculis (BLHer), Cepheids (CEP), Delta Scuti (DSCT), long period variables (LPV), RR Lyrae (RRLyr), RV Tauri (RVTau), W Virginis (WVir)---eclipsing binaries (ECL), and sources of ambiguous type (tagged as miscellaneous, MISC). Example light curves of B1 variable stars in the multiple passbands, for the three most-populated labels (ECL, LPV and RRLyr) are shown in Fig.~\ref{fig:egs}.

\section{Method} \label{sec:method}

There are two steps to the categorization algorithm. The first step is the determination and extraction of features of the variable sources from their light curve data, and the second step comprises identification of the outliers in the feature space. The latter is relatively straightforward when dealing with a low-dimensional feature space.

\begin{figure*}
\includegraphics[width=60mm]{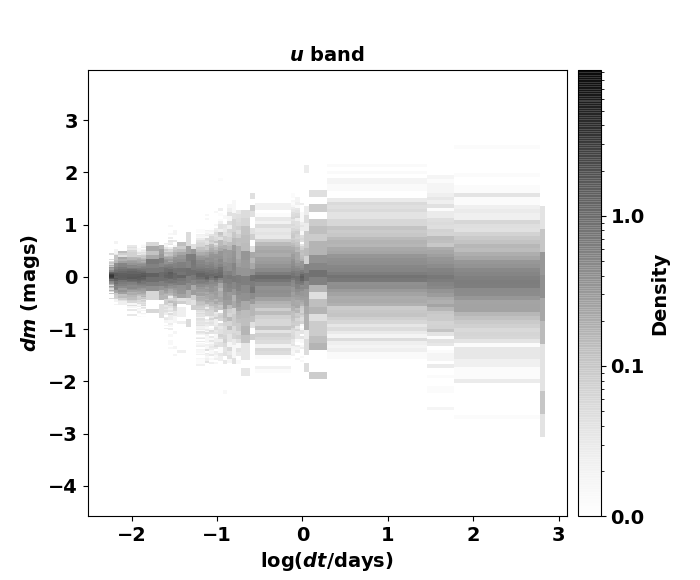}\hfill\includegraphics[width=60mm]{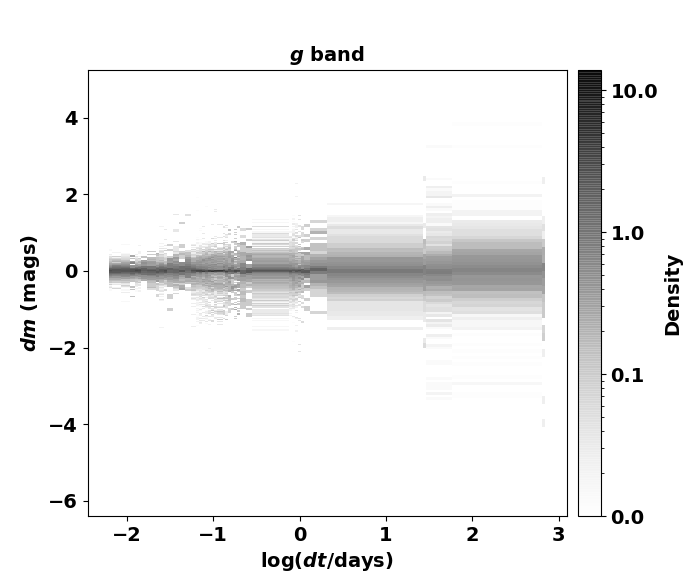}\hfill\includegraphics[width=60mm]{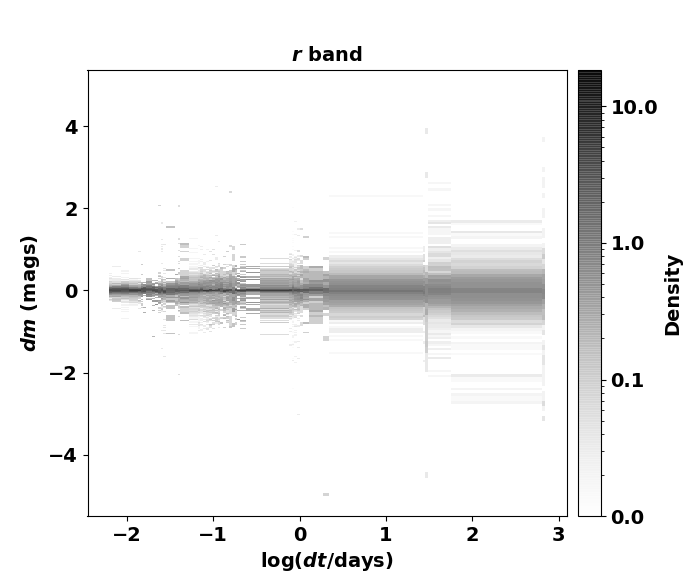}\\
\includegraphics[width=60mm]{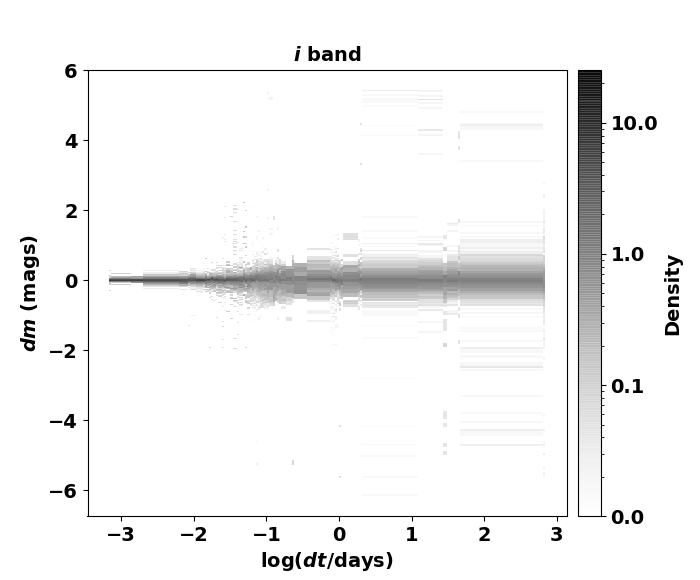}\hfill\includegraphics[width=60mm]{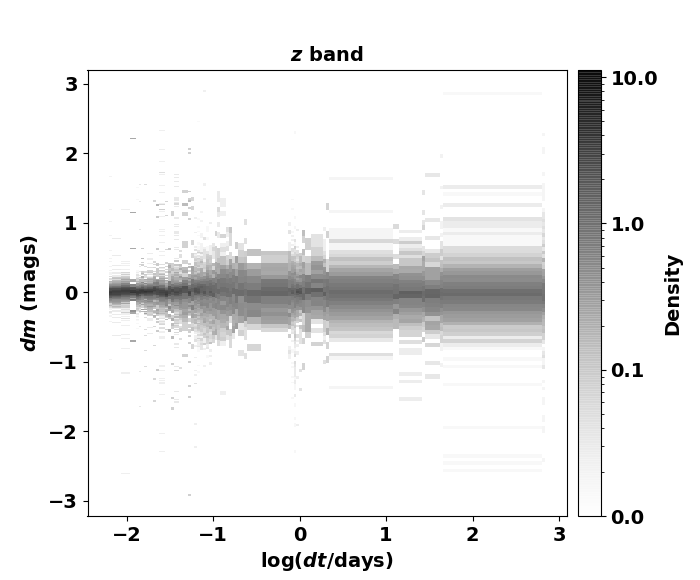}\hfill\includegraphics[width=60mm]{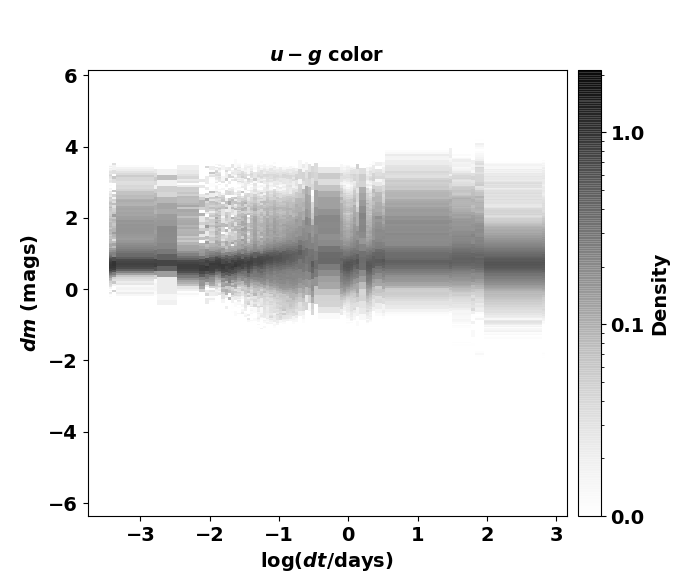}\\
\includegraphics[width=60mm]{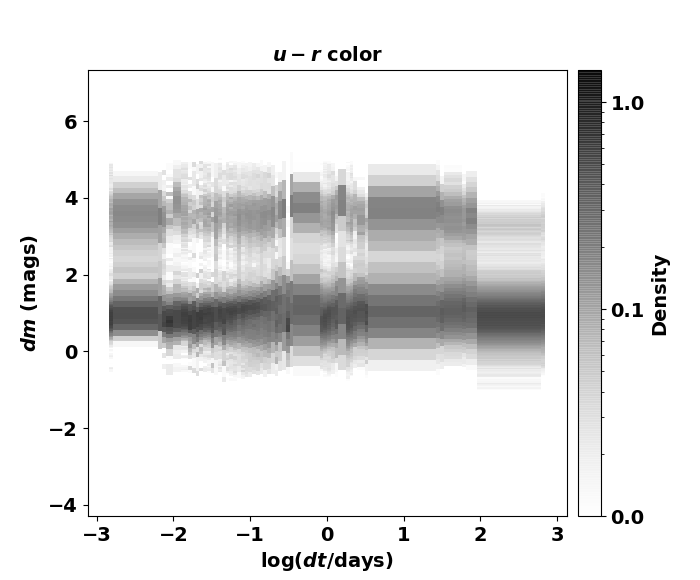}\hfill\includegraphics[width=60mm]{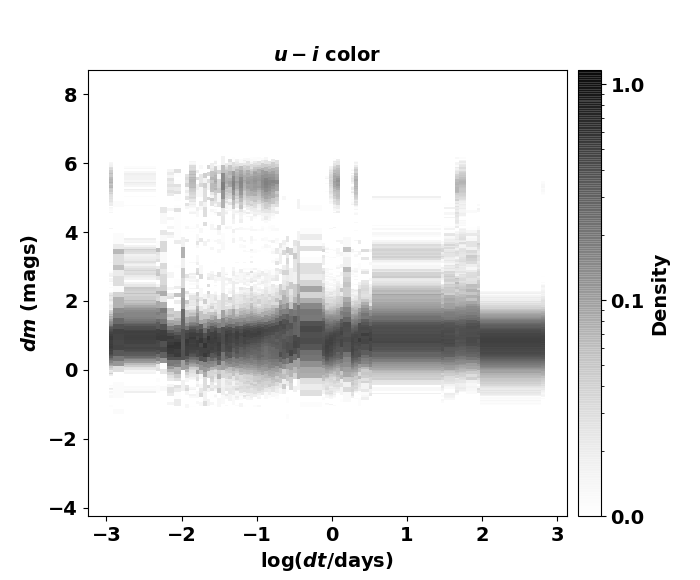}\hfill\includegraphics[width=60mm]{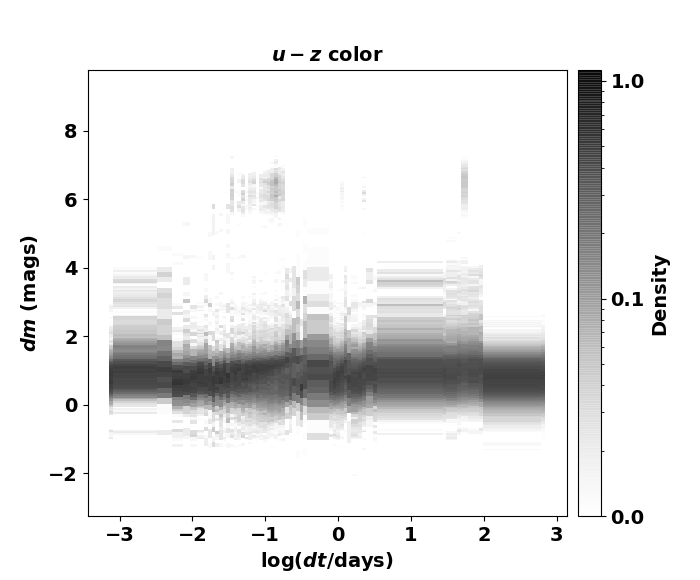}\\
\caption{Distribution of all variable sources in the $\log dt\mbox{--}dm$ spaces. $dm$ represents differential change in magnitudes when considering individual passbands, and asynchronous colors when considering combinations of passbands. The title of each panel indicates the corresponding interpretation of $dm$. It is to be noted that the distributions have been normalized along one axis, specifically with respect to $dm$. For any given set of two passbands (e.g., $u$ and $g$), only one pseudo-color distribution (e.g.,  $u-g$ here) is shown as the other is a mirror image of the first.}\label{fig:dmdt_all}
\end{figure*}

\begin{figure*}
\ContinuedFloat
\includegraphics[width=60mm]{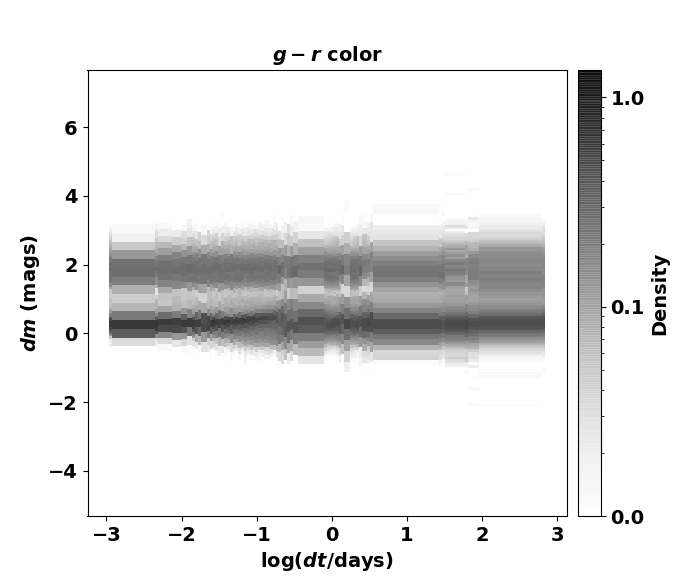}\hfill\includegraphics[width=60mm]{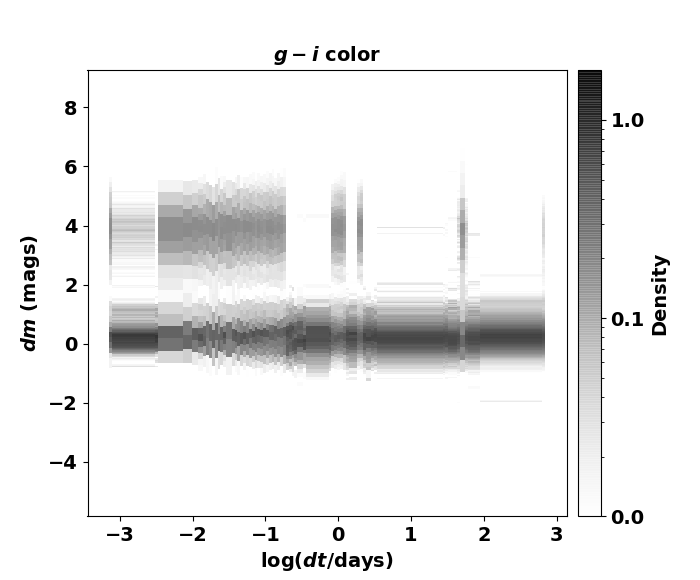}\hfill\includegraphics[width=60mm]{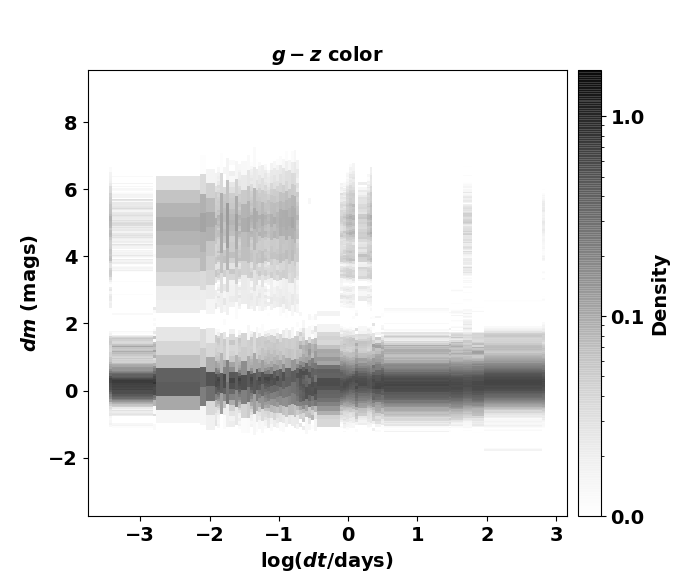}\\
\includegraphics[width=60mm]{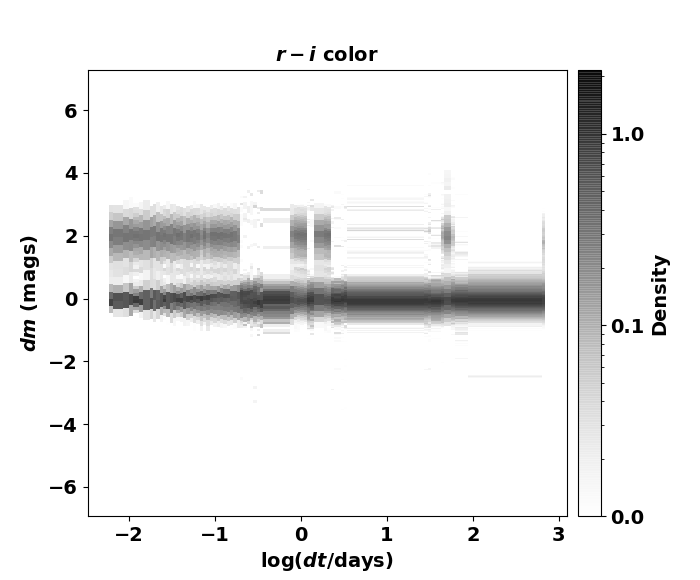}\hfill\includegraphics[width=60mm]{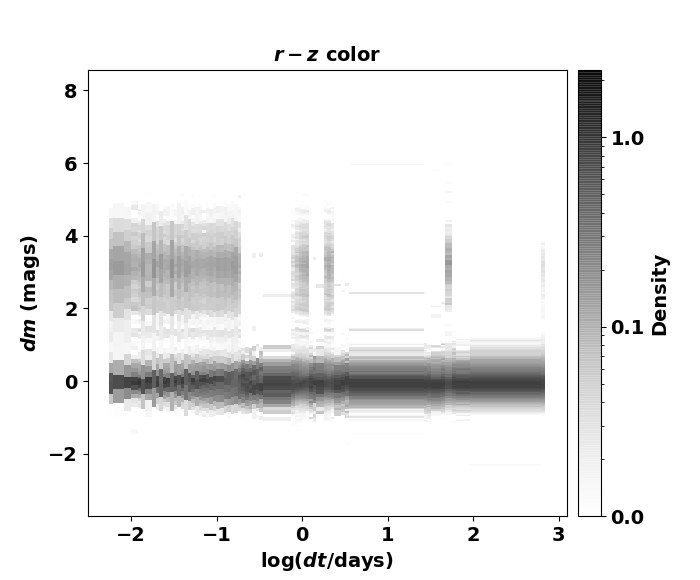}\hfill\includegraphics[width=60mm]{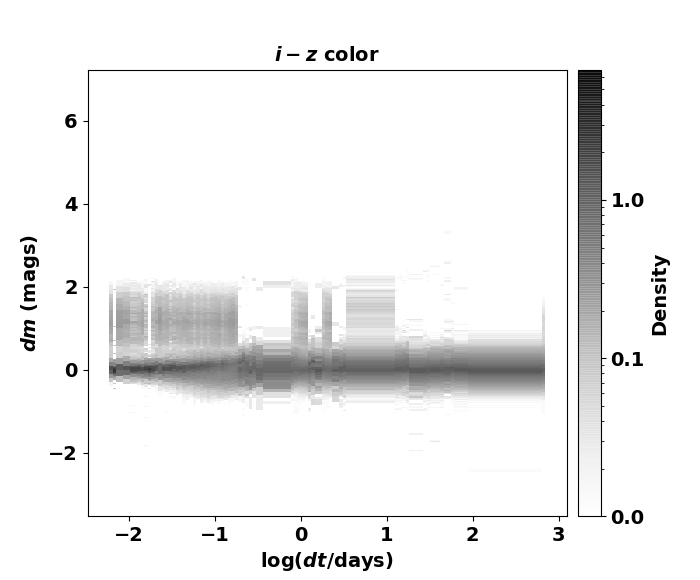}
\caption{Contd.}
\end{figure*}

\subsection{Feature determination}\label{sec:dc-dt}

Our feature engineering step is guided by our choice of parameters that are easily available/derivable and computationally inexpensive; this choice is prudent given the data rate that we will have to keep up with when we apply our algorithm to LSST alerts in real-time.  We use features based on (corresponding) magnitude and time differences, i.e., $dm$ and $dt$, derived from the light curve of a given source. A formalism based on probability distributions $p(dm, dt)$ for the classification of variable stars was introduced by \citet{Mahabal-2011}, and recently \citet{Mahabal-2017} extended it to explore deep-learning techniques, specifically convolutional neural networks (CNN). They mapped the distribution $p(dm,dt)$ of a given variable star's light curve to an image, trained a CNN using such $dm\mbox{--}dt$ images of a sample of labelled variable stars from the Catalina Real-Time Transient Survey, and obtained classification accuracy applying the trained CNN on their test sample similar to that of random forest classifiers. We take a complementary approach that is more easily interpretable and hence less affected by the absence of labelled examples to train the algorithm. We let the data decide for themselves--in a sense our algorithm is unsupervised, leading to two categories of unusual and mundane sources. This is the key difference between the two approaches.

A light curve with measurements in multiple epochs allows us to calculate a number of magnitude differences $dm$ over corresponding time differences $dt$. In the following, we consider the likelihood $(\mathcal{L})$ of the set of these pairs $(dm, dt)$ for a test source to be drawn from the distribution $p(dm, dt)$ assembled from a collection of training sources. Since $dt$ is determined by the cadence of the survey whose data we are analyzing, it is a deterministic variable and thus we consider instead the conditional probability $p(dm|dt)$.  As we have multi-band ($u,g,r,i,z$) data for our variable stars, we construct distributions $p(dm|dt)$ for each passband, as well as across passbands. In the latter, $dm$ is the relative magnitude between two passbands (for example, $g-r$) separated by time $dt$. For small $dt$ values, the relative magnitudes are approximately the colors of the variable sources, however for $dt$ values much greater than the correlation length of variability, they will be uniformly distributed around the relative mean-magnitudes of the two passbands of the given star. The errors in individual magnitude measurements do not enter in the construction of the $p(dm|dt)$ distribution because they only affect the uncertainty in $dm$ and not the estimate for $dm$.

\begin{figure}
\centering
\includegraphics[width=88mm]{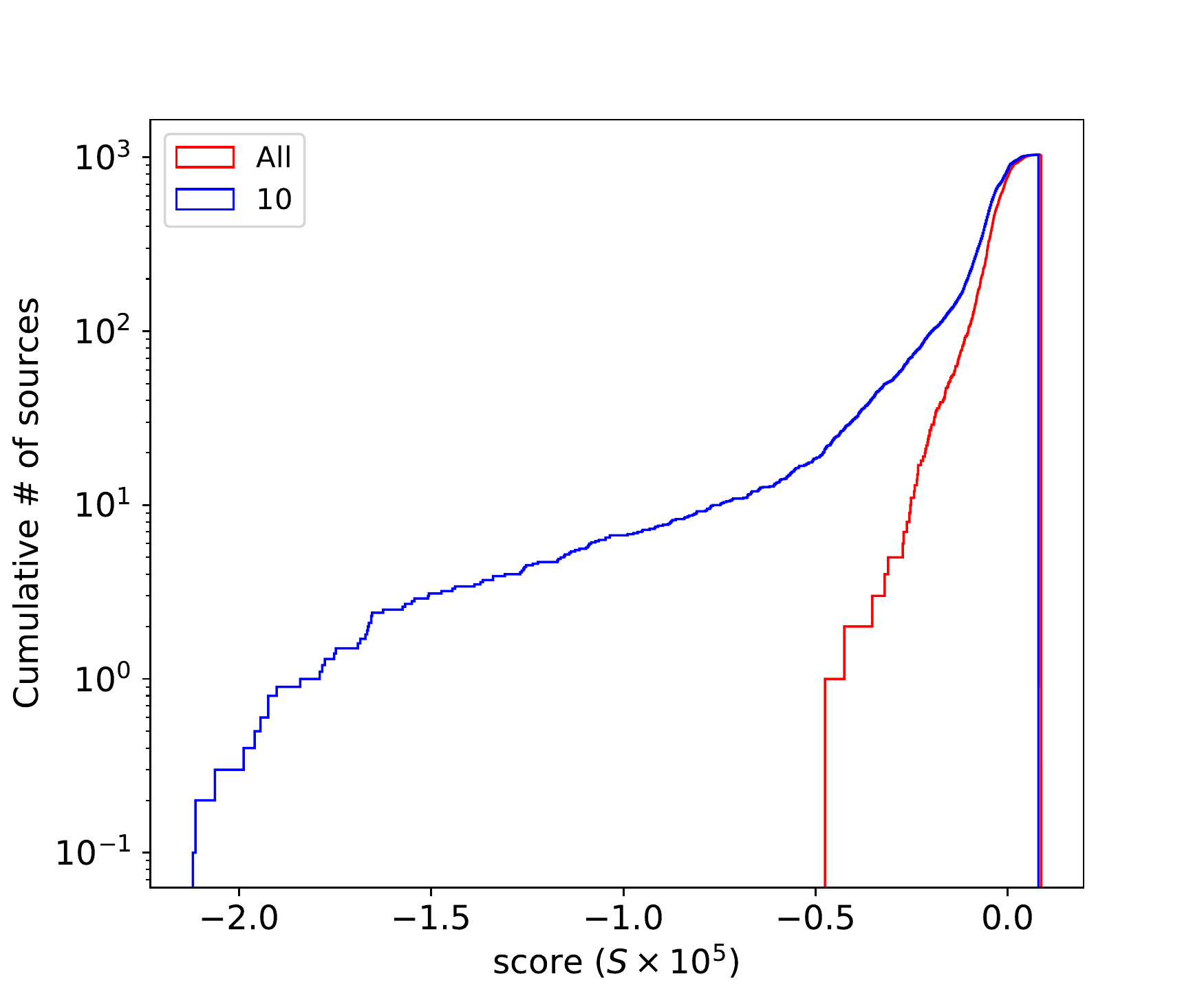}
\caption{Cumulative distribution of scores $S$ (in log unit) for the unlabeled B1-field sources computed using P-all (in red) and P-labeled iterating 10 times (in blue). In the latter case, the distribution has been normalized by the number of iterations.}
\label{fig:Sdist}
\end{figure}

\begin{figure*}
\centering
\includegraphics[width=88mm]{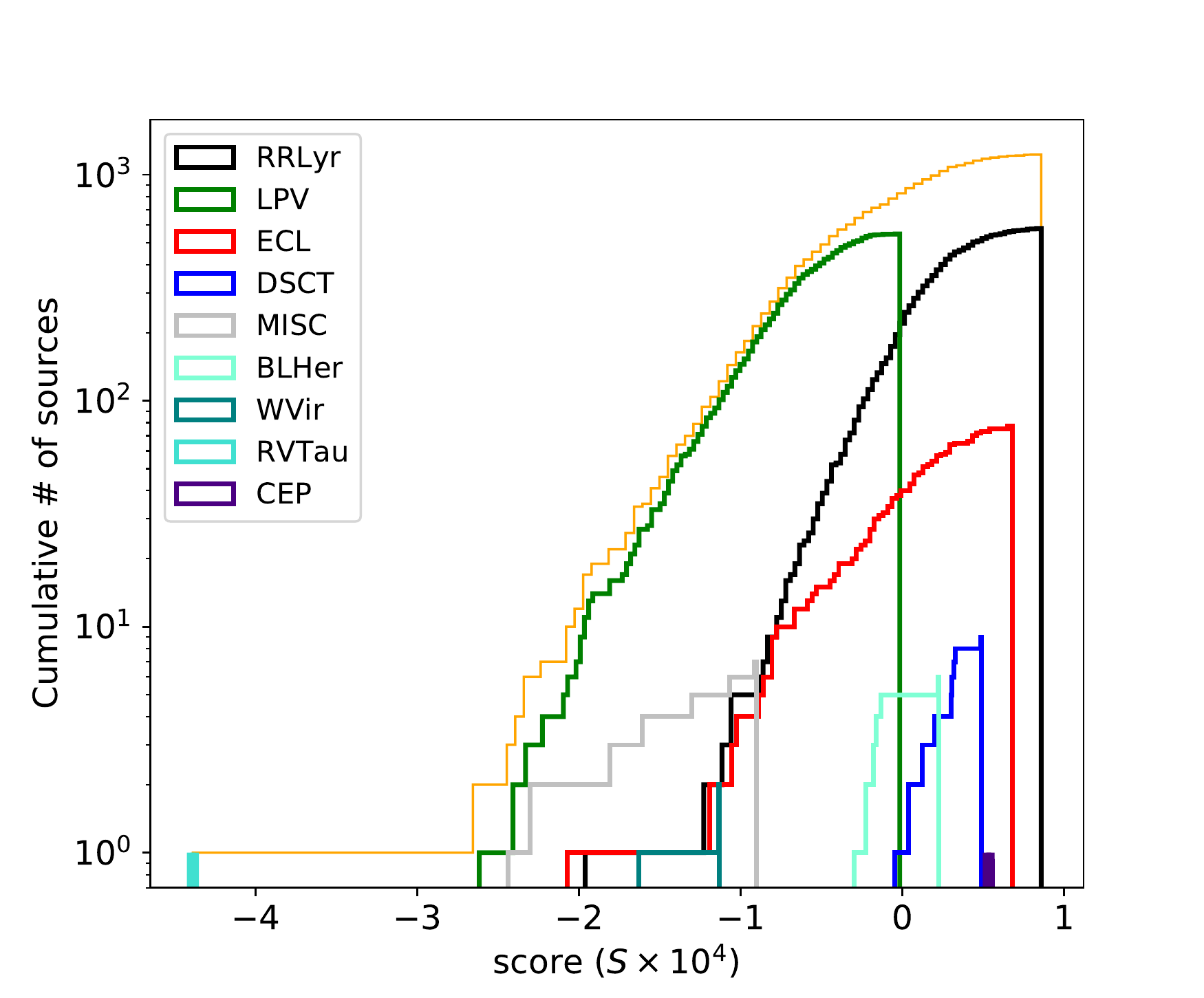}\hfill\includegraphics[width=88mm]{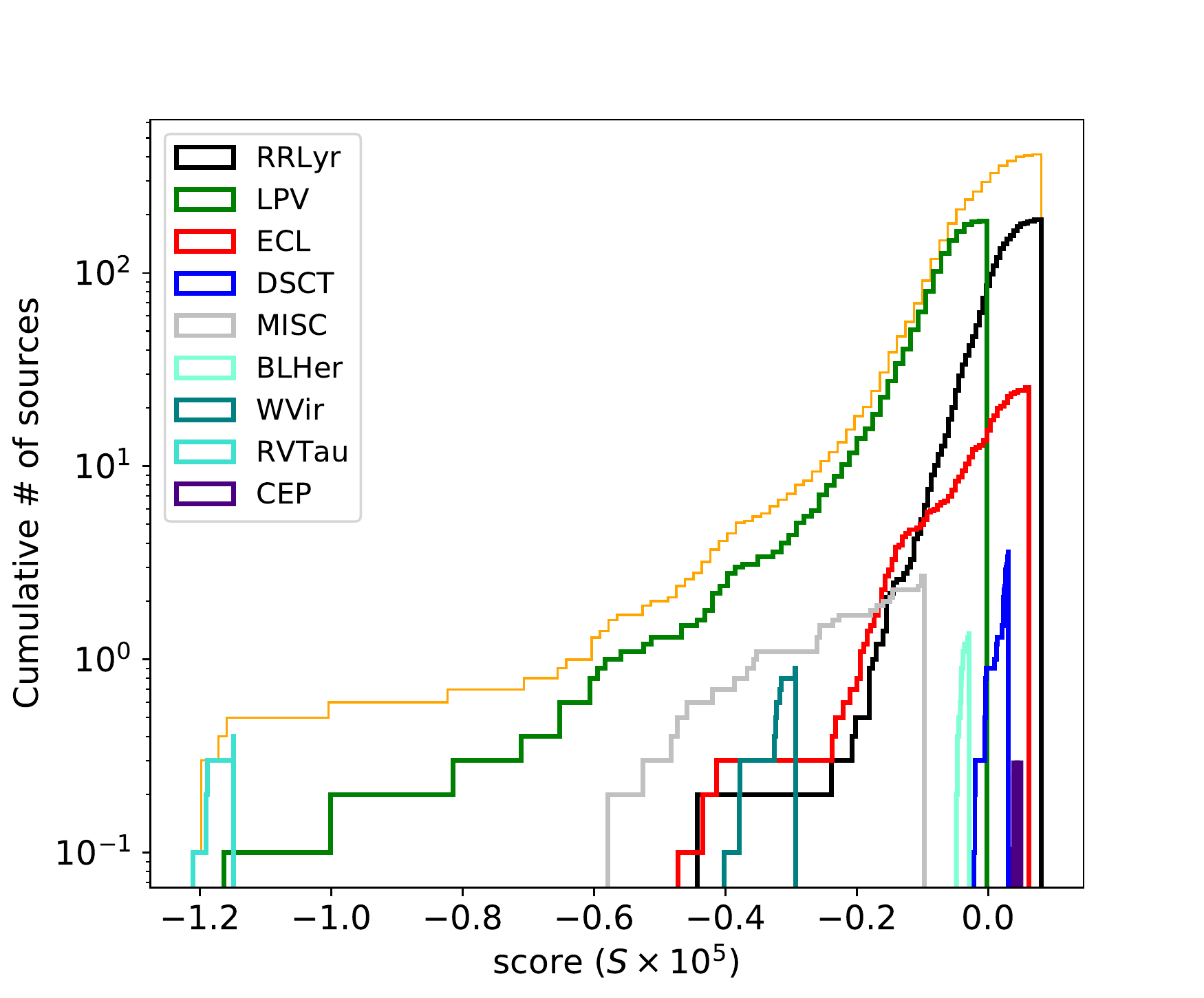}
\caption{Distribution of scores $S$ (in log unit) for the labeled B1-field sources computed using P-all ({\it left}) and P-labeled iterating 10 times ({\it right}). The orange distribution in both panels indicates the population distribution, while the distributions for the various labels are color-coded similar as in Fig.~\ref{fig:dif_ogle_abi} right panel.}
\label{fig:Sdist_lab}
\end{figure*}

We perform two experiments. In one case, we use our entire data set of 2266 sources to construct the $p(dm|dt)$ distribution, which we will hereafter refer to as P-all. In the second case, we split our sample of labelled variable stars into two subsamples of size $2/3$ and $1/3$ of the total size. We use the larger subsample to construct the $p(dm|dt)$ distributions (called P-labeled hereafter). 
We then use these distributions to obtain the corresponding likelihoods for the test sources, via
\begin{equation}
\mathcal{L}=\prod_{f,j} p_{f}(dm_{f,j}|dt_{f,j}),
\label{eq:L}
\end{equation}
where $f$ denotes the passband  $f=u,g,r,i,z$ and cross-passband $f=u-g,u-r,u-i, {\rm etc.}$ generated from the permutation of the five passbands taking two of them at a time, and $j$ is the index for the $dm\mbox{--}dt$ pair obtained from the observed light curve of the test source. 
 Although  it  is reasonable to assume that magnitude differences $dm$ may be correlated, we ignore such correlations here for the sake of simplicity, turning the likelihood into  a  simple  product.  All  test  and  training  sources  are  subject  to this simplifying assumption in the same way.

Note that in the case of the individual passbands, we consider only consecutive time differences (and corresponding magnitude differences) and not all possible $dm\mbox{--}dt$ pairs, since the consecutive differences have sufficient information to generate the rest under our assumption of independence. For example, in the case of a survey characterized by a cadence strategy that was uniform initially, from which we construct $p(dm|dt)$ with a single value of $dt$, and heterogeneous later on, the same $p(dm|dt)$ will be applicable to generate the corresponding density values for, say $n\times dt$, by convolving $p(dm|dt)$ with itself $n-1$ times. For non-integer $n$, one could interpolate.  For the cross-passbands (e.g., $g-r$), we generate all possible pairs of observation times $t_{g}, t_{r}$ where $t_{g}>t_{r}$, and simply compute the corresponding time and magnitude differences.

Finally, we compute features, called scores $S=\log\left(\mathcal{L}/\left<\mathcal{L}\right>\right)$. The normalization of the likelihood by its expectation value takes care of the differences in the number of $dm\mbox{--}dt$ pairs between test sources and also any missing passband(s) for a source. $\left<\mathcal{L}\right>$ can be easily derived as
\begin{equation}
\left<\mathcal{L}\right>=\prod_{f,j}\int^{\infty}_{-\infty}p_{f}(dm|dt_{f,j})\cdot p_{f}(dm|dt_{f,j})d(dm).
\label{eq:Lexp}
\end{equation}

\begin{table*}
\caption{Features \& Derived Quantities}\label{tab:feats}
\renewcommand\arraystretch{1.0}
\centering
\begin{tabular}{m{0.15\textwidth}|m{0.15\textwidth}|m{0.45\textwidth}}
\hline
Features		& $dm_{f}, dt_{f}$		& corresponding magnitude and time differences for a given (cross-)passband $f=u, g, r, i, z, u-g, u-r, u-z$, etc.\\
\hline
\multirow{2}{*}{Derived Quantities}  	& $p_{f}(dm_{f}|dt_{f})$		& probability density function of $dm$ given $dt$ for a given (cross-)passband $f$\\
\cline{2-3}
&  $S=\log\left(\mathcal{L}/\left<\mathcal{L}\right>\right)$	& likelihood score of being consistent with the population of variable sources for a given test source, where $\mathcal{L}$ and $\left<\mathcal{L}\right>$ are given by Eqs.~\ref{eq:L} and \ref{eq:Lexp}\\
\hline
\end{tabular}
\end{table*}

In constructing the probability distribution we do not differentiate the source types/labels. In other words, for a given $f$ we have a {\em population} $p_{f}(dm|dt)$. These distributions are shown in Fig.~\ref{fig:dmdt_all} taking all 2266 sources (P-all case). The motivation here is that for the sample observed by the same survey and containing galactic variable sources at similar distances, the probability densities for different classes of variable stars contained within this sample will be weighted proportionally to their intrinsic population sizes. Similar distributions taking only a single label at once, specifically ECL, LPV and RRLyr, are shown in the Appendix section (Figs.~\ref{fig:dmdt_ecl}, \ref{fig:dmdt_lpv}, and \ref{fig:dmdt_rrlyr}, respectively). 

Comparing these distributions to that in Fig.~\ref{fig:dmdt_all}, it is evident that the prominent features, for example the two high-density stripes seen in the $u-r$ and $g-r$ distributions, arise from the two dominant populations in our sample (cf.~Fig.~\ref{fig:dif_ogle_abi} right panel). In the distribution of scores for the test sources, the outliers at the low-score tail will thus be the unusual/rare ones with respect to the given population of sources used in constructing the $dm\mbox{--}dt$ distributions. 

A summary of the features used in our algorithm is given in Table~\ref{tab:feats}.

\subsection{Analysis of the B1 field variable sources}\label{sec:cluster}
We use the P-all and P-labeled distributions derived above to compute the scores $S$ for the 1038 unlabeled sources in the B1 field. In the P-labeled case, we make ten iterations, randomly sampling the labeled populations to construct $p_f(dm|dt)$ and evaluating the scores for the remaining sources. The resulting cumulative distributions of scores are shown in Fig.~\ref{fig:Sdist}; for the P-labeled case, the distribution is normalized by the number of iterations. Given the formulation of our algorithm above, we expect that in the P-all case, whereby all the sources indiscriminately enter the construction of $p_f(dm|dt)$, the score distribution for the unlabeled sources will be more concentrated toward larger values than in the P-labeled case, since each source has contributed to $p_f(dm|dt)$--an unusual source contributing a smaller fraction than one that is more common. Fig.~\ref{fig:Sdist} confirms our expectation.

\begin{table*}
\caption{Selected Anomalous sources}\label{tab:anomalies}
\renewcommand\arraystretch{1.0}
\renewcommand{\tabcolsep}{8pt}

\centering
\begin{tabularx}{0.8\textwidth}{cllll}
\hline
Source \#	& 	Name	&RA (deg)		&DEC (deg)	&Previous classification \& identification\tablenotemark{a}\\ 
\hline
1 & B1-3380631 & 271.21048 & -29.89543  &  Mira (V1552 Sgr)\\ 
2 & B1-6279497 & 270.2922 & -30.47849  & ECL (OGLE BLG-ECL-251298)\\ 
3 & B1-5444128 & 270.28276 & -30.32556   & Mira (OGLEII DIA BUL-SC38 V0489)\\ 
4 & B1-4649187 & 271.2637 & -30.1651   & ECL (OGLE BLG-ECL-293424)\\ 
5 & B1-8952163 & 271.31679 & -29.32614   & ECL (OGLE BLG513.26 52832)\\ 
6 & B1-2045914 & 270.32522 & -29.76824   & --\\ 
7 & B1-9466963 & 269.86428 & -29.65758   & ECL (OGLE BLG-ECL-231756)\\ 
8 & B1-9298299 & 270.58479 & -29.4486   & Mira (OGLE BLG-LPV-186488)\\ 
9 & B1-2912086 & 270.21482 & -29.94788   & Dwarf nova (OGLE BLG-DN-450)\\
10 & B1-2426396 & 271.52742 & -29.82479   & RS~CVn (MACHO 120.21264.476)\\ 
11 & B1-2892228 & 270.02525 & -29.88416   & RS~CVn (MACHO 118.18793.359)\\ 
12 & B1-657160 & 269.97327 & -29.63352   & --\\ 
13 & B1-6138192 & 269.85953 & -30.44154   & ECL (OGLE BLG-ECL-231568)\\
14 & B1-3224523 & 270.7501 & -29.90653   & Microlensing event (MOA 2013-BLG-402)\\ 
15 & B1-9585329 & 269.84486 & -30.29035   & ECL (OGLE BLG-ECL-230940)\\
16 & B1-8221965 & 270.99724 & -30.71808   & --\\ 
17 & B1-8509754 & 270.99031 & -29.3217   & Transient (OGLEII DIA BUL-SC2 V116)\\ 
18 & B1-451244 & 271.39569 & -29.39931   & 	ECL (OGLE-BLG-ECL-299272)\\ 
\hline
\end{tabularx}
\tablenotetext{a}{References are given in text (Sect.~\ref{sec:char}).}
\end{table*}

We also evaluate the scores for all the labeled sources using P-all and for the corresponding $1/3$ fraction of labeled sources in the P-labeled case that are not used in constructing $p_f(dm|dt)$ in each iteration. The resulting score distributions are shown in Fig.~\ref{fig:Sdist_lab}. As can be seen, the score distributions for the labeled sources appear similar in the two panels. In principle, as discussed above, we expect the scores of the sources to scale with the population size of their corresponding variable star type, i.e., the more populated a label is, the higher is the typical score of a source belonging to this label, while sources from the minority populations (either because of survey sensitivity or the populations are intrinsically rare) are expected in the low-score tail. This should be true as long as the distributions $p_f(dm|dt)$ are distinct enough for the different labels. In Fig.~\ref{fig:Sdist_lab}, we see an almost complete overlap between the distributions for the two most populous labels after LPV, i.e., RRLyr and ECL. This overlap also envelopes other pulsators--DSCT, BLHer, including the lone Cep. Furthermore, there is significant overlap between the distribution of LPV and that of the other types already discussed above, specifically toward higher score values. In fact, the LPV scores are distributed leftward of score $S=0$, at lower score values. Though LPV is the second-most populous (Fig.~\ref{fig:dif_ogle_abi} right panel), it appears the collection of other labels are similar enough to each other to form a majority
together against which the LPV sources appear as the minority and hence the likelihoods ($\mathcal{L}$) of sources belonging to the LPV label are lower than their expected likelihoods. The distributions of the longer-period systems like WVir (PopII Cepheids with period greater than 10~days) are also enveloped by the LPV distribution. The lone RVTau, though, exhibits a lower score than most of the sources belonging to the different labels.

\begin{figure*}
\centering
\includegraphics[width=88mm]{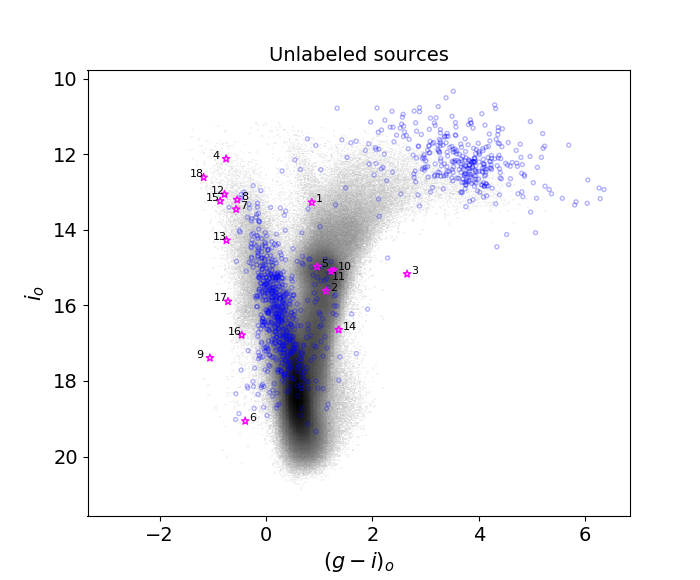}\hfill\includegraphics[width=88mm]{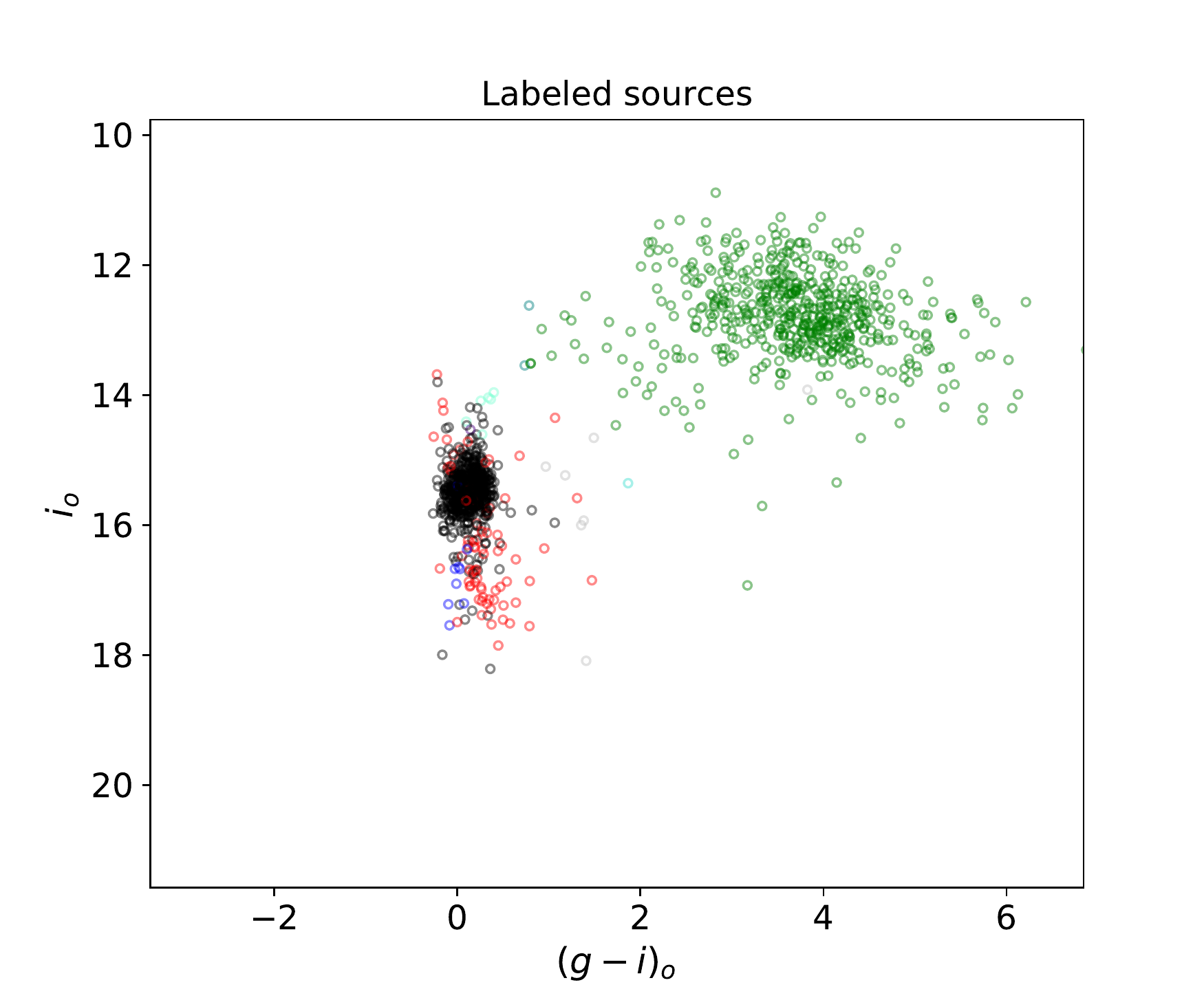}
\caption{Color-magnitude diagram (CMD) of the variable sources in the B1 field without OGLE labels ({\it left}) and those with OGLE labels available ({\it right}). The color-coding for the various labels in the right panel follows that of Fig.~\ref{fig:dif_ogle_abi} right panel. The magenta stars in the left panel mark the identified 18 candidate anomalous sources from Sect.~\ref{sec:cluster} along with their corresponding ID numbers, while the grey background shows the full CMD of the B1 field from \citet{Saha-2019}.}
\label{fig:cmd}
\end{figure*}

It is evident from Fig.~\ref{fig:Sdist_lab} that a clean separation of the different labels is not achieved with our algorithm. Our goal, however, is detection of unusual sources as opposed to classification. 
These are the sources having significantly lower scores compared to the others in the composite population. We identify them simply by setting a threshold, which we define as the second-percentile score. In each of our experiments (P-all and ten iterations of P-labeled), we find 21 {\em outlier} unlabeled sources having scores lower than the respective threshold. Many of these sources are flagged multiple times. We compile a final sample that includes only those sources flagged more than $\mathcal{N}$ times, where $\mathcal{N}={\rm Med}(x_{j})-{\rm MAD}$, where $x_{j}$ is the number of times the outlier source $j$ is flagged and MAD is the median absolute deviation. The final sample includes 18 sources, as listed in Table~\ref{tab:anomalies}, and are discussed in the following section.

For real-time applications, i.e., as a processing stage within a broker like ANTARES, the distributions $p_f(dm|dt)$ can be computed for a given field of view from the data gathered within the first 0.5--1~year of operation -- typically one such distribution for each field covered by the survey (e.g., LSST) in one snapshot. This will ensure a relatively stable, statistical sampling of the characteristics of the different variable stars and transients contained within the field to identify the most unusual events of all. A distribution built from just the first few days of observations will be less representative. However, the distributions can be dynamically updated, for example in day-time, during the course of the survey. This will also incorporate any evolution in the populations of the variable stars and transients, including the peculiar types.

We note that we are using the full light curves of the test sources in assessing their peculiarity. The sampling of the Galactic Bulge data used in the present study was optimized for the discovery of a specific type of variables, namely RR~Lyrae, and hence the data set is not ideal for experimenting with shorter baseline coverage to investigate the effectiveness of our algorithm at different levels of recovery of the light curve of a test source. We defer such an analysis to a future work (see Sect.~\ref{sec:summary}).

\section{Characteristics of the selected sources}\label{sec:char}

\begin{figure*}
\centering
\includegraphics[width=60mm]{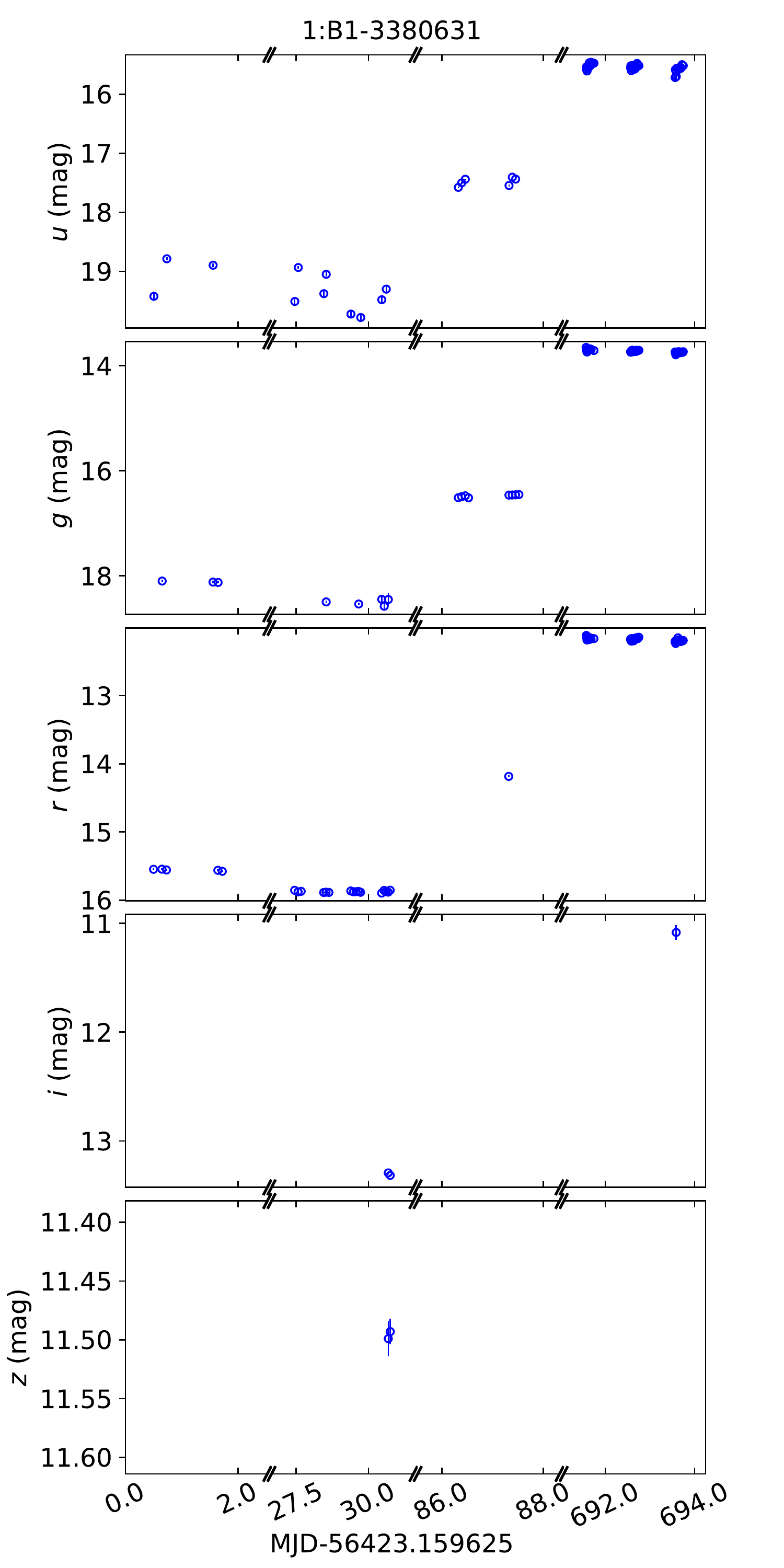}\hfill\includegraphics[width=60mm]{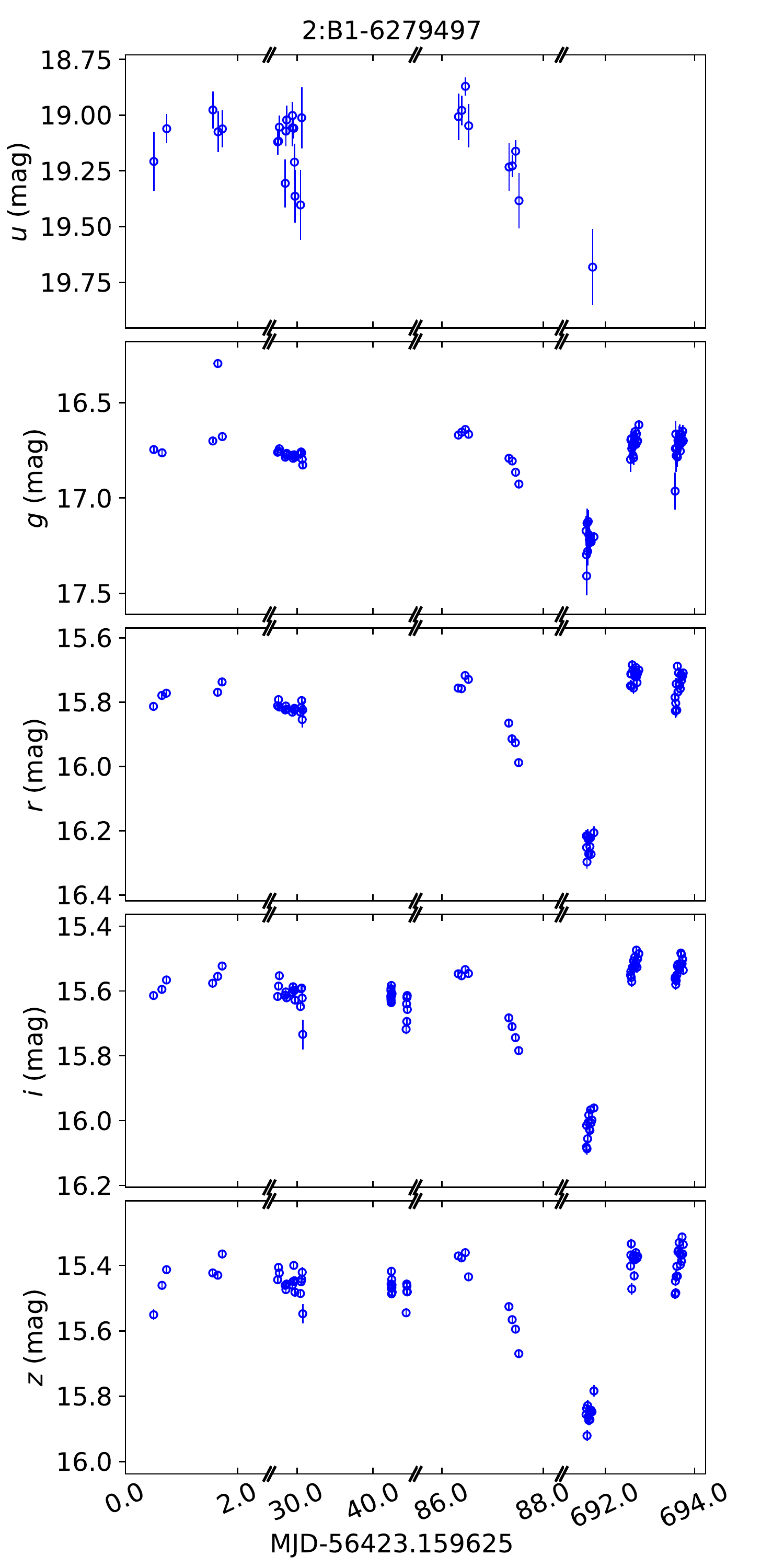}\hfill\includegraphics[width=60mm]{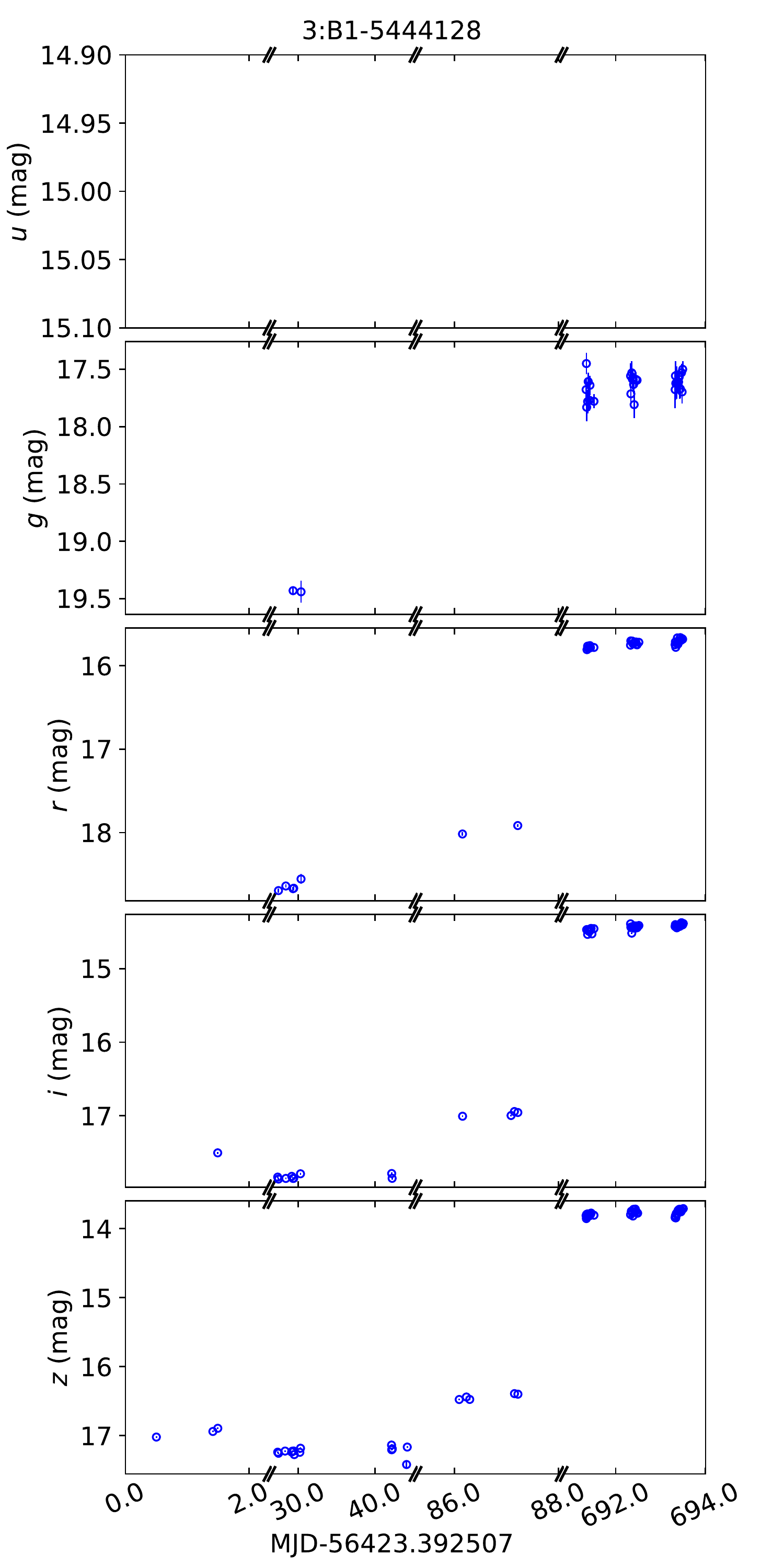}
\caption{Multi-wavelength light curves of the 21 selected sources in a format similar to that of Fig.~\ref{fig:egs}. The ID numbers of the stars shown on top of each plot are the same as those displayed on Fig.~\ref{fig:cmd}. There are no $u$-band observations for source 3, hence its uppermost panel is empty.}\label{fig:lcs}
\end{figure*}

\begin{figure*}
\centering
\ContinuedFloat
\includegraphics[width=60mm]{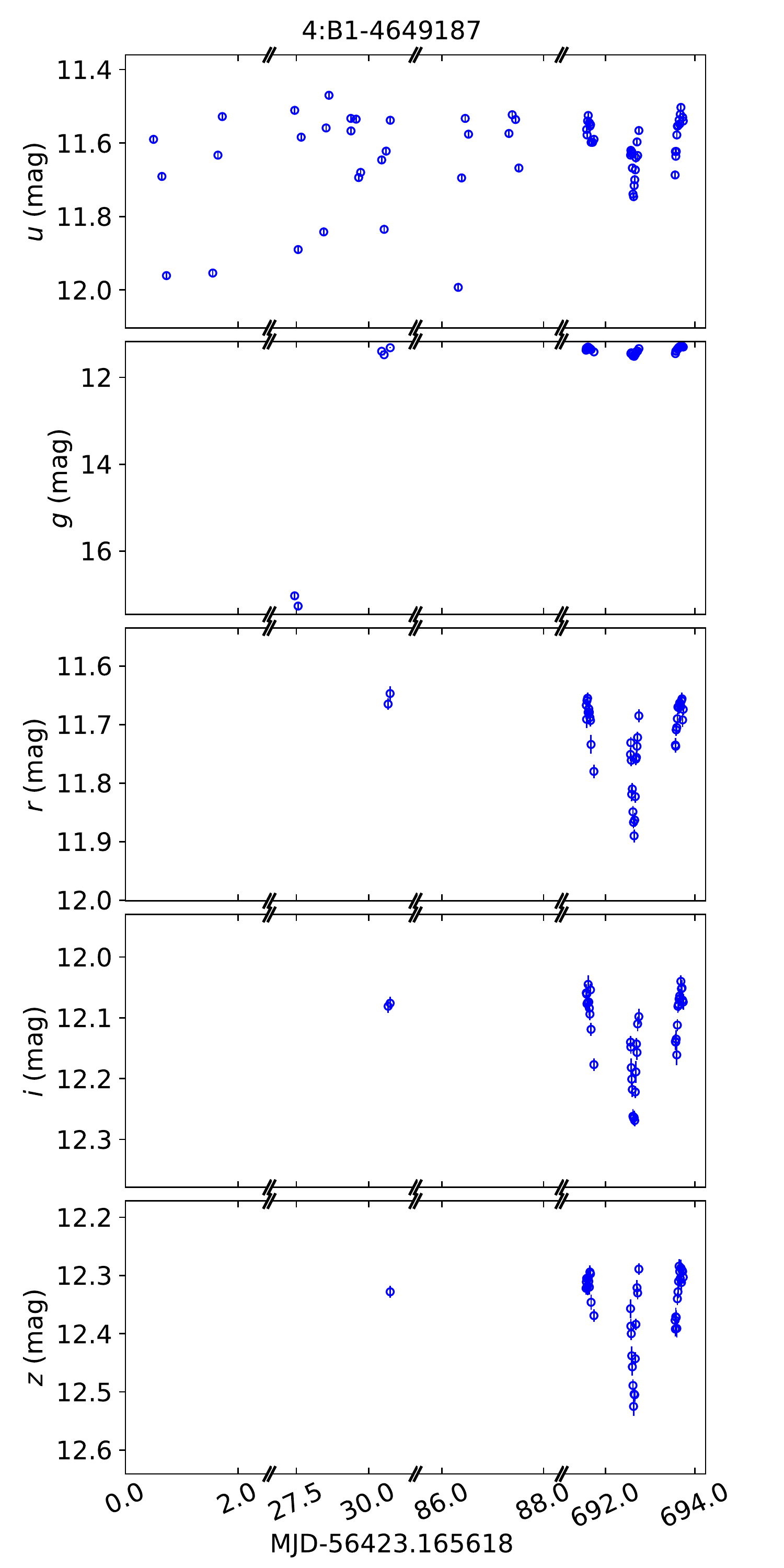}\hfill\includegraphics[width=60mm]{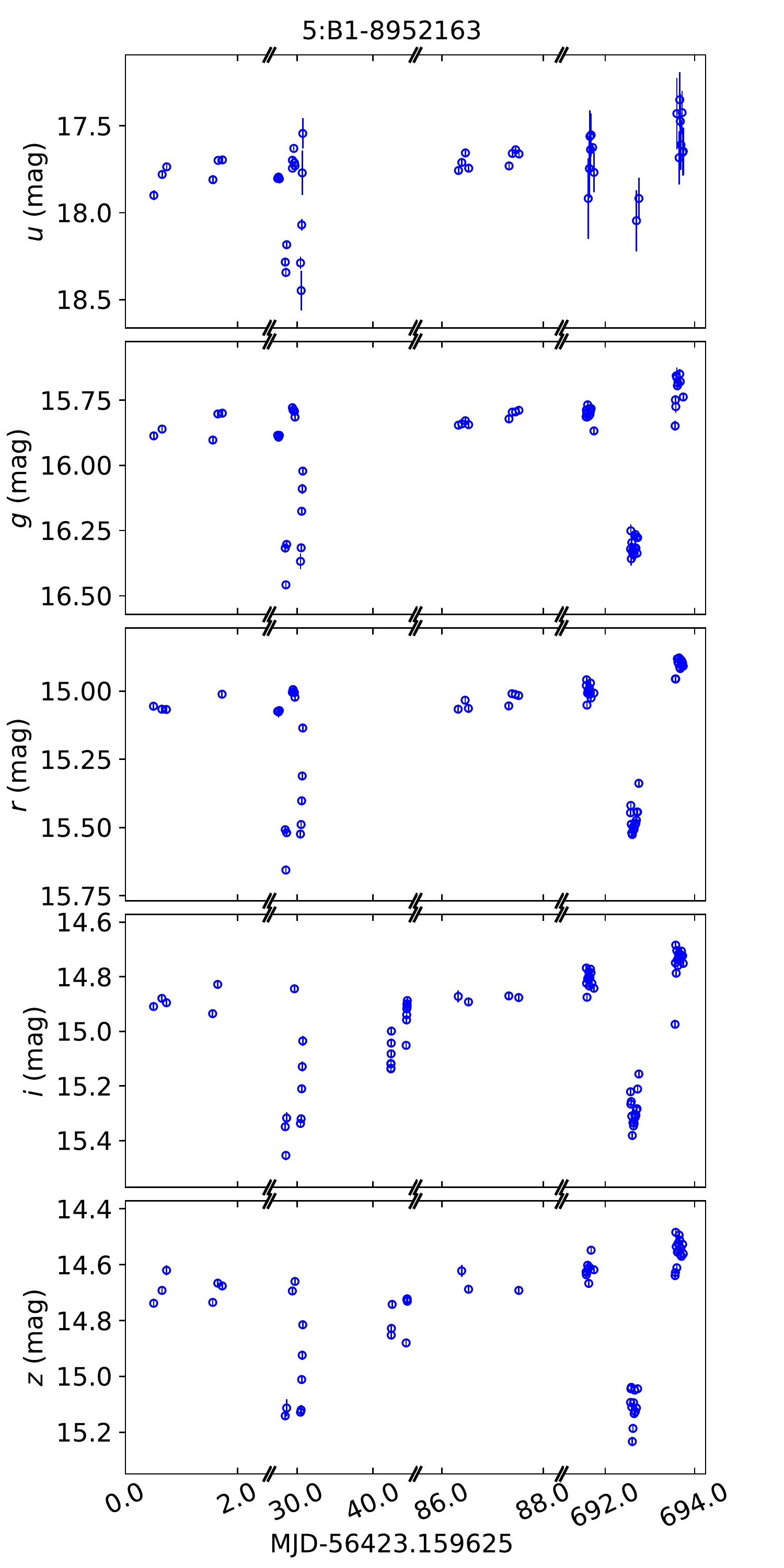}\hfill\includegraphics[width=60mm]{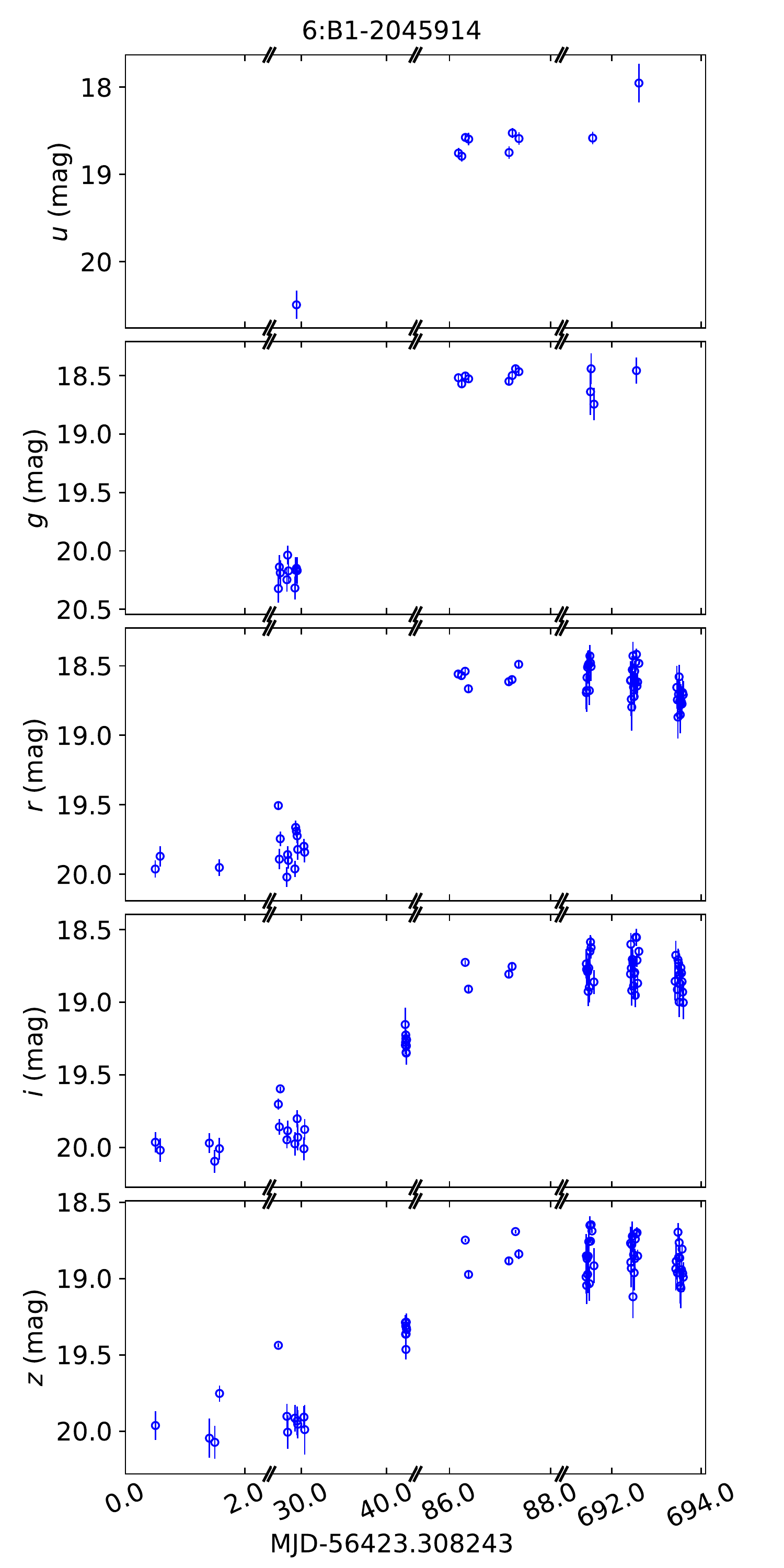}
\includegraphics[width=60mm]{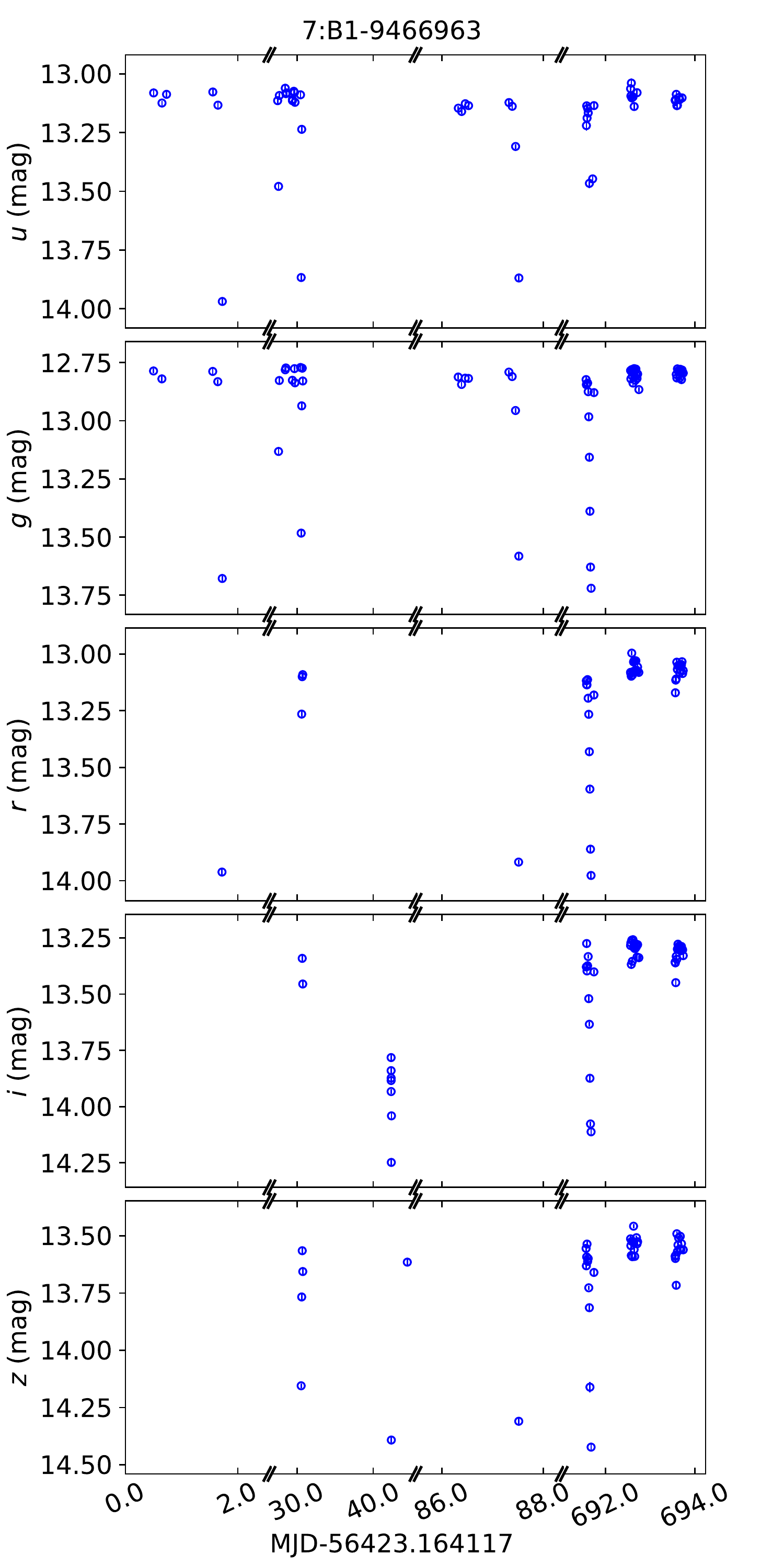}\hfill\includegraphics[width=60mm]{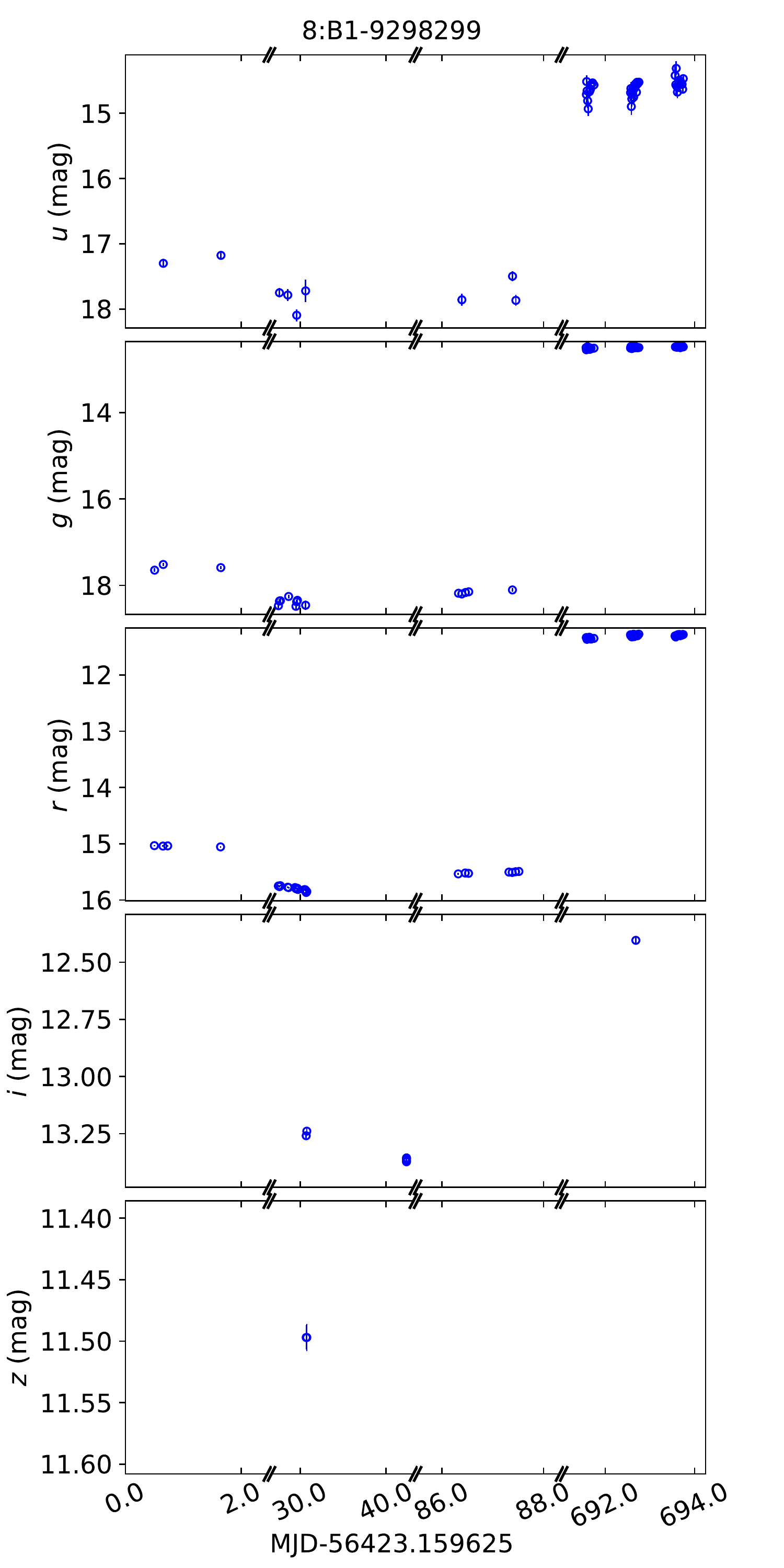}\hfill\includegraphics[width=60mm]{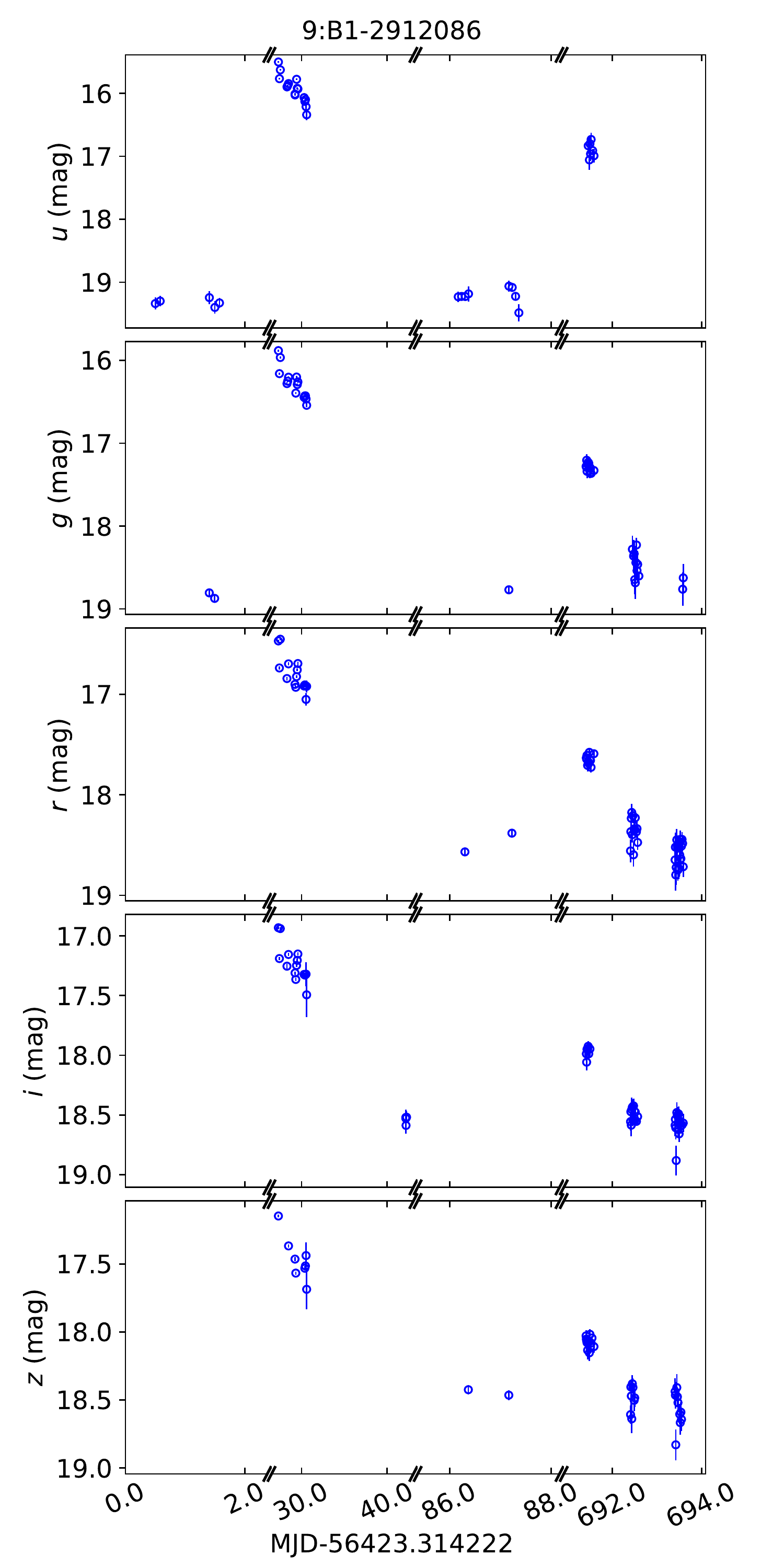}
\caption{Contd.}
\end{figure*}

\begin{figure*}
\centering
\ContinuedFloat
\includegraphics[width=60mm]{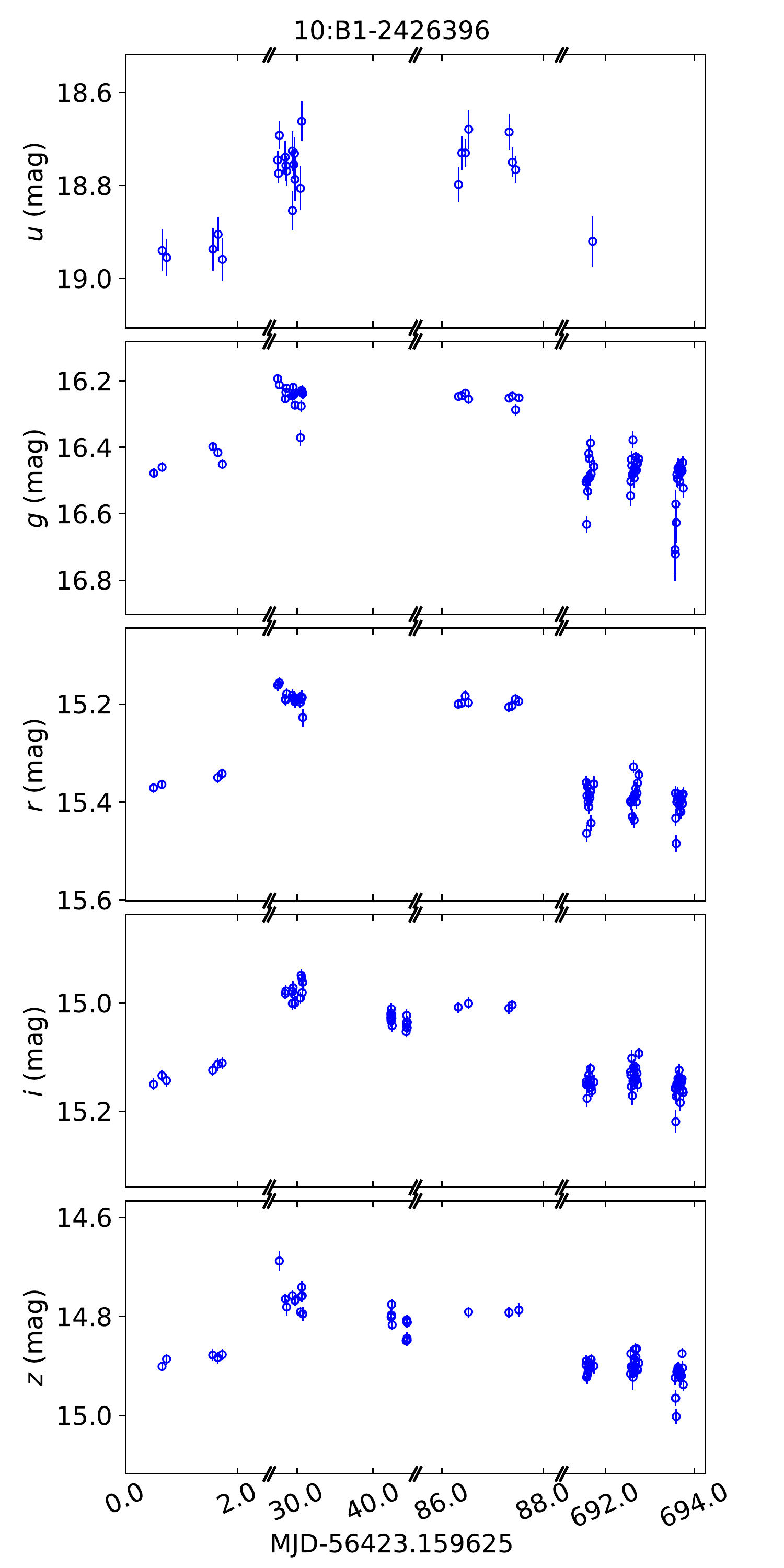}\hfill\includegraphics[width=60mm]{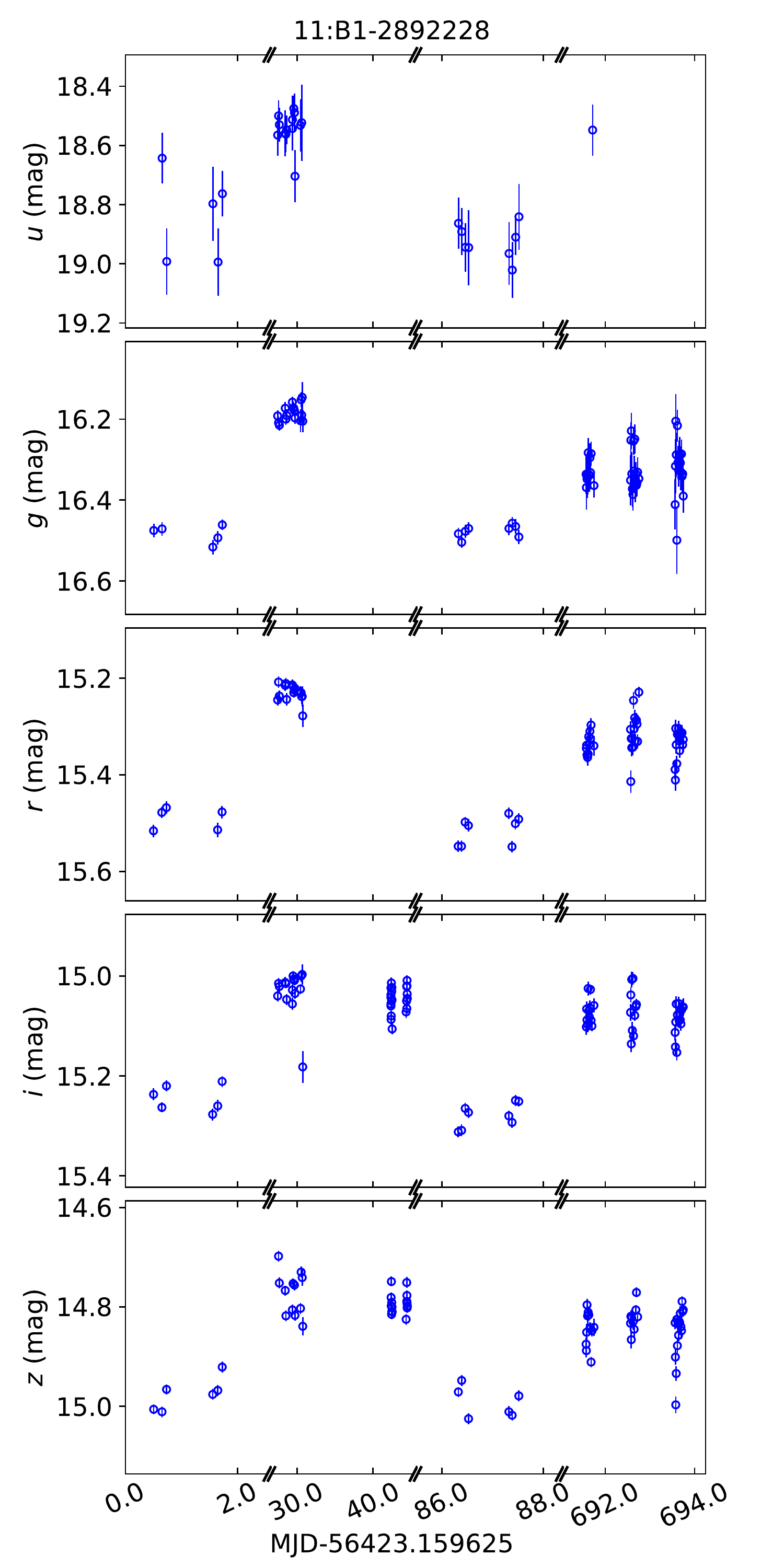}\hfill\includegraphics[width=60mm]{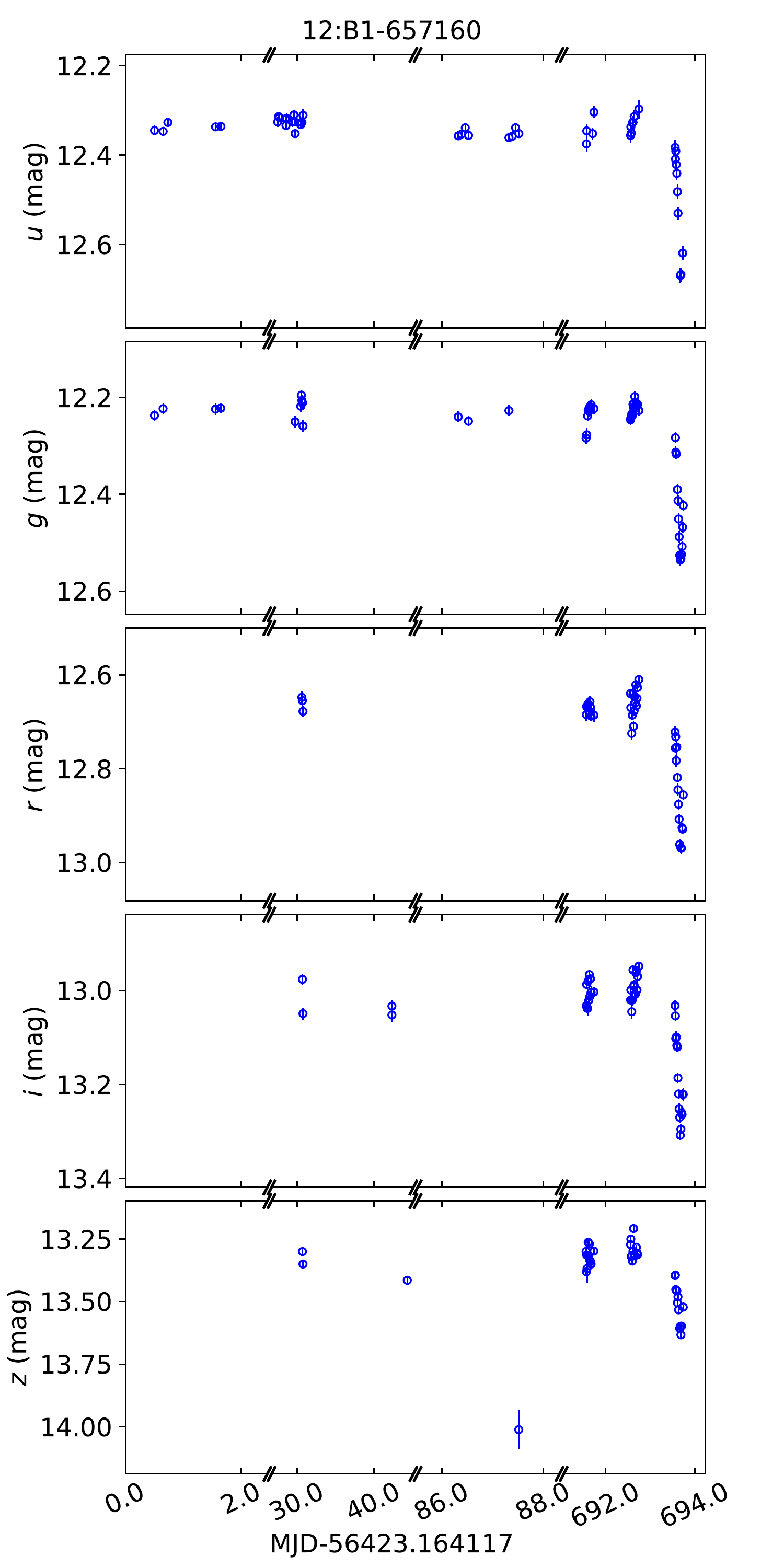}
\includegraphics[width=60mm]{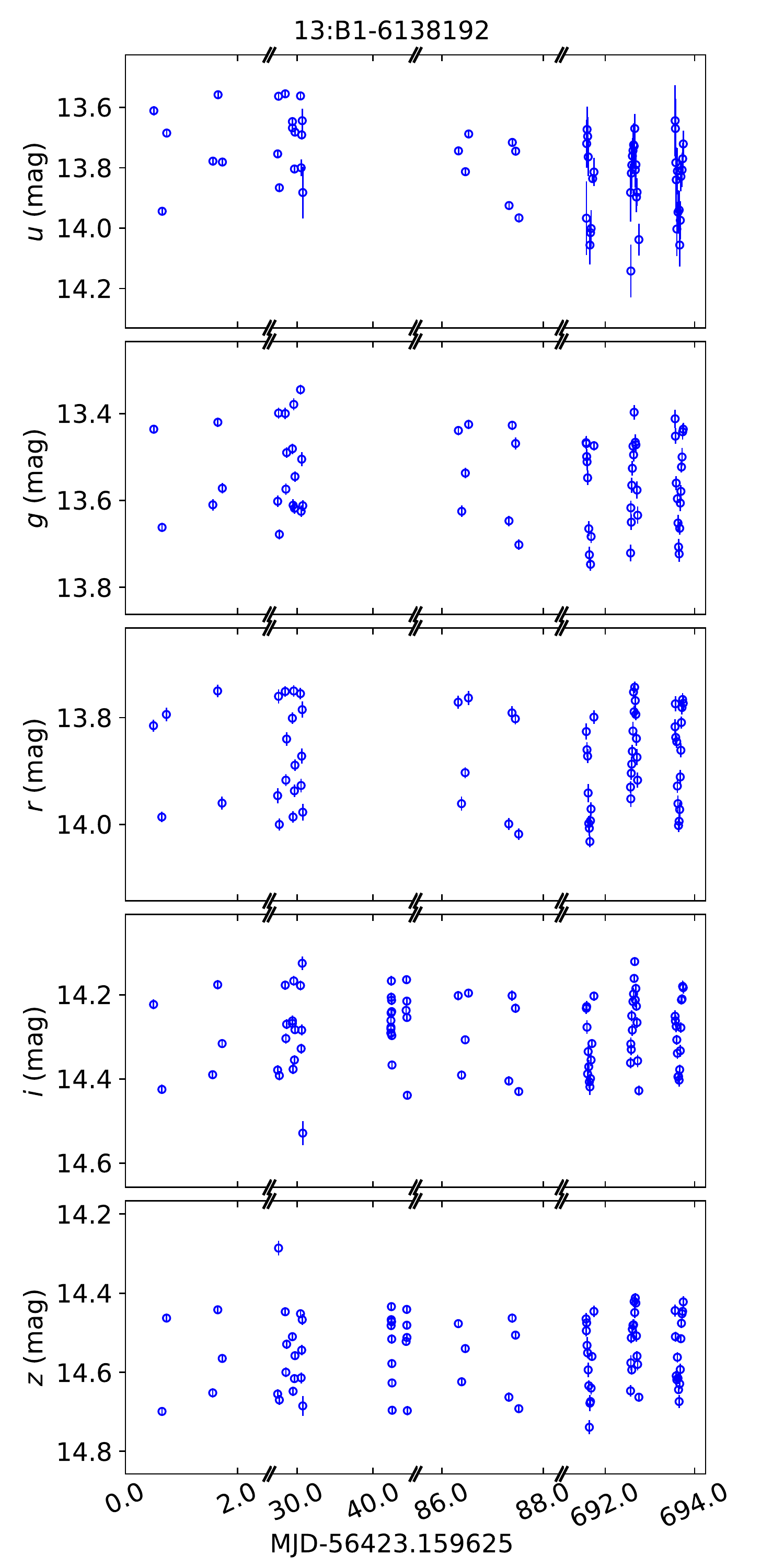}\hfill\includegraphics[width=60mm]{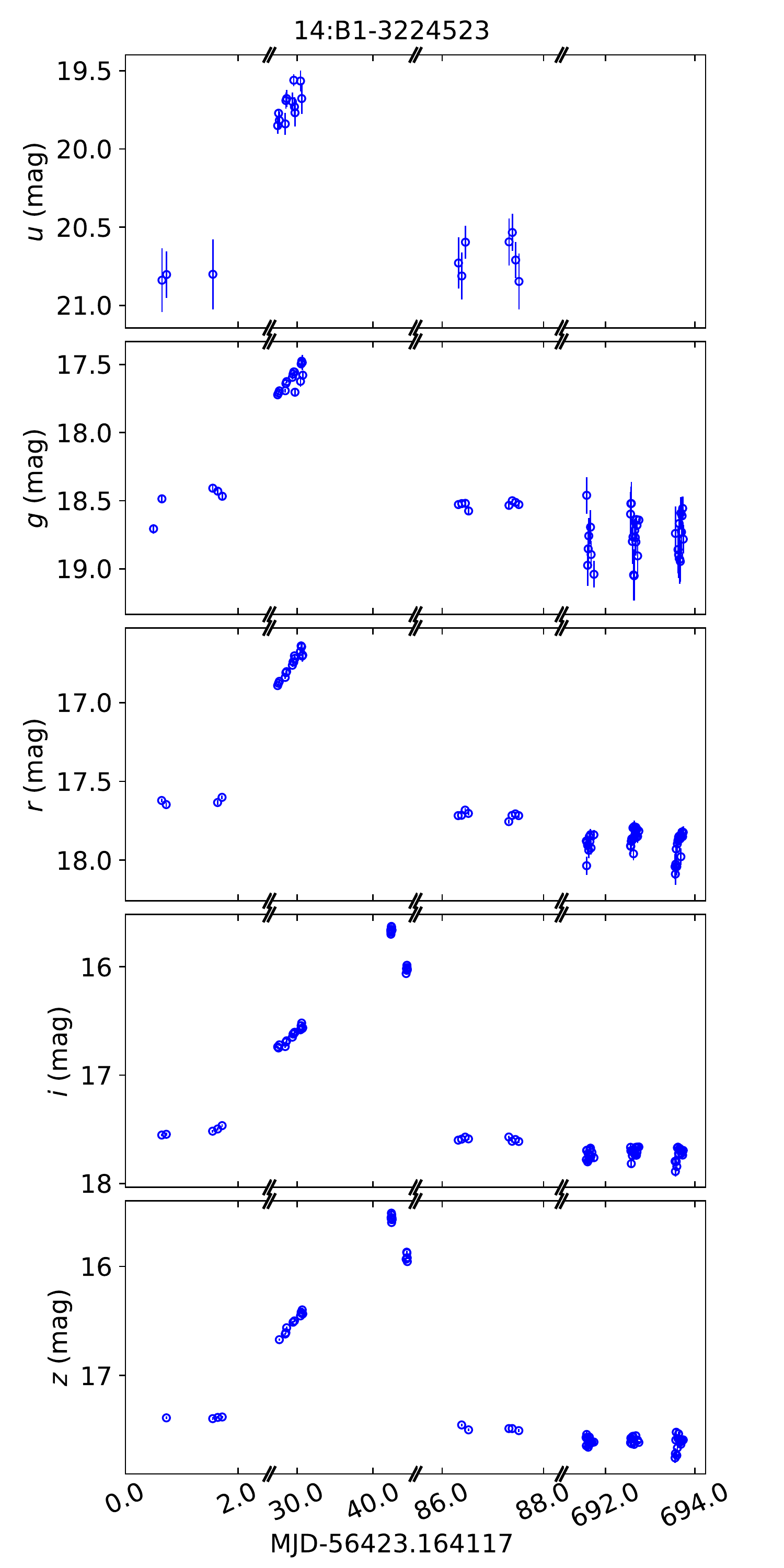}\hfill\includegraphics[width=60mm]{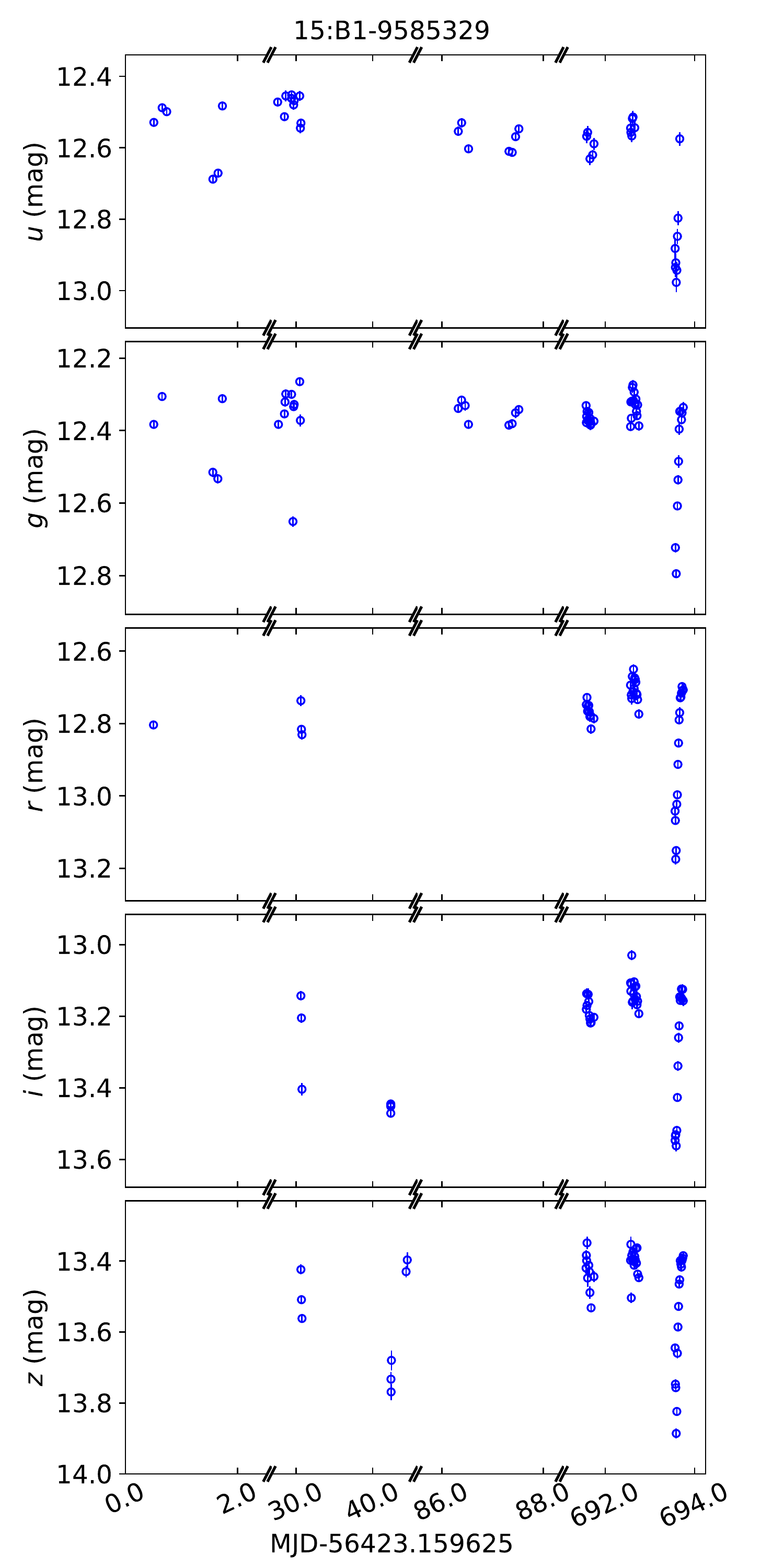}
\caption{Contd.}
\end{figure*}

\begin{figure*}
\centering
\ContinuedFloat
\includegraphics[width=60mm]{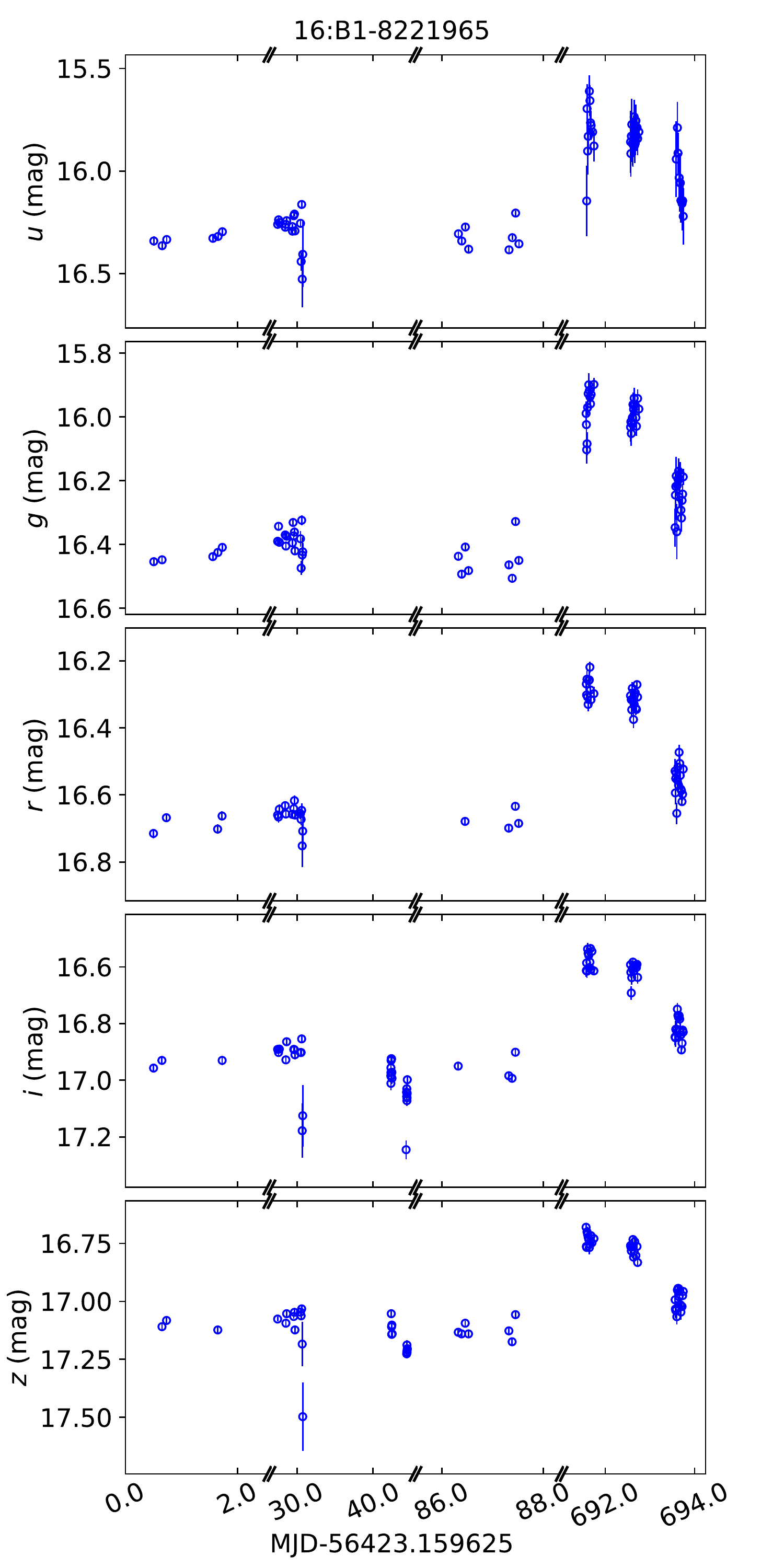}\hfill\includegraphics[width=60mm]{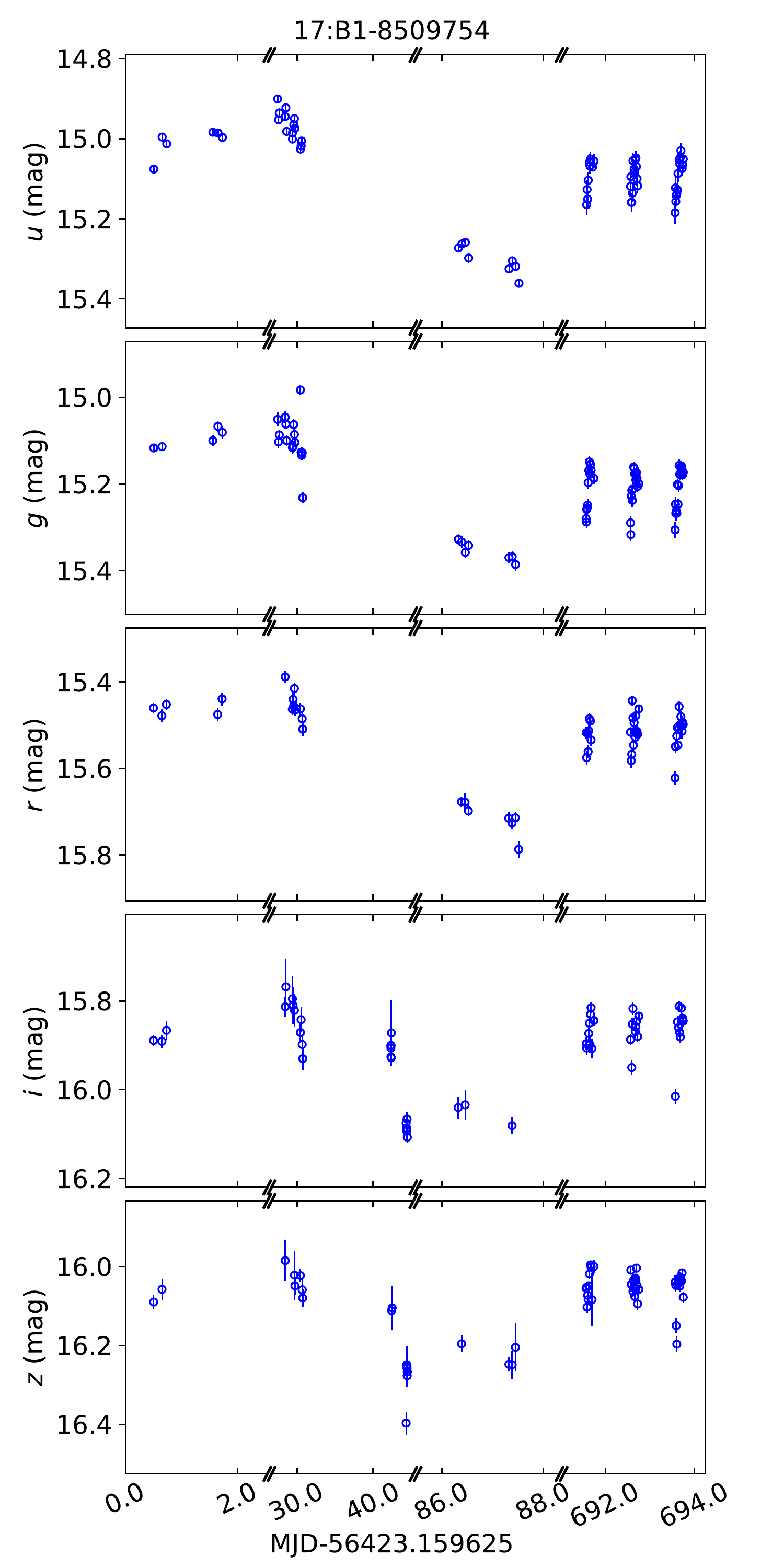}\hfill\includegraphics[width=60mm]{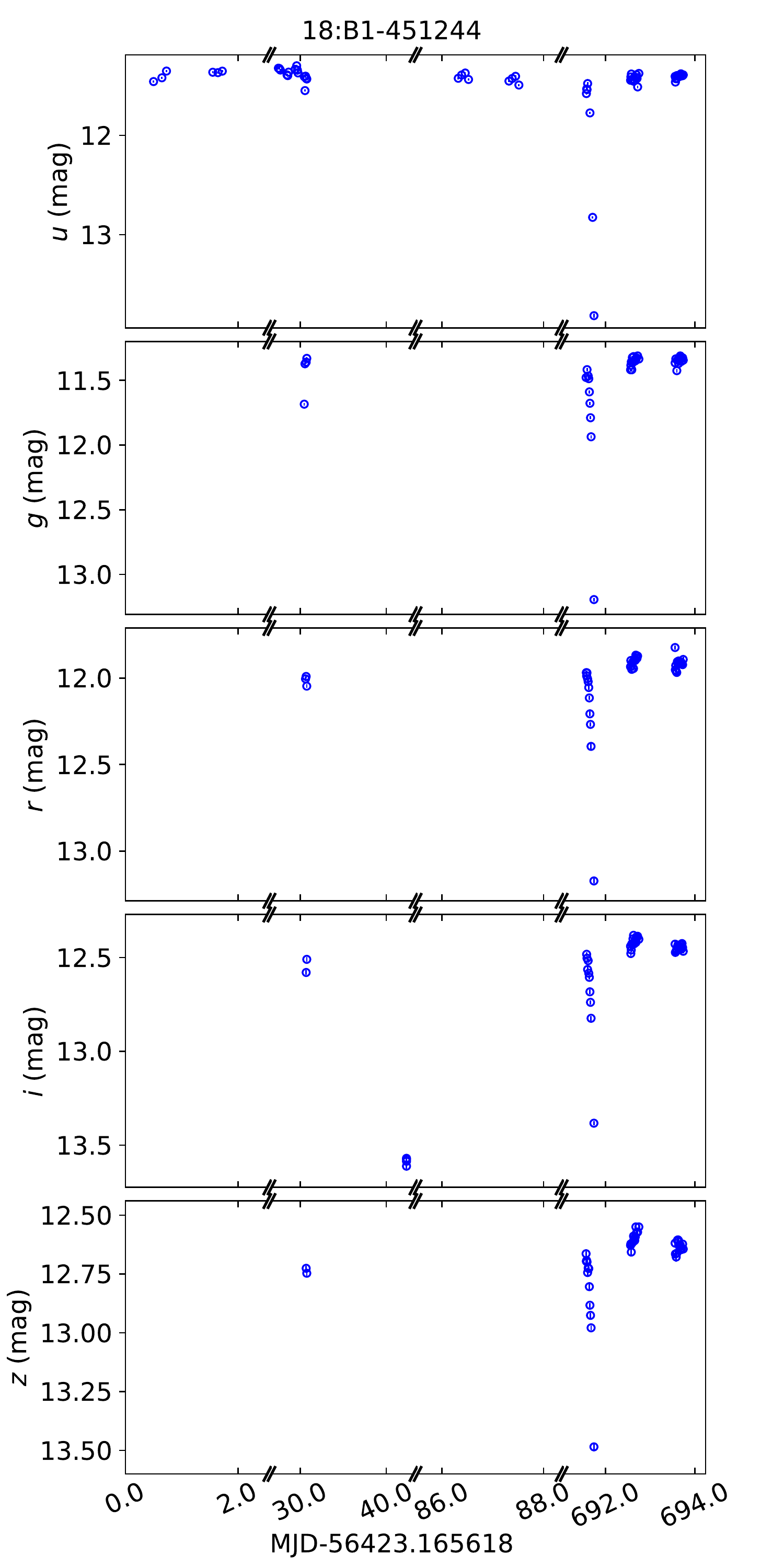}
\caption{Contd.}
\end{figure*}

\begin{figure*}
\includegraphics[width=88mm]{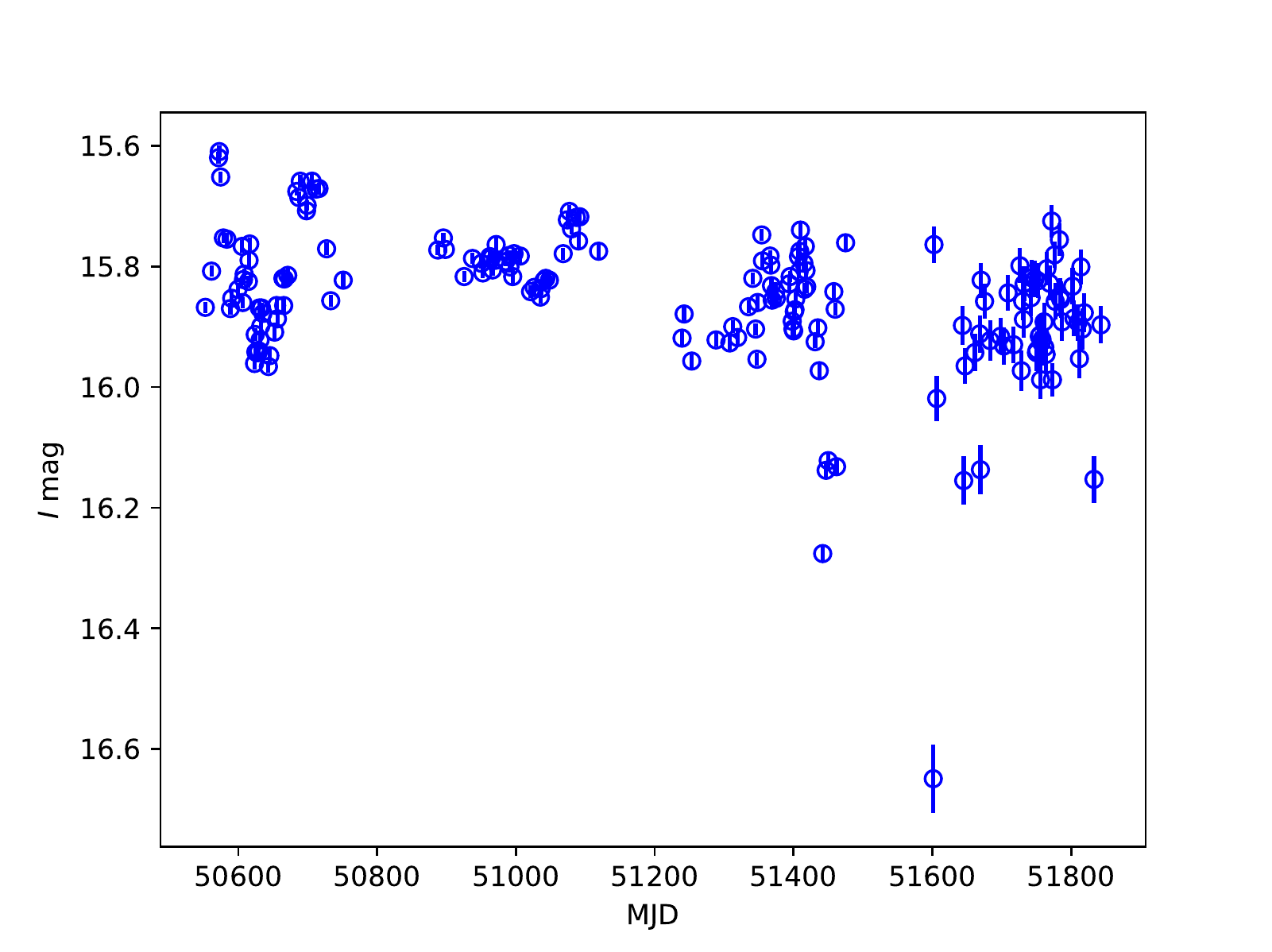}\hfill\includegraphics[width=88mm]{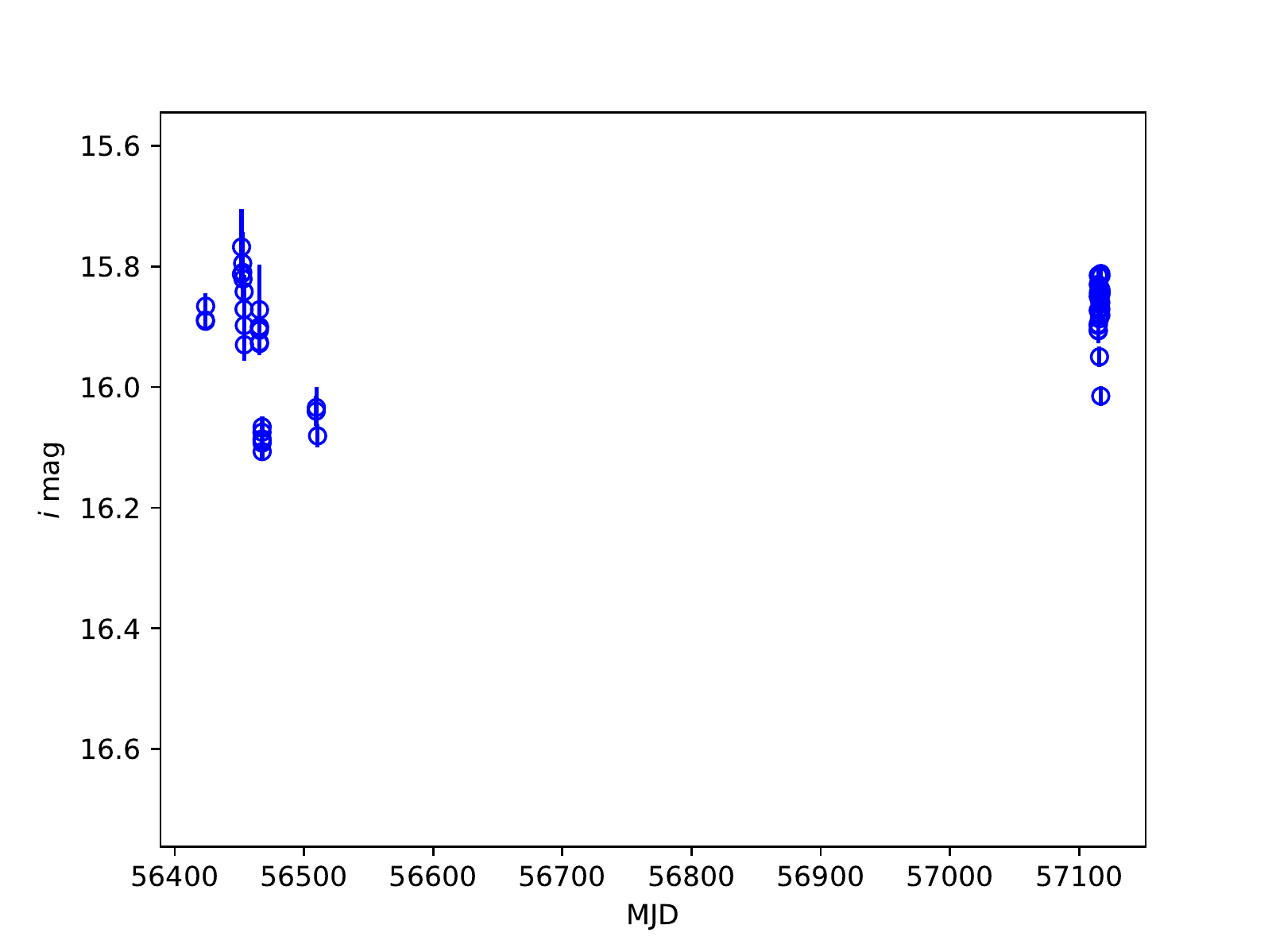}
\caption{Light curves of source 12 (B1-8509754) based on OGLE $I$-band data ({\it left}) and DECam $i$-band data ({\it right}).\label{fig:src13}}
\end{figure*}

The 18 unlabeled outlier sources are shown annotated by their ID numbers (first column in Table~\ref{tab:anomalies}) in the $(g-i)_o$ vs $i_o$ color-magnitude diagram (CMD) in Fig.~\ref{fig:cmd} (left panel). The right panel of the figure also shows the distribution in the color-magnitude space for the labeled sources, which are color-coded following the scheme in Fig.~\ref{fig:dif_ogle_abi} right panel. We use the average magnitudes of the time series in the corresponding passbands to construct the CMDs.  
Indeed, the labels showing strong overlap in Fig.~\ref{fig:Sdist_lab} are also tightly clustered together in the CMD, particularly the RRLyr and ECL groups, and appear well-separated from the LPVs.

As discussed in Sect.~\ref{sec:data}, the greater sensitivity of the DECam observations at the bright and faint end is also reflected comparing the two CMDs, with the extended cluster of points tracing fainter than the main sequence turn-off (see Fig.~15 of \citealt{Saha-2019}) in the left panel as compared to the right panel. The outlier sources can be seen clustered in three regions---at the extreme blue edge ($(g-i)_{o}<0$) and at the red edges, with the single source ID\#3 located at the extreme red edge ($(g-i)_{o}>2$) of $i>15$~mag. It is evident that our sources are outliers in the CMD itself (Fig.~\ref{fig:cmd}). Their multi-wavelength light curves are shown in Fig.~\ref{fig:lcs}.

{\em Red ($(g-i)_{o}>2$) anomalous source(s):}
Source 3 (B1-5444128) is the reddest of the outlier. It is a known Mira identified by \citet{Matsunaga-2005}. These authors found an I-band peak-to-peak amplitude of 3.76~mag, which appears to be consistent with our DECam light curve. For Miras, the amplitude increases with the wavelength and source 3 has the highest amplitude in the reddest ($z$) passband of all sources; its $z$ central 90\% range is greater than 3.5~mag (cf.~Fig.~\ref{fig:epochs} right panel) and hence it is truly an outlier in our sample.  Also, it can be seen that source 3 is not located within the LPV region of the CMD (upper right part in Fig.~\ref{fig:cmd}), which is constructed using the average magnitude of the light curve in the corrsponding passband for a given source. This may be due to incomplete phase sampling for source 3, given the long timescales of Miras and the very large amplitude exhibited by it.

{\em Red ($1<(g-i)_{o}<2$) anomalous sources:}
Sources 14 (B1-3224523), 2 (B1-6279497), 11 (B1-2892228), 10 (B1-2426396), 5 (B1-8952163), and 1 (B1-3380631) constitute another group of red outliers. 
					 
Source 14 exhibits an outburst-like profile with an amplitude of 2~mag lasting around 86~days with an approximately equal rise and decline timescale of 45 days. It does not show any color-evolution and remains constant at $g-i\approx1$~mag, thus excluding cataclysmic variable outbursts like novae and dwarf novae. A microlensing event, MOA 2013-BLG-402 alerted by the Microlensing Observation in Astrophysics (MOA; \citealt{Bond-2001}) and confirmed by \citet{Bensby-2017}, is $1.76''$ away from source 14 but its light curve matches that of source 14, including the peak time and event timescale. In fact, this astrometric offset can be accounted for by the relatively large seeing value of around $2.5''$ for MOA. Hence, source 14 is indeed a confirmed microlensing event. Based on the microlensing event rate toward the Galactic Bulge estimated by \citet{Sumi-2013}, we can make a very crude estimate of the number of microlensing events expected for the DECam observations of the B1 field used in this study. We expect fewer than five such events, which is in line with our detection of source 14 as the only microlensing event.

Most characteristics of the sources with IDs 2, 11, and 10 appear not out of the ordinary, apart from their very red colors in the CMD. Their variability amplitudes are small (less than around $0.5\mbox{--}0.6$~mag). However, the amplitudes reach 0.8~mag and 4~mag, respectively, for sources 5 and 1. For all of them, the amplitudes appear to increase with bluer passbands.  Sources 2 and 5 are identified as eclipsing binaries in the Simbad database \citep{Wenger-2000} based on OGLE data classified by \citet{Soszynski-2016} that was not used in our cross-matching in Sect.~\ref{sec:data}. Sources 10 and 11  have counterparts from MACHO at around $1''$ (MACHO 120.21264.476) and $0.9''$ (MACHO 118.18793.359), respectively, which are typed as RS~Canum~Venaticorum (RS~CVn) binary stars by \citet{Drake-2006}; these RS~CVn stars contain chromospherically active primary components that are typically giants of late spectral type located in the Red Giant Clump region of the CMD, in agreement with the locations of these sources. Source 1 also has a known Mira counterpart from \citet{Matsunaga-2005} at around $1''$, however, its amplitude appears to increase with decreasing wavelength atypical for Miras. Moreover, it is located close to the region of the Blue Loop Stars in Fig.~15 of \citet{Saha-2019}.

{\em Blue ($(g-i)_{o}<0$) anomalous sources:}                                                      
The remaining 11 outlier sources---IDs 6 (B1-2045914), 9 (B1-2912086), 16 (B1-8221965), 17 (B1-8509754), 13 (B1-6138192), 7 (B1-9466963), 8 (B1-9298299), 15 (B1-9585329), 12 (B1-657160), 18 (B1-451244), and 4 (B1-4649187)---are at the extreme blue edge of the CMD. 

Sources 4, 7, 13, 15 and 18 are known eclipsing binaries, classified by \citet{Soszynski-2016}. 
However, sources  13, 7, 8, 15, 12, 18 and 4 are in the foreground main sequence star region of the CMD of \citet{Saha-2019}. Hence, the extinction corrections{\footnote{The reddening map in \citet{Saha-2019} estimates reddening at the distance of the Bulge.}} applied to them, assuming they are located in the Bulge, are too large. Nevertheless, they are rightly flagged by our algorithm because of their extreme blue colors resulting from the overestimated extinction corrections. Source 8 has a large amplitude and it is, in fact, $1.22''$ away from an OGLE LPV, which is missed in our cross-matching in Sect.~\ref{sec:data} due to the smaller search radius. This lends support to the hypothesis that the above sources got flagged because of the erroneous extinction corrections.

Source ID 6 appears to have a bluer color evolution and shows an amplitude $\gtrsim1$~mag. It does not have any known variable star or transient counterpart in the Simbad database. 
Source 9 shows quite a large amplitude, especially in the blue bands ($u$ central 90\% range greater than 3.5~mag) and also redder color evolution, and is a dwarf nova already known in Simbad based on the OGLE data classification by \citet{Mroz-2015}, which was not used in our cross-matching. Source 17 shows no color evolution and its variability amplitudes are small ($\lesssim 0.5$). \citet{Wozniak-2002} tagged it as a ``transient'' (sources showing episodic variations) in their difference imaging analysis of the OGLE II $I$-band data covering $1997\mbox{--}1999 $. The OGLE $I$-band along with the DECam $i$-band light curves of this source are shown in Fig.~\ref{fig:src13}. It appears the light curve variation seen in the DECam data is similar in character to that seen in the OGLE observations taken more than 15~years earlier. Source 16 is characterized by an amplitude $\lesssim 0.6$ and appears to show no color evolution. Interestingly, both of these stars (IDs 16 and 17) are located in the Blue Horizontal Branch region of the CMD \citep{Saha-2019}.

\section{Performance comparison}\label{sec:eff}

The purpose of our algorithm is to detect outliers, because outliers will correspond to astrophysically distinct (i.e., novelty) or otherwise interesting sources. That the detected outliers are interesting was shown using the DECam data (cf.~Sect.~\ref{sec:char}). Here, we shall try to assess how good our algorithm is at detecting outliers in the first place, as compared to others.

\subsection{Strategy}
Our algorithm falls into the category of Naive Bayes classifiers, whose underlying principle is based on Bayes' theorem. Such a classifier assigns membership of a given source to a class by maximizing its posterior probability evaluated using the likelihood with respect to the different classes and prior information on the classes. When the prior probabilities are assumed to be equal for all classes (as done in our case), the classifier becomes a maximum-likelihood classifier.  Given $N$ observed features for each source, the likelihood is an $N$-dimensional joint probability distribution, conditioned on the source belonging to a given class, thus incorporating the correlations among the features. However, by assuming independence of the features, the joint likelihood is reduced to a simple product of 1-D likelihoods. Even though independence of features is frequently violated in many applications (including ours, where the multi-passband light curves are correlated), Naive Bayes classifiers have been found to perform well and the ease of use makes them quite popular. Such classifiers have been successfully used in the astronomy literature \citep[e.g.,][]{Broos-2011, Oszkiewicz-2014, Visscher-2015, Lochner-2016}.

In our algorithm, we have essentially classified a source into two classes -- outliers and inliers, by thresholding their likelihoods, since a higher likelihood corresponds to a higher probability of the source being an inlier. Additionally, in our formulation, we use a normalized likelihood (i.e., $\mathcal{L}/\left<\mathcal{L}\right>$, Sect.~\ref{sec:dc-dt}), which automatically accounts for missing features for a given source. For multivariate methods using a fixed set of features on the other hand, one will need to invest time imputing the missing features before the classification can be performed.

The simplifying, and incorrect, assumption made in our algorithm (termed IND hereafter) is that all features are independently distributed. There are other algorithms, such as Kernel Density Estimator (KDE) and Isolation Forest (iForest) implemented in \texttt{scikit-learn} \citep{scikit-learn}, that do not make this simplifying assumption. By comparing these three algorithms, we investigate whether our simplification is a problem.

While we can, in some sense, define what an interesting astrophysical source is, many of those will not manifest themselves as outliers in a time-domain dataset, due to various effects (e.g., a distant supernova in a dusty environment can get swamped by nearby events such as Galactic novae/dwarf novae). To define what an outlier is, we need to employ some sort of mathematical definition and apply it on the data, in other words, apply an algorithm. In the following, we will use the Photometric LSST Astronomical Time Series Classification Challenge \citep[PLAsTiCC;][]{Kessler-2019} data set of realistic LSST-type simulated sources. We will use continous/persistent variables as our set of uninteresting/common sources and transients as astrophysically interesting sources.

We will use each of the three studied algorithms to select simulated transients with outlier-like characteristics semi-randomly as described below and then see how well the other algorithms fare at recovering them.

\subsection{Monte Carlo setup}\label{sec:MC}

PLAsTiCC used astrophysical models of various variable and transient sources, coupled with realistic volumetric rates for the transients and LSST observing strategy to simulate multiband ($u$, $g$, $r$, $i$, $z$, $y$) light curves of Galactic and extragalactic sources in the Southern sky as would be observed by LSST in three~years of operation. The project simulated more than 100 million sources for the LSST Wide-Fast-Deep (WFD) survey and ten-thousands of sources for the LSST Deep Drilling Field (DDF) mini survey. Rates for the Galactic models, which include RRLyr, ECL, Miras, flares from M-dwarf stars, and microlensing events, were selected arbitrarily such that they made up $10\%$ of the WFD PLAsTiCC sample, while the DDF sample contained 0.083\% of the number of Galactic sources in WFD. As compared to the WFD survey, where the cadence will be around 1-2 weeks, the DDF survey will have a better cadence of a few days \citep{Ivezic-2019}. In the PLAsTiCC simulation, the DDF temporal sampling was a factor $\sim 2.5$ better than the WFD. For time-domain studies, DDF is better suited than the WFD survey. We thus use the PLAsTiCC DDF dataset.

We work in magnitude units, whereby the differential fluxes in the PLAsTiCC dataset for the transients are directly converted to magnitudes using a zero-point of 27.5 for each passband. For the persistent variables, the subtracted fluxes are added before the conversion. We clean the data to remove bad measurements, i.e., those with large errors ($>0.1$~mag) in any given passband, and require sources to have measurements in two or more passbands and with a total of three or more data points. We use the persistent variables to represent the `common' population. They include AGN, ECL, RRLyr, and M-dwarf flares, with each class comprising a few hundred sources (there are only 10 Miras in the dataset, so we do not use them). The sample size of the common sources is more than 1000. On the other hand, the cleaned extragalactic transient dataset numbers more than 12000 sources and includes different types of supernovae, tidal disruption events, kilonovae, Calcium-rich transients, and intermediate-luminosity optical transients (cf.~\citealt{Kessler-2019}). We use this dataset to select transients with outlier-like characteristics relative to the variables.

We then perform Monte Carlo simulation following the steps enumerated below.

\begin{enumerate}
\item \label{step:training-test-split}We split the sample of persistent variables 2:1 into training dataset and test dataset. We make 10 such random splits.
  
\item Similar to the DECam data analysis, for each split, probability densities $p_{f}(dm|dt)$ for the five passbands $u$, $g$, $r$, $i$, $z$ (and 20 combinations $u-g$, $u-r$, etc.) are set up using the training dataset.
  
\item Of the simulated transient sample, we randomly select 1000 sources. We do this 1000 times for each of the ten splits described in Step~(\ref{step:training-test-split}).

\item \label{step:pick-transients}For the 1000 transient sources, we use KDE to rank their similarity to the variable sources randomly selected in Step~(\ref{step:training-test-split}). The input features for each source for the KDE (also for iForest) algorithm are the 25 likelihood scores from the (cross-)passbands $S_{f}=\log(\mathcal{L}_{f}/\left<\mathcal{L}_{f}\right>)$. In cases of missing features for sources, we impute them using KNNImputer implemented in \texttt{scikit-learn}, which substitutes the values of the missing features using the mean values from K (which we choose to be three) nearest neighbors in the transient subsample. The distance between a pair of samples is evaluated using only features that both samples have in common.

We pick the five transient sources with the least similarity to our variable-source training data, according to KDE, and add them to the test sample of Step~(\ref{step:training-test-split}).

\item Now we apply all three algorithms IND, KDE, and iForest to select outliers from the complete test sample (1/3 of the variable sources plus five semi-randomly selected transients) and see which sources are flagged by each of the algorithms.
\end{enumerate}

We perform two additional sets of simulation in the manner described above using iForest and IND to select transients in Step~(\ref{step:pick-transients}).

\begin{figure}
\includegraphics[width=88mm]{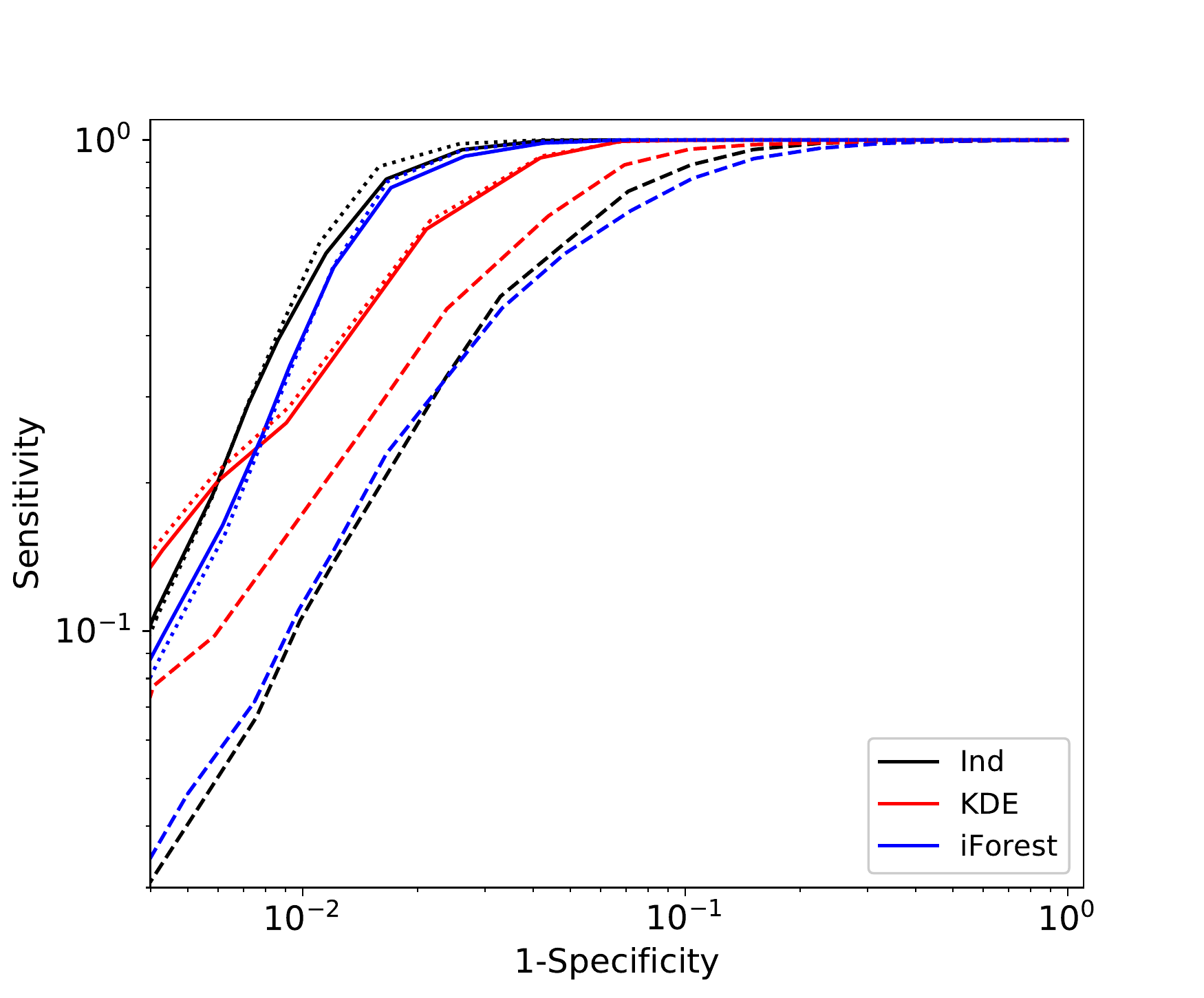}
\caption{ROC curves for the three outlier-detection algorithms (IND, KDE, and iForest) applied to the PLAsTiCC LSST DDF dataset. The threshold for flagging outliers is varied along the curves. The solid lines are the results for recovery of iForest-selected transients, dotted lines for IND-selected transients, and dashed lines for KDE-selected transients (see text). \label{fig:rocs}}
\end{figure}

\subsection{Results}
For each set of Monte Carlo simulation, we compute the sensitivity, i.e., the ratio of true positives (ingested transients flagged as outliers) to the total number of ingested transients, and $1-\textrm{specificity}$, i.e., the ratio of false positives (variable sources flagged as outliers) to the total number of variable sources in the test sample, at different decision rules. The decision rules comprise different percentiles (for IND and KDE) or contamination (for iForest) thresholds, which we take to be between 0.1 and 100 in logspace. Finally, for each threshold, we take the average of the $10\times 1000$ sensitivity values for a given algorithm as its final sensitivity, and similarly for the specificity values to compute its final specificity, at the given threshold. The resulting Receiver Operating Characteristic (ROC) curves for the three algorithms for each set of simulation are shown in Fig.~\ref{fig:rocs}.    

As is evident from the plot, the results of the three algorithms for the iForest-selected and IND-selected transients (cf.~Step~\ref{step:pick-transients} in Sect.~\ref{sec:MC}) are similar. The area under the ROC curve (AUC) is 98.5\% (98.7\%) for IND, 98.0\% (98.0\%) for iForest, and 97.7\% (97.8\%) for KDE for the iForest-selected (IND-selected) transients. The performance drops slightly for all three algorithms when using the KDE-selected transients, with AUC values of 94.3\%, 92.8\%, and 95.6\%, respectively for IND, iForest and KDE.

Our algorithm thus performs very similar to the other two tested algorithms. Furthermore, for a reasonable threshold that flags 10-30 outliers, IND has a sensitivity greater than 80-90\%. Despite the simplicity of our algorithm, it is highly competitive with the other multivariate methods.

\section{Summary and Future work} \label{sec:summary}
Using the extinction-corrected DECam multi-band ($u$, $g$, $r$, $i$, and $z$) time-series data of 2266 variable stars identified in Baade's Window by \citet{Saha-2019}, we have developed a statistically motivated algorithm for identifying novelties or unusual events within a given population of variable stars and transients. It relies on features that are computationally inexpensive, specifically the probability density distributions $p_f(dm|dt)$ of magnitude differences ($dm$) over a given time interval $dt$ for the passband $f$, and the normalized likelihood (or score) for a test source to belong to the overall population (Sect.~\ref{sec:method}), computed based on these distributions.  We categorize the test sources with the lowest scores as the most unusual ones from the bulk. The threshold score for the categorization can be tuned according to the capacity for analysis or even follow-up in the case of real-time applications.

The DECam data set used in this study is dominated by long-period variables (Miras, semi-regular variables, etc.), pulsating stars of different types, such as RR~Lyrae, Delta Scuti stars, etc. and eclipsing binaries. This is confirmed by cross-identification with the classified variable stars in the same field from the OGLE survey for which we obtain matches for around half of our sample size with a conservative search radius of $1''$.  Applying our algorithm, we identified 18 peculiar/outlier sources from the remaining subsample of more than 1000 variables and transients without OGLE-matches, for a threshold score value corresponding to the lower 2nd percentile of the score distribution. We have demonstrated that the flagged sources are indeed outliers in the CMD, for example, being located at the extreme blue and red edges of the CMD. Among others, our outlier set includes sources in the Blue Horizontal Branch region of the CMD without any known counterparts, chromospherically active RS~CVn stars, a microlensing event, and a dwarf nova, confirmed by other authors, which are indeed rare sources for the given population of variable stars in terms of their numbers and rates.

The availability of multi-band information for the data set used in this study and its subsequent incorporation in our algorithm enhances the efficacy of the latter. Furthermore, the characteristics of the data set, particularly the multiple passbands, are similar to those expected from LSST and thus the present study lays the groundwork for an efficient identification of peculiar sources in a given population of variable stars and transients. 

In the future, we plan to expand our analysis to variables in different host environments. We are gathering mutli-band time-series data spanning two years for two other interesting nearby galaxies, M83 and Centaurus~A, with a better temporal sampling than that of the Galactic Bulge data, which were originally designed for the discovery of RR~Lyrae. We will also use the M83 and Centaurus~A data to assess the performance of our algorithm at different time baselines covering less complete light curves of the test sources. In the immediate future, we plan to deploy our algorithm in the ANTARES broker for real-time processing of the ZTF public alert data by computing the distributions $p_f(dm|dt)$ using ZTF archival data of variable stars.

Our procedure is promising for harvesting interesting and novel variable phenomena after the light curve of the source has been populated to some extent. The novelties may include interesting new less-common variable stars, or relatively long-duration transients. We also plan to investigate other techniques to exploit correlated variation patterns to make predictions at early time.

\acknowledgments
{\it Acknowledgments}: We thank the referee for suggestions, which have helped improve the paper. MDS is supported by the Illinois Survey Science Fellowship of the Center for Astrophysical Surveys at the University of Illinois at Urbana-Champaign.   
This project used data obtained with the Dark Energy Camera (DECam), which was constructed by the Dark Energy Survey (DES) collaboration. Funding for the DES Projects has been provided by the U.S. Department of Energy, the U.S. National Science Foundation, the Ministry of Science and Education of Spain, the Science and Technology Facilities Council of the United Kingdom, the Higher Education Funding Council for England, the National Center for Supercomputing Applications at the University of Illinois at Urbana-Champaign, the Kavli Institute of Cosmological Physics at the University of Chicago, the Center for Cosmology and Astro-Particle Physics at the Ohio State University, the Mitchell Institute for Fundamental Physics and Astronomy at Texas A\&M University, Financiadora de Estudos e Projetos, Funda{\c c}{\~a}o Carlos Chagas Filho de Amparo {\`a} Pesquisa do Estado do Rio de Janeiro, Conselho Nacional de Desenvolvimento Cient{\'i}fico e Tecnol{\'o}gico and the Minist{\'e}rio da Ci{\^e}ncia, Tecnologia e Inovac{\~a}o, the Deutsche Forschungsgemeinschaft, and the Collaborating Institutions in the Dark Energy Survey. 

The Collaborating Institutions are Argonne National Laboratory, the University of California at Santa Cruz, the University of Cambridge, Centro de Investigaciones En{\'e}rgeticas, Medioambientales y Tecnol{\'o}gicas-Madrid, the University of Chicago, University College London, the DES-Brazil Consortium, the University of Edinburgh, the Eidgen{\"o}ssische Technische Hoch\-schule (ETH) Z{\"u}rich, Fermi National Accelerator Laboratory, the University of Illinois at Urbana-Champaign, the Institut de Ci{\`e}ncies de l'Espai (IEEC/CSIC), the Institut de F{\'i}sica d'Altes Energies, Lawrence Berkeley National Laboratory, the Ludwig-Maximilians Universit{\"a}t M{\"u}nchen and the associated Excellence Cluster Universe, the University of Michigan, {the} National Optical Astronomy Observatory, the University of Nottingham, the Ohio State University, the OzDES Membership Consortium, the University of Pennsylvania, the University of Portsmouth, SLAC National Accelerator Laboratory, Stanford University, the University of Sussex, and Texas A\&M University.

Based on observations at Cerro Tololo Inter-American Observatory, National Optical Astronomy Observatory (2013A-0719; PI: A.~Saha), which is operated by the Association of Universities for Research in Astronomy (AURA) under a cooperative agreement with the National Science Foundation.

This research has made use of the SIMBAD database, operated at CDS, Strasbourg, France. This research has made use of the VizieR catalogue access tool, CDS, Strasbourg, France (DOI : 10.26093/cds/vizier). The original description  of the VizieR service was published in A\&AS 143, 23. EO is partially supported by NSF grant AST-1815767.

\software{\texttt{numpy} \citep{numpy}, \texttt{scipy} \citep{scipy}, \texttt{astropy} \citep{astropy}, \texttt{matplotlib} \citep{matplotlib}, \texttt{scikit-learn} \citep{scikit-learn}.}

\bibliography{references}

\appendix
\section{{Probability distributions \MakeLowercase{{\it p}({\it dm}}$\vert$\MakeLowercase{{\it dt})} for different classes of variable stars in the Galactic Bulge sample}}\label{append}

\subsection{Eclipsing Binaries}
\begin{figure*}[h]
\includegraphics[width=60mm]{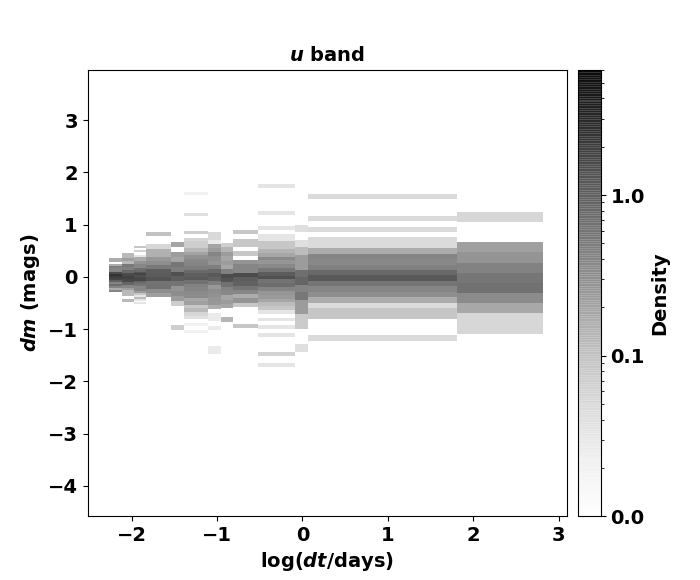}\hfill\includegraphics[width=60mm]{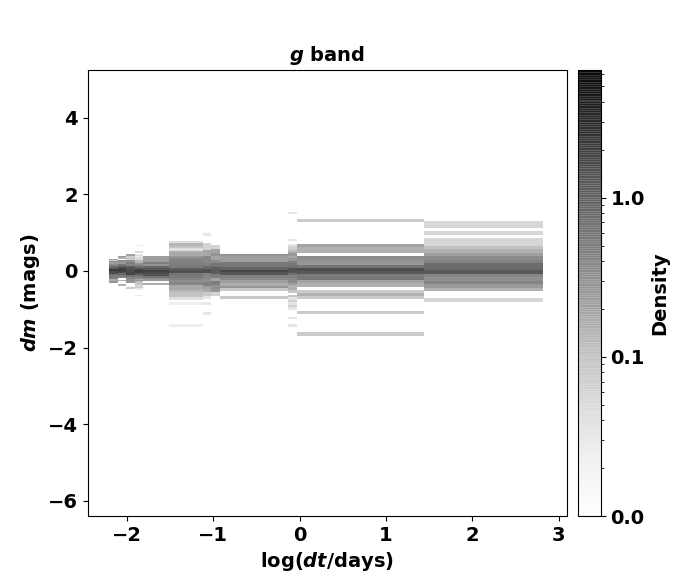}\hfill\includegraphics[width=60mm]{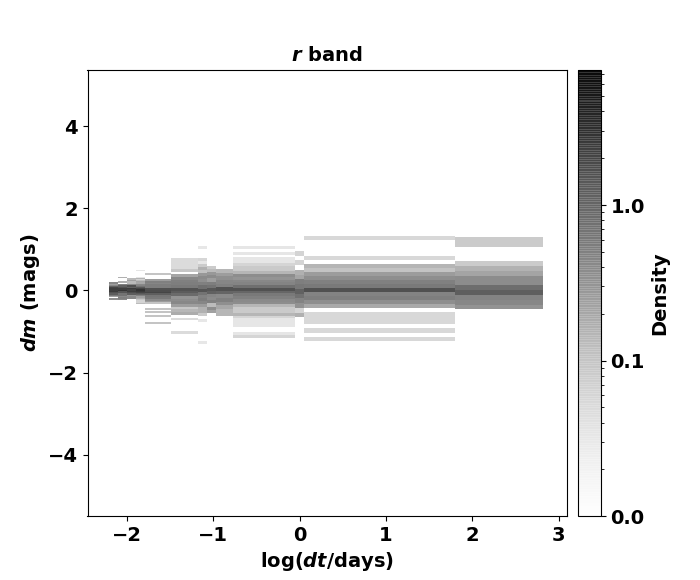}\\
\includegraphics[width=60mm]{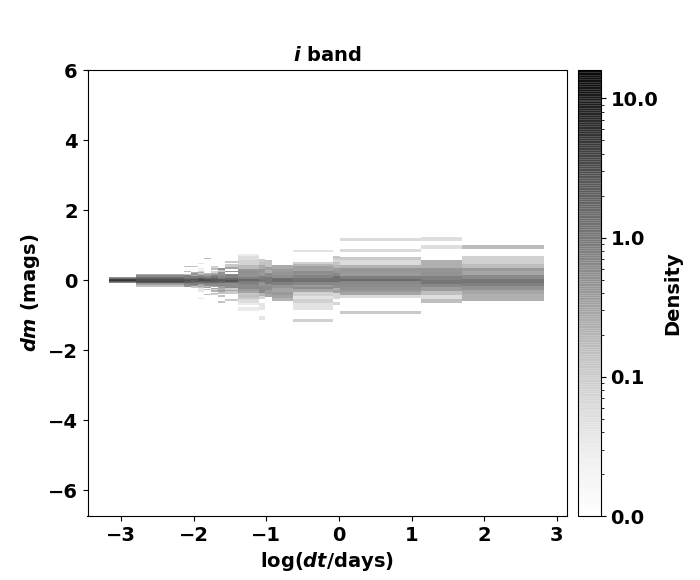}\hfill\includegraphics[width=60mm]{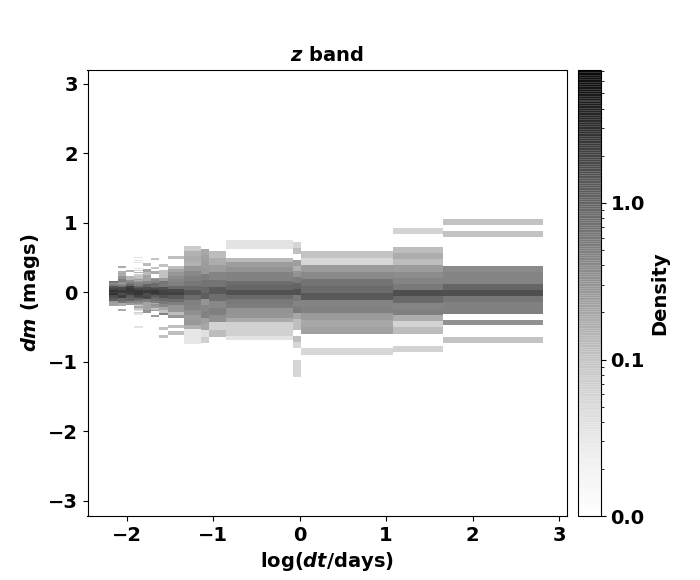}\hfill\includegraphics[width=60mm]{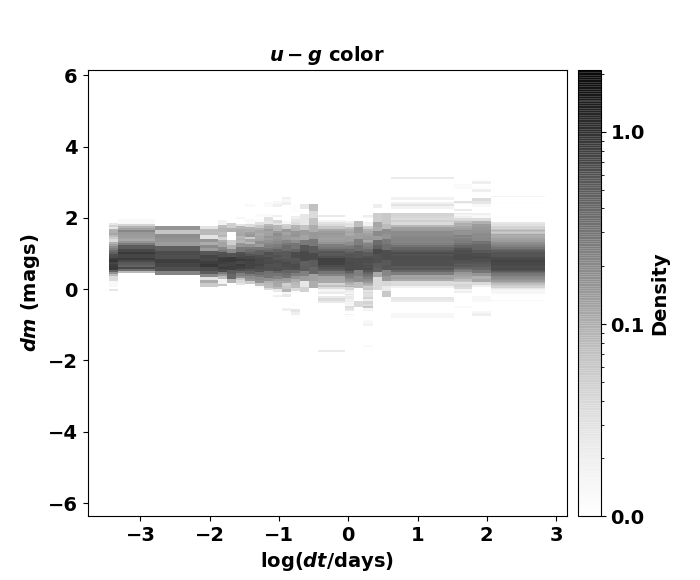}\\
\includegraphics[width=60mm]{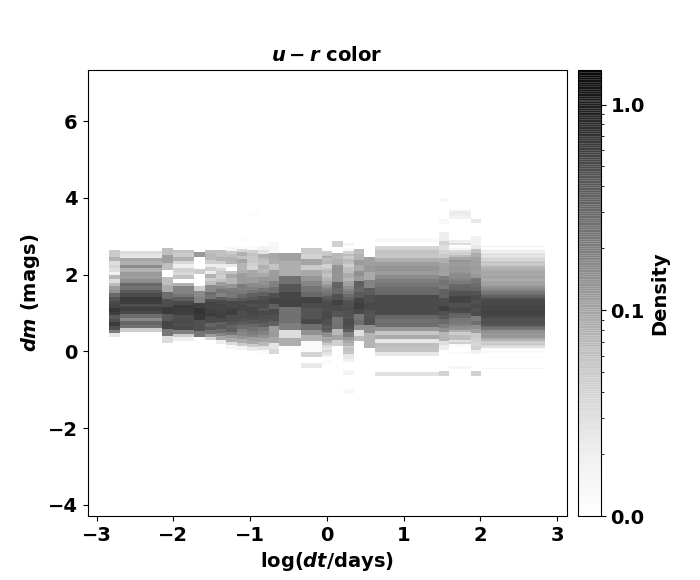}\hfill\includegraphics[width=60mm]{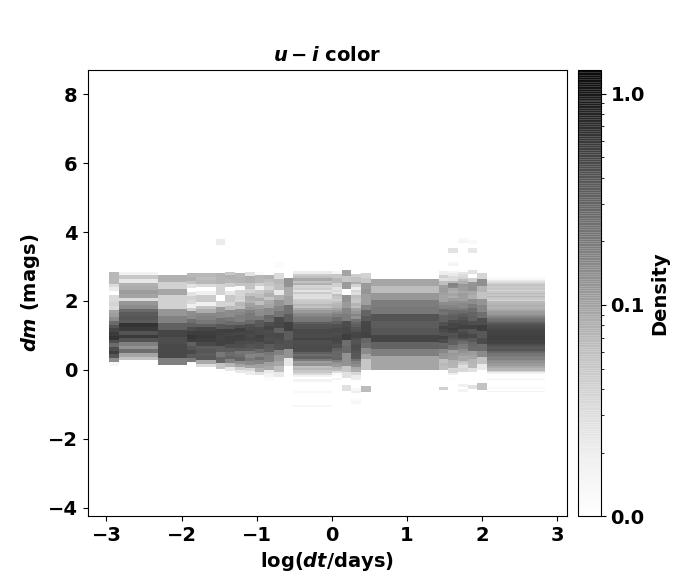}\hfill\includegraphics[width=60mm]{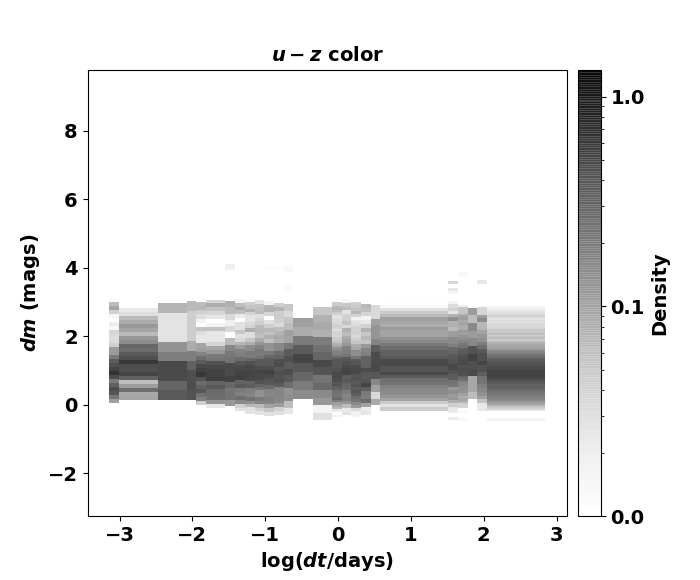}
\caption{Similar plots as Fig.~\ref{fig:dmdt_all}, but for sources that are labeled as ECL in OGLE.}\label{fig:dmdt_ecl}
\end{figure*}

\begin{figure*}
\ContinuedFloat
\includegraphics[width=60mm]{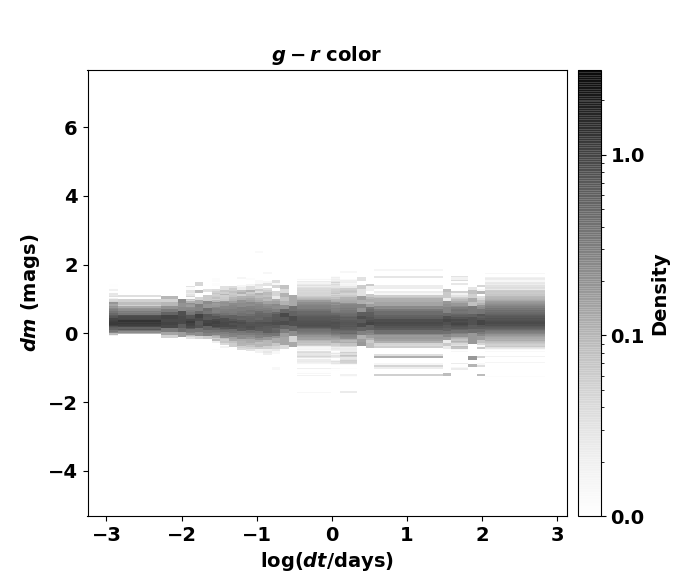}\hfill\includegraphics[width=60mm]{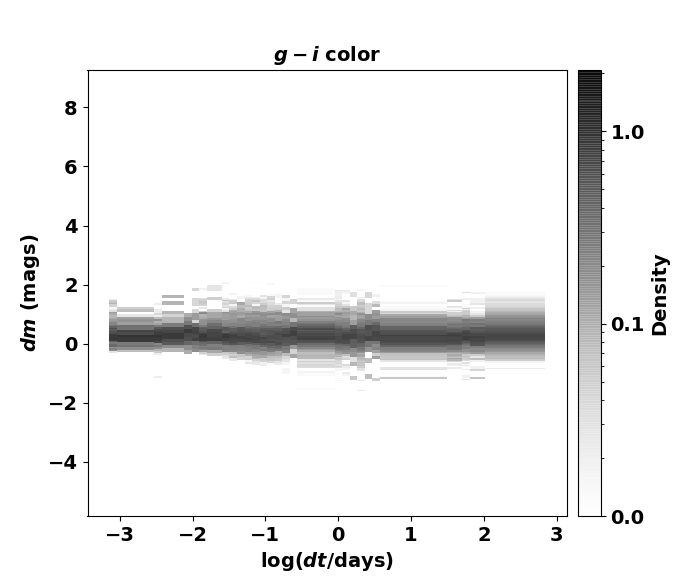}\hfill\includegraphics[width=60mm]{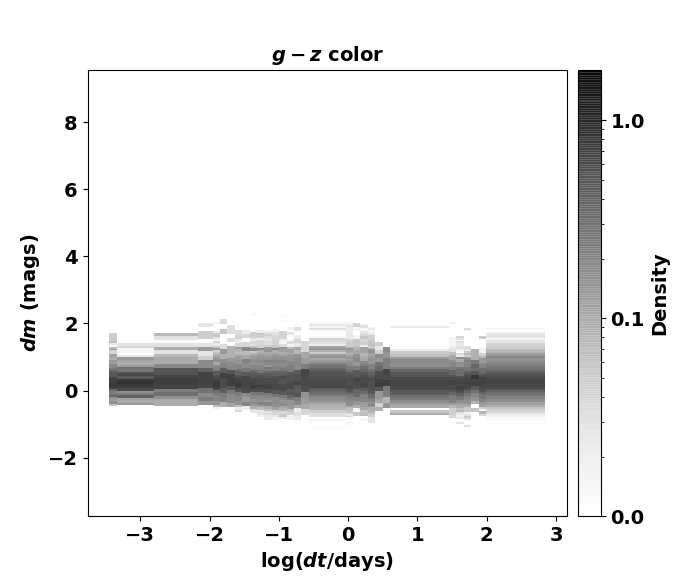}\\
\includegraphics[width=60mm]{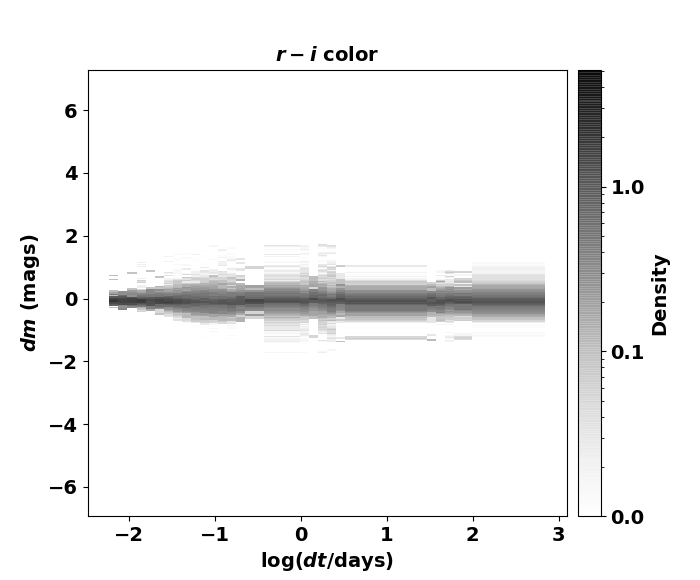}\hfill\includegraphics[width=60mm]{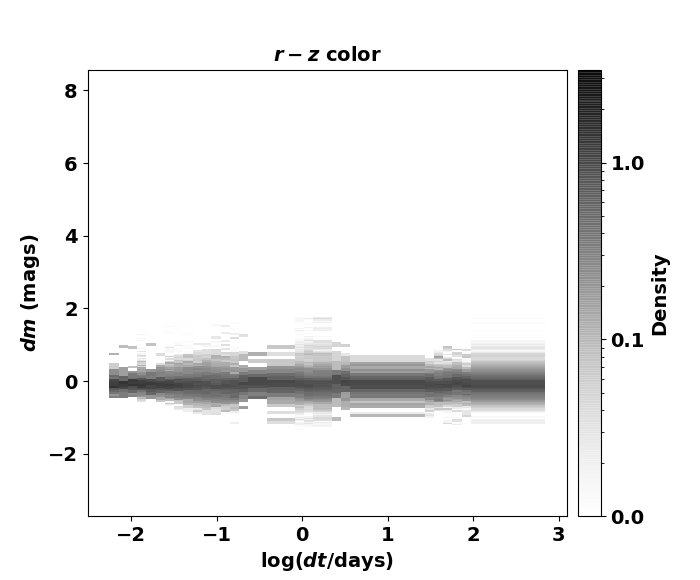}\hfill\includegraphics[width=60mm]{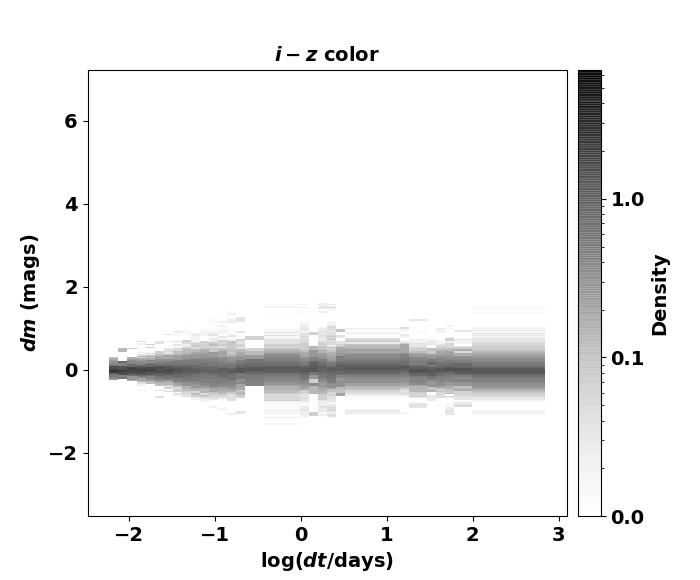}
\caption{Contd.}
\end{figure*}

\clearpage
\subsection{Long Period Variables}
\begin{figure*}[h]
\includegraphics[width=60mm]{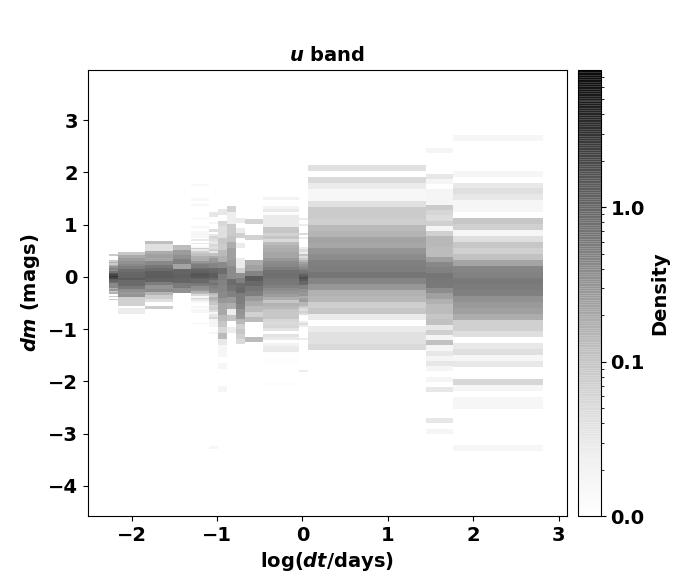}\hfill\includegraphics[width=60mm]{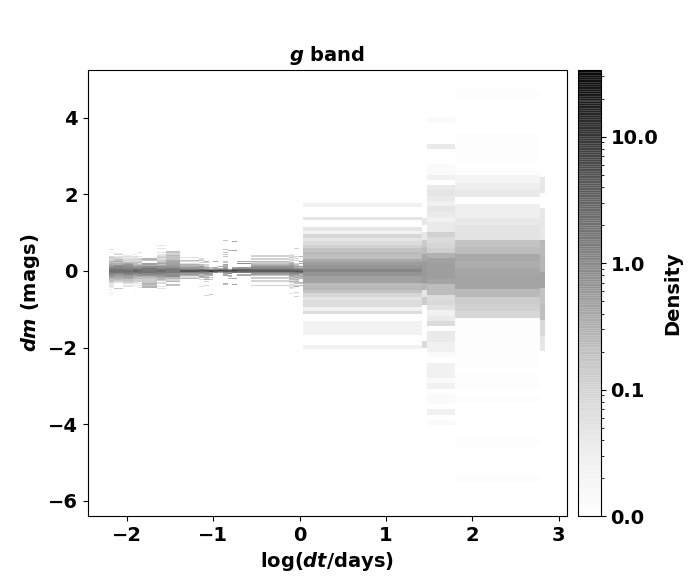}\hfill\includegraphics[width=60mm]{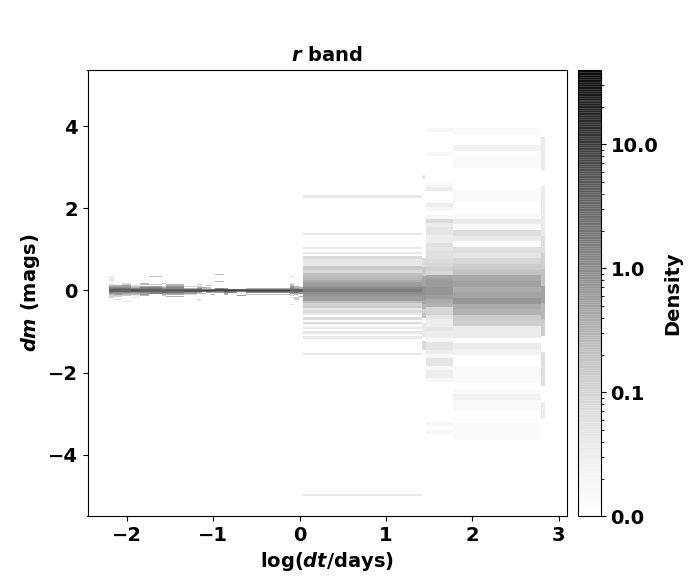}\\
\includegraphics[width=60mm]{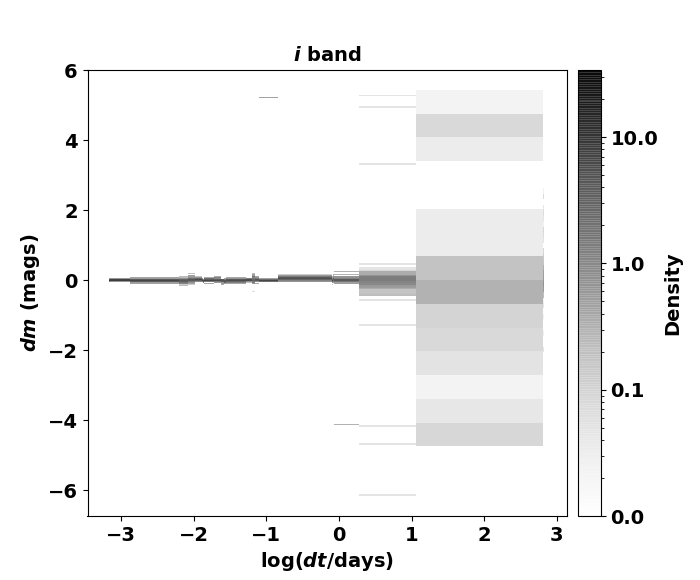}\hfill\includegraphics[width=60mm]{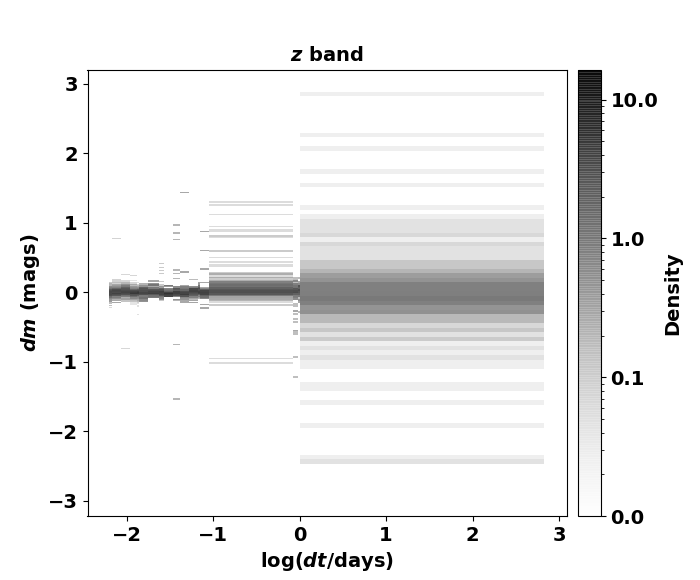}\hfill\includegraphics[width=60mm]{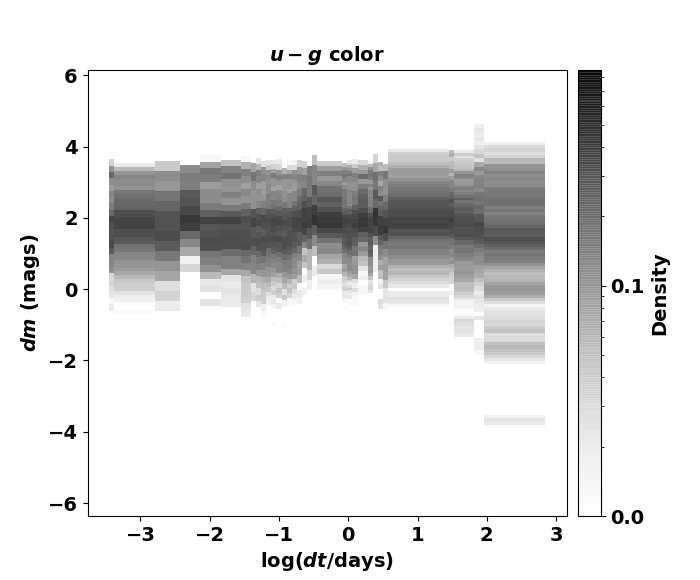}\\
\includegraphics[width=60mm]{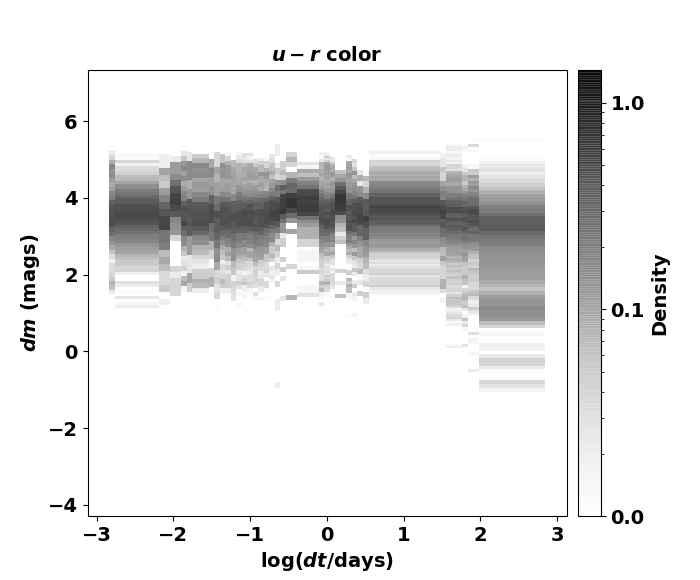}\hfill\includegraphics[width=60mm]{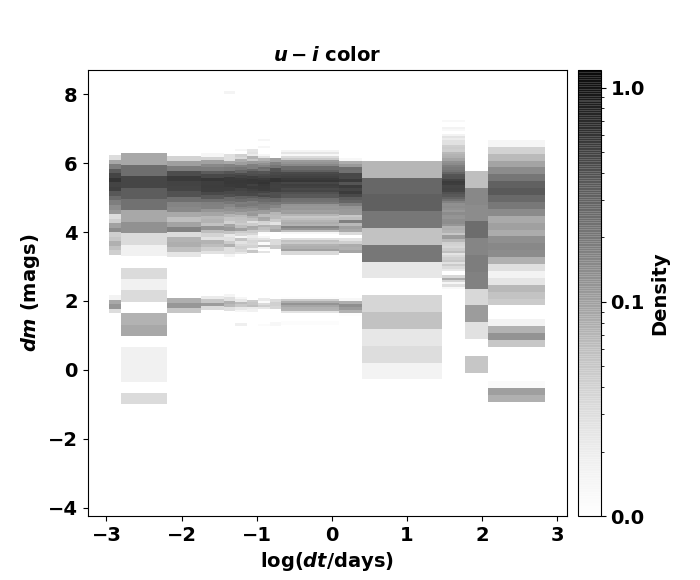}\hfill\includegraphics[width=60mm]{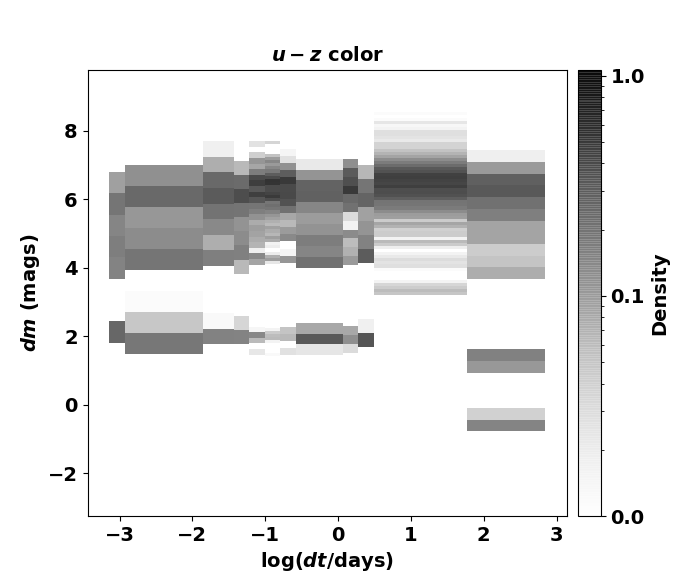}
\caption{Similar plots as Fig.~\ref{fig:dmdt_all}, but for sources that are labeled as LPV in OGLE. 
}\label{fig:dmdt_lpv}
\end{figure*}

\begin{figure*}[h]
\ContinuedFloat
\includegraphics[width=60mm]{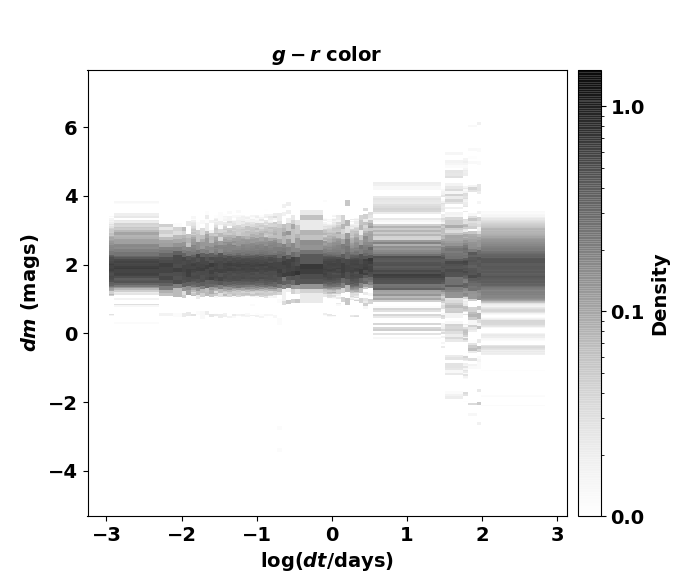}\hfill\includegraphics[width=60mm]{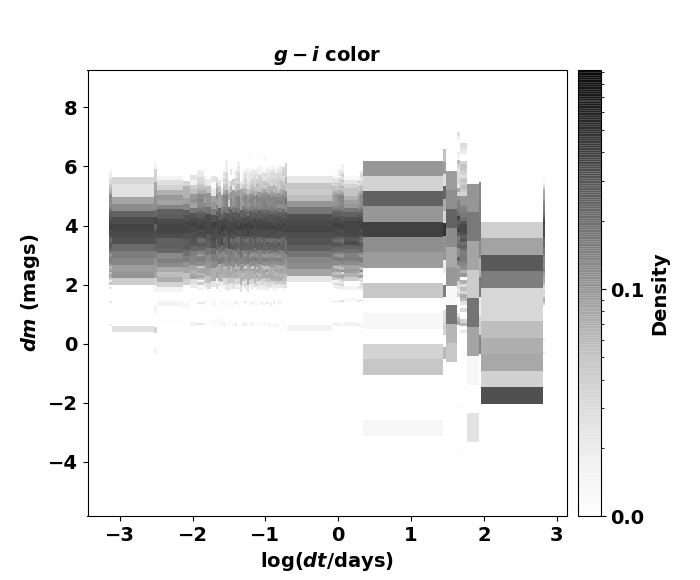}\hfill\includegraphics[width=60mm]{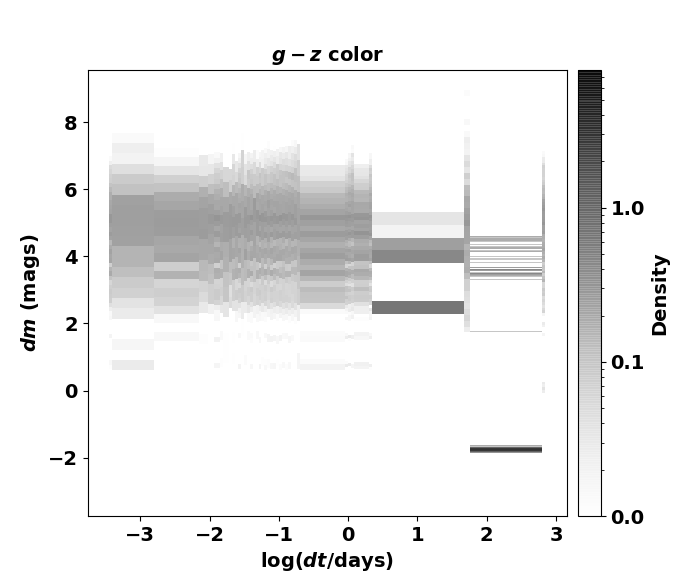}
\includegraphics[width=60mm]{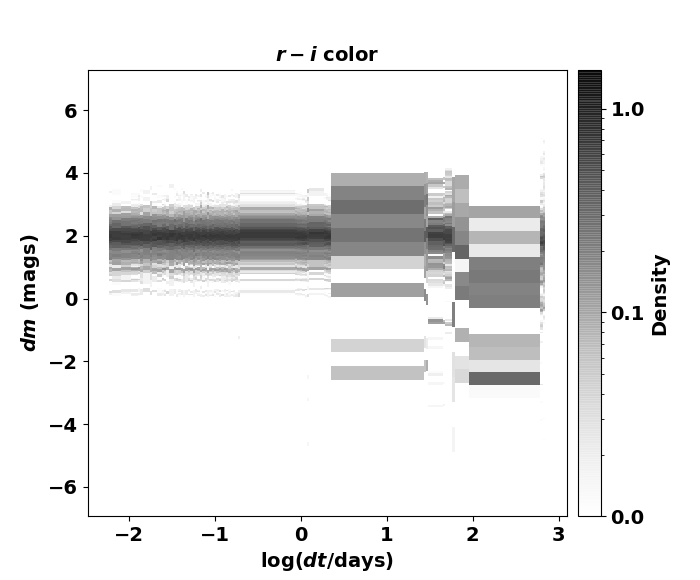}\hfill\includegraphics[width=60mm]{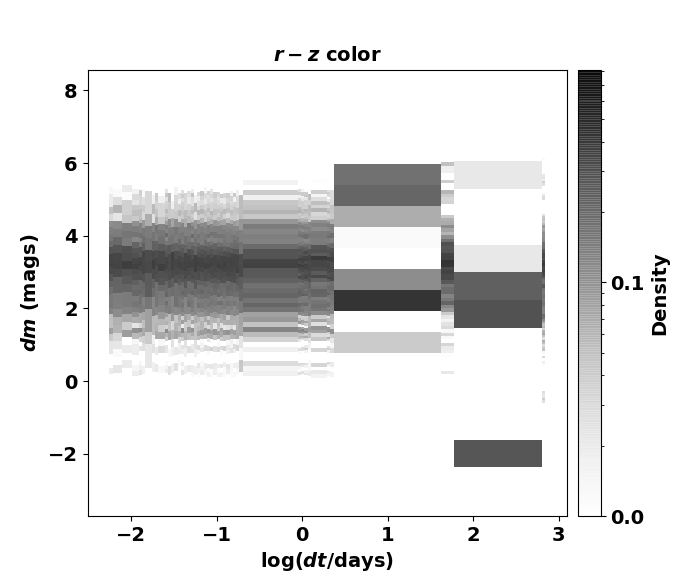}\hfill\includegraphics[width=60mm]{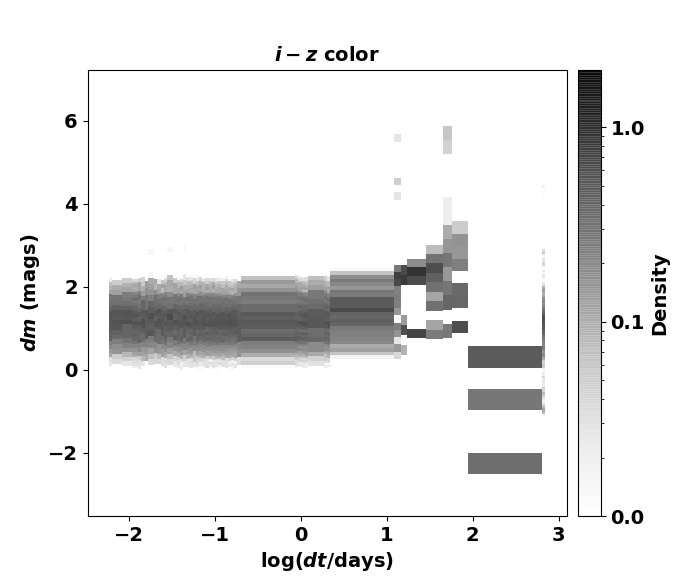}
\caption{Contd.}
\end{figure*}

\clearpage
\subsection{RR Lyrae}
\begin{figure*}[h]
\includegraphics[width=60mm]{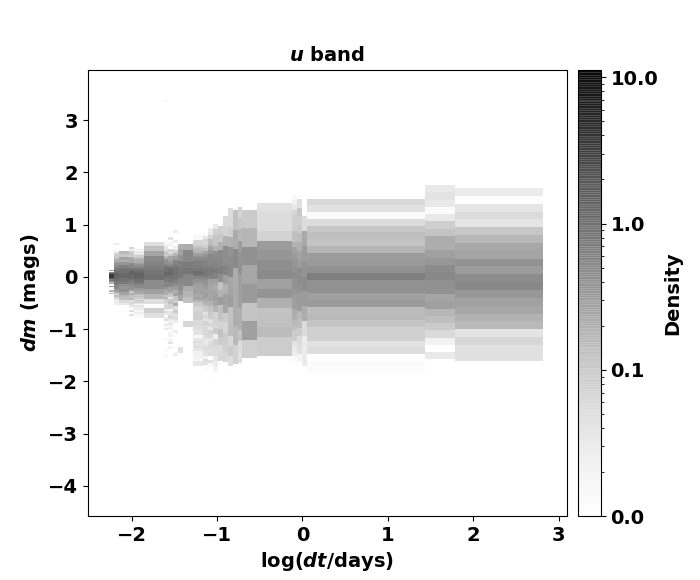}\hfill\includegraphics[width=60mm]{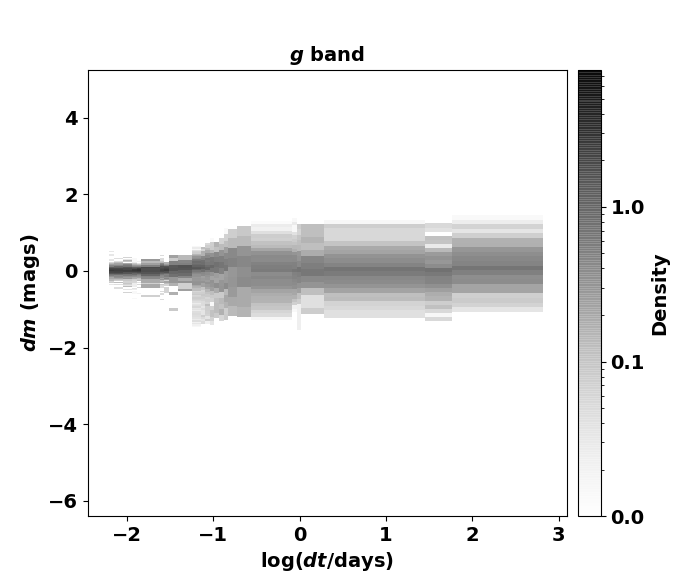}\hfill\includegraphics[width=60mm]{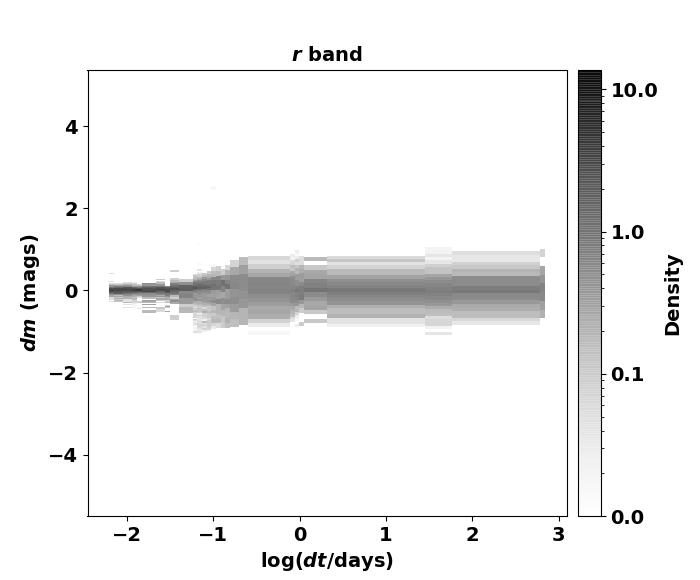}\\
\includegraphics[width=60mm]{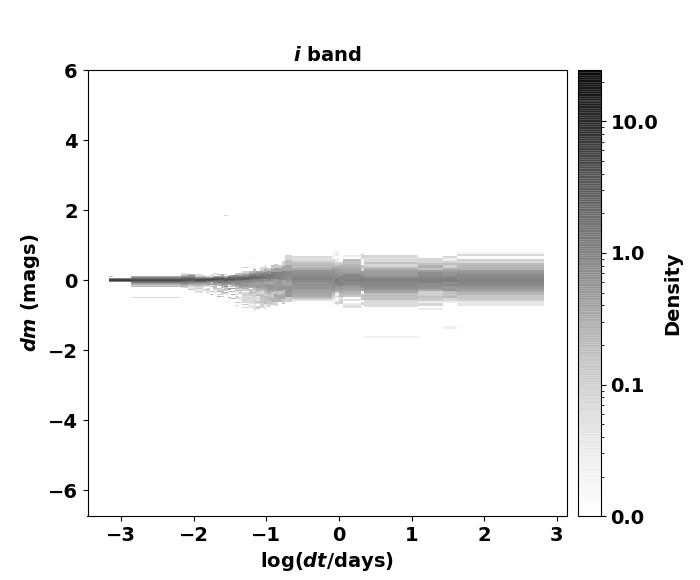}\hfill\includegraphics[width=60mm]{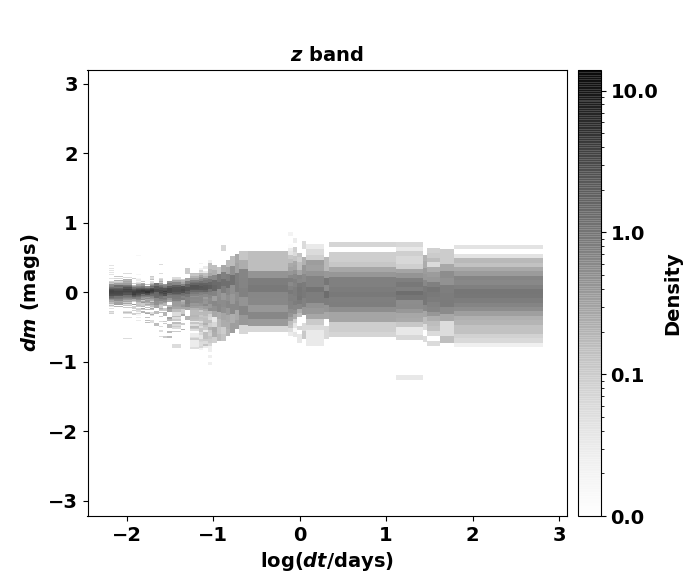}\hfill\includegraphics[width=60mm]{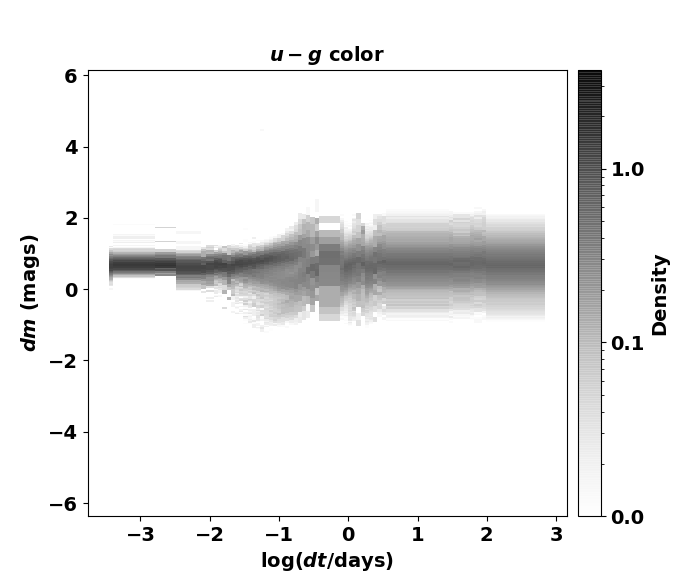}\\
\includegraphics[width=60mm]{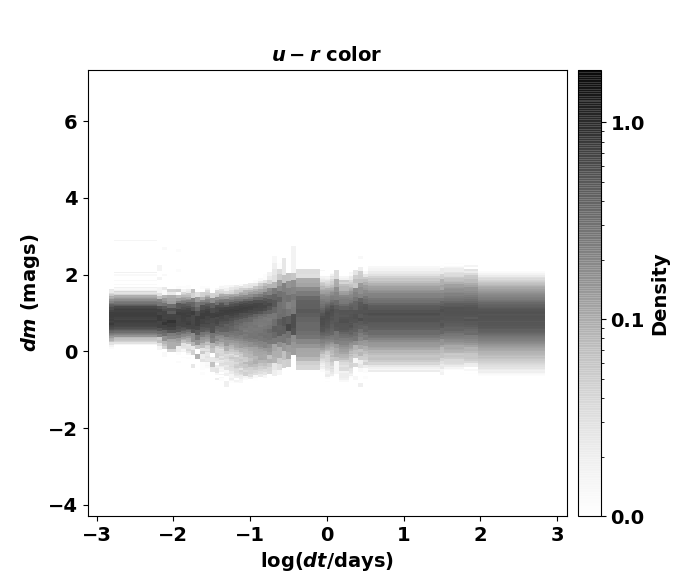}\hfill\includegraphics[width=60mm]{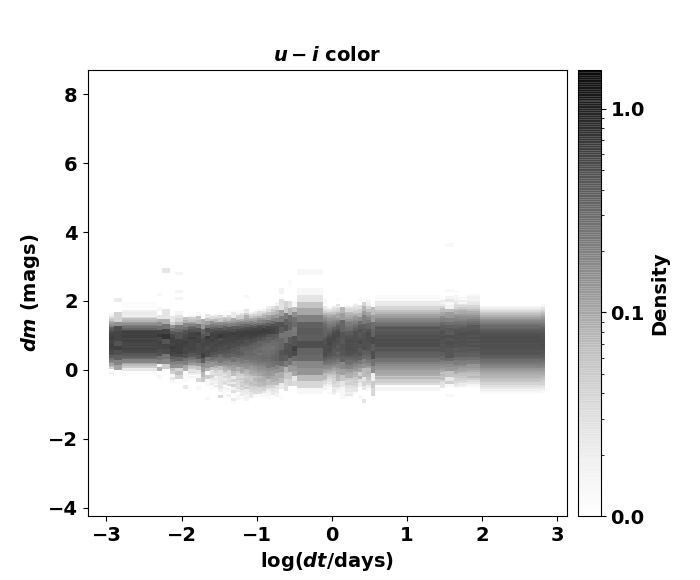}\hfill\includegraphics[width=60mm]{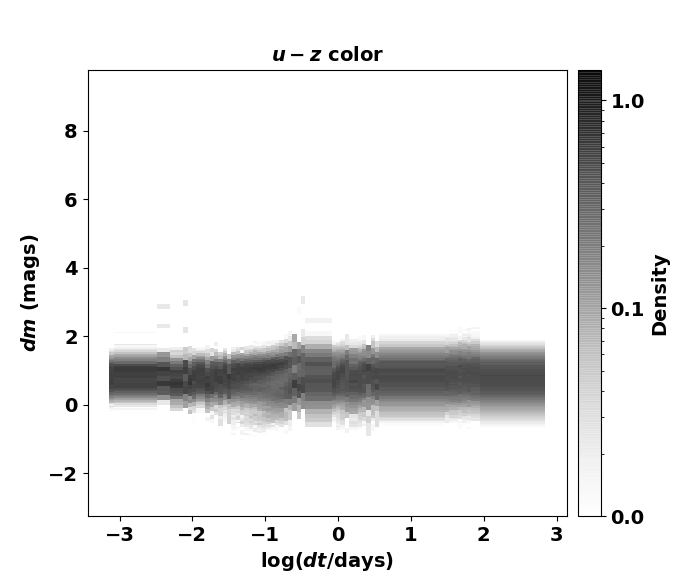}
\caption{Similar plots as Fig.~\ref{fig:dmdt_all}, but for sources that are labeled as RRLyr in OGLE.}\label{fig:dmdt_rrlyr}
\end{figure*}

\begin{figure*}
\ContinuedFloat
\includegraphics[width=60mm]{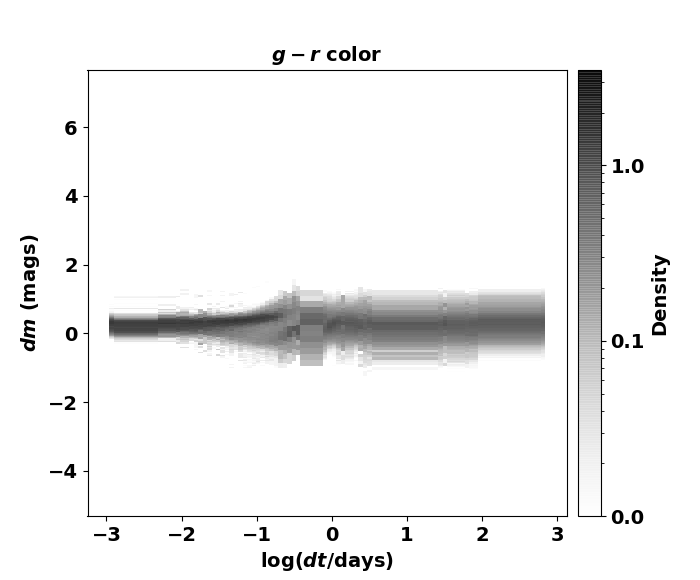}\hfill\includegraphics[width=60mm]{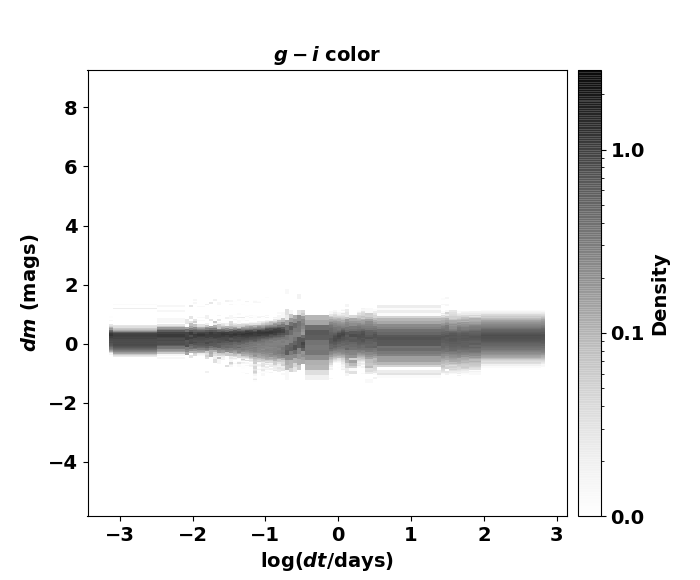}\hfill\includegraphics[width=60mm]{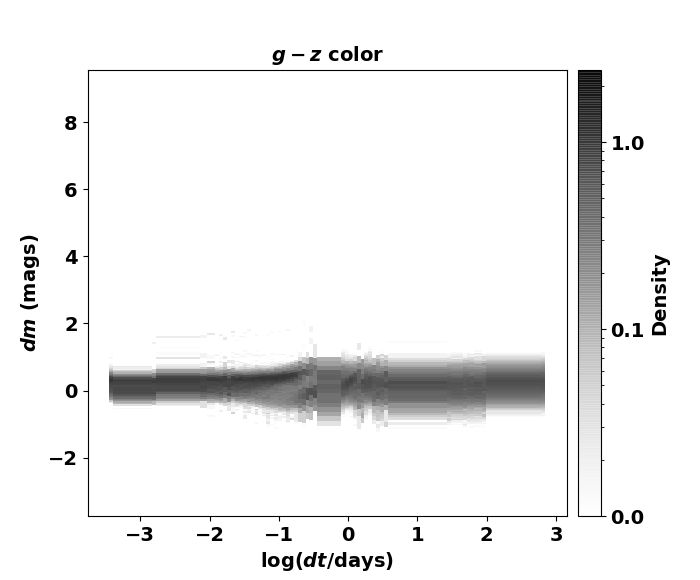}\\
\includegraphics[width=60mm]{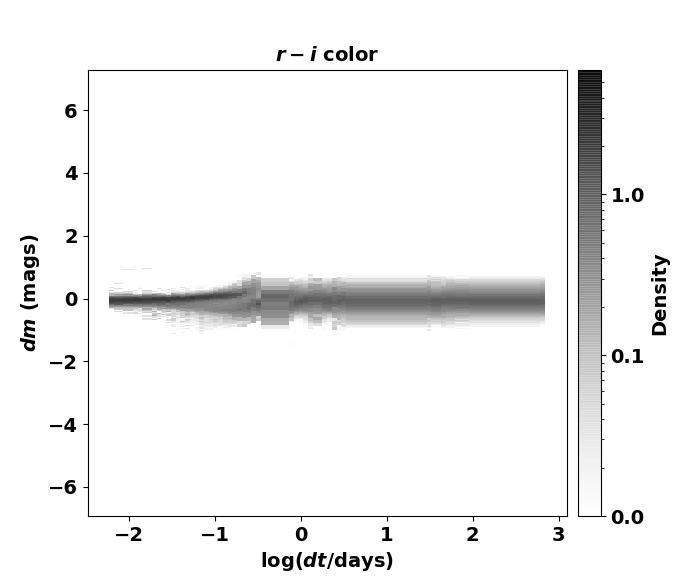}\hfill\includegraphics[width=60mm]{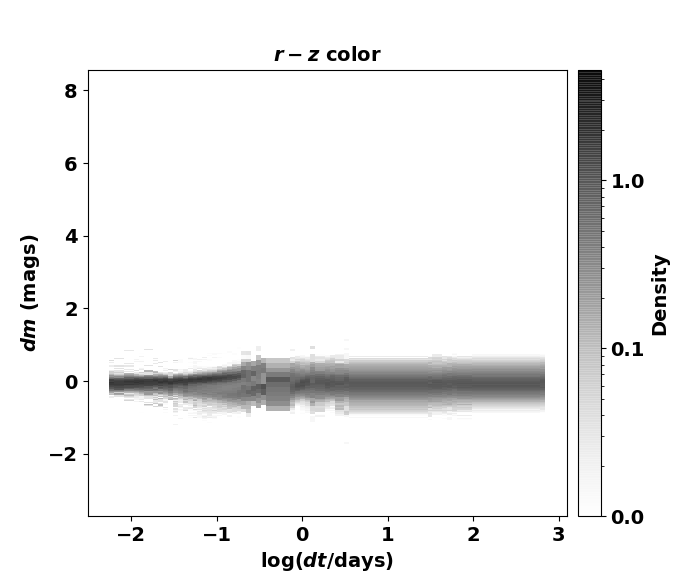}\hfill\includegraphics[width=60mm]{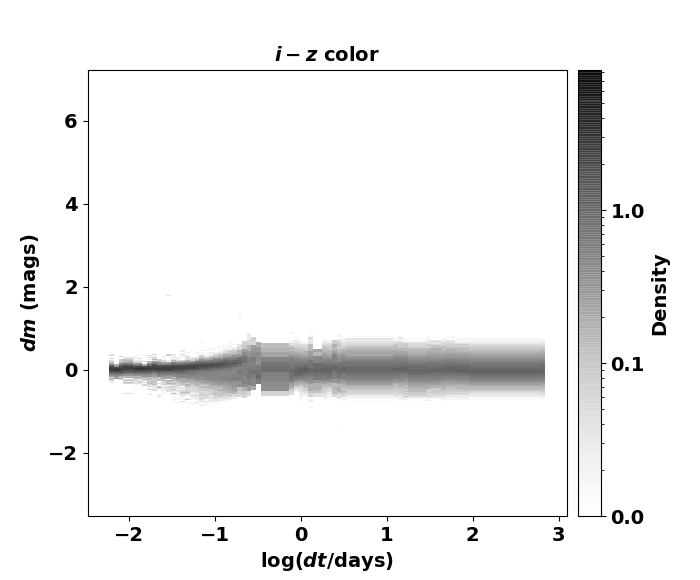}
\caption{Contd.}
\end{figure*}

\end{document}